\documentclass[11pt]{article}
\pdfoutput=1
\usepackage{epsfig,graphicx}
\usepackage{amsmath,amsfonts,amssymb} 
\makeatletter
\def\maketag@@@#1{\hbox{\m@th\normalfont\normalsize#1}}
\makeatother
\usepackage{color}
\usepackage{cite}
\usepackage{slashed}
\usepackage{hyperref}
\usepackage{caption}
\usepackage{subcaption}
\usepackage{graphicx}
\usepackage[normalem]{ulem}
\usepackage{soul}
\usepackage{float}
\usepackage{cancel}

 \usepackage[section]{placeins}
\usepackage{afterpage}

\newcommand{\unit}{\leavevmode\hbox{\small1\kern-3.6pt\normalsize1}}

\def\lsim{\raise0.3ex\hbox{$\;<$\kern-0.75em\raise-1.1ex\hbox{$\sim\;$}}}
\def\gsim{\raise0.3ex\hbox{$\;>$\kern-0.75em\raise-1.1ex\hbox{$\sim\;$}}}

\allowdisplaybreaks

\definecolor{vmlorange}{rgb}{1.0, 0.49, 0.0}

\definecolor{vmlorange}{rgb}{1.0, 0.49, 0.0}
\definecolor{emerald}{rgb}{0.2,0.5,0.1}

\definecolor{tobycolour}{rgb}{.6,.0,.4}

\definecolor{mkgreen}{rgb}{0.2,.70,.3}

\definecolor{mkblue}{rgb}{0.2,0.2,0.7}

\begin{document}

\thispagestyle{empty}
\begin{flushright}
  BONN-TH-2018-09\\ 
  IFT-UAM/CSIC-18-103\\ 

\vspace*{2.mm} \today
\end{flushright}

\newcommand{\AddrBonn}{%
Bethe Center for Theoretical Physics \& Physikalisches Institut der 
Universit\"at Bonn,\\ Nu{\ss}allee 12, 
 53115 Bonn, Germany
}

\newcommand{\AddrSA}{
National Institute for Theoretical Physics,\\
School of Physics and Mandelstam Institute for Theoretical Physics,\\
University of the Witwatersrand, Johannesburg,
Wits 2050, South Africa  
}

\newcommand{\AddrMadrid}{
Instituto de F\'isica Te\'orica (UAM/CSIC), Universidad Aut\'onoma de Madrid,\\
Cantoblanco, E--28049 Madrid, Spain
}

\newcommand{\AddrSantander}{
Instituto de F\'isica de Cantabria (CSIC-UC),
E--39005 Santander, Spain
}

\begin{center}
  {\Large \textbf{
  Updating Bounds on \boldmath $R$-Parity Violating Supersymmetry from Meson Oscillation Data}
  }  
  
  \vspace{0.5cm}
      Florian~Domingo${}^{a,b,c}$\footnote{florian.domingo@csic.es}, Herbert~K.~Dreiner${}^{a}$\footnote{dreiner@uni-bonn.de}, Jong~Soo~Kim${}^{d}$\footnote{jongsoo.kim@tu-dortmund.de}, Manuel~E.~Krauss${}^{a}$\footnote{mkrauss@th.physik.uni-bonn.de}, V\'ictor~Mart\'in~Lozano${}^{a}$\footnote{lozano@th.physik.uni-bonn.de} and Zeren~Simon~Wang${}^{a}$\footnote{wzeren@physik.uni-bonn.de}\\[0.2cm] 
      
    {\small \textit{ 
      ${}^a$ 
      \AddrBonn\\[0pt] 
      }} 
  {\small \textit{ 
      ${}^b$ 
      \AddrMadrid\\[0pt] 
      }}
  {\small \textit{ 
      ${}^c$ 
      \AddrSantander\\[0pt] 
      }}
   {\small \textit{ 
      ${}^d$ 
      \AddrSA\\[0pt] 
      }} 
\vspace*{0.7cm}

\begin{abstract}
We update the bounds on $R$-parity violating supersymmetry originating from meson oscillations in the $B^0_{d/s}$ and $K^0$ systems. 
To this end, we explicitly calculate all corresponding contributions from $R$-parity violating operators  at the one-loop level, thereby completing 
and correcting existing calculations. We apply our results to the derivation of bounds on $R$-parity violating couplings, based on up-to-date 
experimental measurements. In addition, we consider the possibility of cancellations among flavor-changing contributions of various origins, 
$\textit{e.g.}$ from multiple $R$-parity violating couplings or $R$-parity conserving soft terms. Destructive interferences among new-physics contributions 
could then open phenomenologically allowed regions, for values of the parameters that are naively excluded when the parameters are varied 
individually.
\end{abstract}

\end{center}

\newpage
	

\section{Introduction}\label{sec:intro}

Several years of operation of the LHC have (as yet) failed to reveal any conclusive evidence for physics beyond the Standard Model (SM) 
\cite{Ichep:2018pdf}. On the contrary, experimental searches keep placing ever stronger limits on  hypothesized  strongly 
\cite{atlasSUS,cmsSUS,atlasEXO,cmsEXO} and even weakly-interacting \cite{Aaboud:2018jiw} particles in the electroweak--TeV range. 
While this situation tends to leave the simpler models in an uncomfortable position, for the so-called ``CMSSM" see for example 
Ref.~\cite{Bechtle:2015nua}, it also advocates for a deeper study of more complicated scenarios, satisfying the central motivations of the 
original paradigm but also requiring more elaborate experimental investigations for testing.
 
Softly-broken supersymmetric (SUSY) extensions of the SM \cite{Nilles:1983ge,Haber:1984rc} have long been regarded as a leading class 
of candidates for the resolution of the hierarchy problem \cite{Gildener:1976ai}, as well as a possible framework in view of understanding the nature of dark 
matter or the unification of gauge-couplings. The simplest of such models, the Minimal Supersymmetric Standard Model (MSSM), has thus 
been the focus of numerous studies in the past decades. An implicit ingredient of the usual MSSM is $R$-parity ($R_p$) \cite{Farrar:1978xj}, 
a discrete symmetry related to baryon and lepton number. In addition to the preservation of these quantum numbers, $R_p$ is also invoked 
in order to justify the stability of the lightest SUSY particle, leaving it in a position of a dark-matter candidate \cite{Jungman:1995df}.

Despite its attractive features, $R_p$ conservation is not essential to the phenomenological viability of a SUSY model. $R_p$ violation 
(RpV) --- see \cite{Dreiner:1997uz,Barbier:2004ez} for reviews --- is viable as well; simply a different discrete (or gauge) symmetry is 
required \cite{Chamseddine:1995gb,Dreiner:2005rd,Dreiner:2007vp,Dreiner:2013ala}. It also leads to a distinctive phenomenology 
which is relevant to LHC searches \cite{Hanussek:2012eh,Dercks:2017lfq}.

With experimental constraints now coming from both 
low-energy physics and the high-energy frontier, it seems justified to give the RpV-phenomenology a closer look, beyond the tree-level 
or single-coupling approximations that are frequently employed in the literature.

In this paper, we consider the most general RpV-model with minimal superfield content. The superpotential of the $R_p$-conserving MSSM is thus 
extended by the following terms \cite{Weinberg:1981wj}:
\begin{eqnarray}
W_{\not{R}_p}=\mu_i H_u \cdot L_i+\frac{1}{2}\lambda_{ijk}L_i\cdot L_j \bar{E}_k+\lambda'_{ijk}L_i\cdot  Q_j \bar{D}_k+\frac{1}{2}\lambda''_{ijk}
\varepsilon_{abc}\bar{U}^a_i\bar{D}^b_j\bar{D}^c_k,
\label{eq:RpVSuperpotential}
\end{eqnarray}
where $Q$, $\bar{U}$, $\bar{D}$, $L$, $\bar{E}$ denote the usual quark and lepton superfields, $\cdot$\ is the $SU(2)_L$ invariant product and 
$\varepsilon_{abc}$ is the 3-dimensional Levi-Civita symbol. The indices $i$, $j$, $k$ refer to the three generations of flavor, while $a$, $b$, $c$ 
correspond to the color index. We note that symmetry-conditions may be imposed on the parameters $\lambda_{ijk}$ and $\lambda''_{ijk}$ without 
loss of generality: $\lambda_{ijk}=-\lambda_{jik}$, $\lambda''_{ijk}=-\lambda''_{ikj}$. The first three sets of terms of 
Eq.(\ref{eq:RpVSuperpotential}) violate lepton-number and the last set of terms violate baryon-number.

The superpotential of Eq.(\ref{eq:RpVSuperpotential}) contains several sources of flavor-violation, in both the lepton and the quark sectors. Such effects are 
steadily searched for in experiments, placing severe bounds on the parameter space of the model. The impact of lepton-flavor violating observables on the 
RpV-MSSM has been discussed extensively in the literature, see \textit{e.g.}\ \cite{Jang:1997jy,Cheung:2001sb,Vicente:2013fya,Carvalho:2002bq,Endo:2009cv,Choi:2000bm,deGouvea:2000cf,Vicente:2015cka,Gemintern:2003gd,Chen:2008gd,Cheng:2012zu,Arhrib:2012ax,Arhrib:2012mg,Cao:2009cp,Gomez:2002kw,Dreiner:2006gu,Dreiner:2012mx,Li:2005rr,Li:2013sga,Cho:2011bd,Bose:2010eb,Grossman:1998py,Dreiner:2011ft,Dreiner:2007uj,Hirsch:2000ef}. In the quark sector, observables such as leptonic $B$-decays or radiative 
$b\to s$ transitions \cite{deCarlos:1996yh,Dreiner:2001kc,Dreiner:2013jta} have been considered. Here, we wish to focus on neutral-meson mixing observables, 
$\Delta M_K$, $\Delta M_d$, $\Delta M_s$, for $K^0$, $B^0_d$ and $B^0_s$ mesons, respectively. Such observables have been discussed in the R-parity 
conserving \cite{Altmannshofer:2007cs,Gabbiani:1996hi} as well as in an RpV context in the past \cite{Agashe:1995qm,Choudhury:1996ia,deCarlos:1996yh,Bhattacharyya:1998be,Saha:2003tq,Kundu:2004cv,Nandi:2006qe,Wang:2006xha,Wang:2010vv}. Yet, diagrams beyond
the tree-level and box contributions as well as sfermion or RpV-induced mixings have been routinely ignored. The purpose of this paper consists in
addressing these deficiencies and proposing a full one-loop analysis of the meson-mixing observables in the RpV-MSSM.

From the experimental perspective, the measurements of $B$-meson oscillations by the ALEPH, DELPHI, L3, OPAL, CDF, D0, BABAR, Belle, ARGUS, CLEO and LHCb collaborations
have been combined by the Heavy-Flavor Averaging Group \cite{Amhis:2016xyh}, leading to the averages:
\begin{subequations}\label{DMBexp}\begin{align}
\Delta M_d^{\mbox{\tiny\em exp}}&=0.5065\pm 0.0019\text{ ps}^{-1},\\
\Delta M_s^{\mbox{\tiny\em exp}}&=17.757\pm 0.021\text{ ps}^{-1}.
\end{align}\end{subequations}
These values are in excellent agreement with the SM computations \cite{Lenz:2010gu,Lenz:2011ti,Artuso:2015swg}, resulting in tight constraints on new physics contributions. However, we note that the latest SM evaluation of $\Delta M_s$ \cite{DiLuzio:2017fdq} is in tension with Eq.~(\ref{DMBexp}). This largely appears as a consequence of the new lattice evaluation of the non-perturbative parameter $f_{B_s}^2B_{B_s}$ by Ref.~\cite{Bazavov:2016nty}, with reduced uncertainties. While this situation interestingly favors effects beyond the SM, we prefer to remain conservative as long as the new value of $f_{B_s}^2B_{B_s}$ is not confirmed by other studies. We thus assume that the uncertainties on the SM prediction are still of the order of the older computations.

For the $K^0-\bar{K}^0$ system, the Particle Data Group \cite{Olive:2016xmw} combines the experimental measurements as:
\begin{equation}\label{DMKexp}
 \Delta M_K^{\mbox{\tiny\em exp}}=(0.5293\pm 0.0009)\cdot10^{-2}\text{ ps}^{-1}.
\end{equation}
Despite the precision of this result, constraints from $K^0-\bar{K}^0$ mixing on high-energy contributions are considerably relaxed by the large theoretical uncertainties due to long-distance effects.
Historically, estimates of the latter have been performed using the techniques of large $N$ QCD --- see \textit{e.g.} 
Ref.~\cite{Bijnens:1990mz} ---  while lattice QCD collaborations such as \cite{Bai:2014cva} 
are now considering the possibility of evaluating these effects in realistic kinematical configurations. Ref.~\cite{Buras:2013raa} settles for a long-distance contribution
at the level of $(20\pm10)\%$ of the experimental value, and we follow this estimate below. Concerning short-distance contributions, Ref.~\cite{Brod:2011ty} performed a NNLO
study of the charm-quark loops, resulting in a SM estimate of $\Delta M_K^{\mbox{\tiny SM, Short Dist.}}=(0.47\pm 0.18)\cdot10^{-2}\text{ ps}^{-1}$.

Beyond the mass differences, CP-violating observables are also available in the meson-mixing system. Although our study is valid for these as well, we will not discuss them in the following, since we do not wish to pay much attention to the new-physics phases.

The computation of the meson oscillation parameters is usually performed in a low-energy effective field theory 
(EFT), where short-distance effects intervene
via the Wilson coefficients of dimension $6$ flavor-changing ($\Delta F=2$) operators \cite{Buchalla:1995vs}. This procedure ensures a resummation of large logarithms 
via the application of the renormalization group equations (RGE) from the matching high-energy (\textit{e.g.}\ electroweak) scale down to the low-energy (meson-mass) scale where 
hadronic matrix elements should be computed \cite{Buras:2001ra}. In this work, we calculate the contributions to the Wilson coefficients arising in the 
RpV-MSSM up to one-loop order. The $\lambda'$ couplings of Eq.(\ref{eq:RpVSuperpotential}) already generate a tree-level diagram. 
Going beyond this, at one-loop order, diagrams contributing to the meson mixings involve both R-parity conserving and R-parity violating couplings. These are furthermore 
intertwined via RpV-mixing effects stemming for example from the bilinear term $\mu_i H_u \cdot L_i$. Our analysis goes beyond the approximations that are 
frequently  encountered in the literature. We also find occasional differences with published results, which we point out accordingly.

In the following section, we present the general ingredients of our full one-loop analytical calculation of the Wilson coefficients of the $\Delta F=2$ EFT (effective field theory) in the RpV-MSSM, referring to the appendices where the exact expressions are provided. In Section \ref{sec:numerical}, we discuss 
our implementation of these results employing the public tools \texttt{SPheno} \cite{Porod:2003um,Porod:2011nf}, \texttt{SARAH} 
\cite{Staub:2008uz,Staub:2009bi,Staub:2010jh,Staub:2012pb,Staub:2013tta,Staub:2015kfa}, \texttt{FlavorKit} \cite{Porod:2014xia} and \texttt{Flavio} 
\cite{david_straub_2017_897989}. Finally, numerical limits on the RpV-couplings are presented in a few simple scenarios, before a short conclusion.

\section{\boldmath Matching conditions for the $\Delta F=2$ EFT of the RpV-MSSM}\label{sec:analytic}

We consider the $\Delta F=2$ EFT relevant for the mixing of $(\bar{d}_id_j)$-$(\bar{d}_jd_i)$ mesons --- $d_{i}$ corresponds to the down-type quark of $i$th generation 
($d$, $s$ or $b$). The EFT Lagrangian is written as
\begin{align}\label{eqn:EFTLag}
\mathcal{L}_{EFT}=\sum_{i=1}^{5}C_i O_i+\sum_{i=1}^{3}\tilde{C}_i \tilde{O}_i,
\end{align}
where we employ the following basis of dimension $6$ operators:
\begin{align}\label{eqn:EffOps}
 & O_1=(\bar{d}_j\gamma^{\mu}P_Ld_i)(\bar{d}_j\gamma_{\mu}P_Ld_i), & & \tilde{O}_1=(\bar{d}_j\gamma^{\mu}P_Rd_i)(\bar{d}_j\gamma_{\mu}P_Rd_i),\nonumber\\
 & O_2=(\bar{d}_jP_Ld_i)(\bar{d}_jP_Ld_i), & & \tilde{O}_2=(\bar{d}_jP_Rd_i)(\bar{d}_jP_Rd_i),\\
 & O_3=(\bar{d}_j^aP_Ld^b_i)(\bar{d}^b_jP_Ld^a_i), & & \tilde{O}_3=(\bar{d}^a_jP_Rd^b_i)(\bar{d}^b_jP_Rd^a_i),\nonumber\\
 & O_4=(\bar{d}_jP_Ld_i)(\bar{d}_jP_Rd_i), & & O_5=(\bar{d}^a_jP_Ld^b_i)(\bar{d}^b_jP_Rd^a_i).\nonumber
\end{align}
The superscripts ($a,b=1,2,3$) refer to the color indices when the sum is not trivially contracted within the fermion product. We have employed the usual four-component spinor notations above, with $P_{L,R}$ denoting the left- and right-handed projectors.

The Wilson coefficients $C_i,\,\tilde{C}_i$ associated with the operators of Eq.(\ref{eqn:EffOps}) in the Lagrangian of the EFT --- Eq.(\ref{eqn:EFTLag}) --- 
are obtained at  high-energy by matching the $d_i\bar{d}_j\to d_j\bar{d}_i$ amplitudes in the EFT and in the full RpV-MSSM. We restrict ourselves to the leading-order coefficients (in a QCD/QED expansion) on the EFT-side. On the side of the RpV-MSSM, we consider only 
short-distance effects, \textit{i.e.}\ we discard QCD or QED loops. Indeed, the photon and gluon are active fields in the EFT, so that a proper processing of the 
corresponding effects would require a NLO matching procedure. Furthermore, both tree-level and one-loop contributions are considered in the RpV-MSSM: 
we stress that this does not induce a problem in power-counting, as the tree-level contribution is a strict RpV-effect, so that $R_p$-conserving (or violating) 
one-loop amplitudes are not (all) of higher QED order. Numerically speaking, one possibility is that the tree-level
is dominant in the Wilson coefficients, in which case, the presence of the one-loop corrections does not matter. This case is essentially excluded if we consider the
experimental limits on the meson-oscillation parameters. If, on the contrary, the tree-level contribution is of comparable (or subdominant) magnitude with the one-loop amplitudes, then 
the electroweak power-counting is still satisfied. Yet, one-loop contributions that are aligned with the tree-level always remain subdominant.

For our calculations in the RpV-MSSM, we employ the Feynman `t Hooft gauge \cite{Fujikawa:1972fe} and dimensional regularization 
\cite{SIEGEL1979193,Capper:1979ns}. For reasons of consistency with the tools that we employ for the numerical implementation, $\overline{DR}$-renormalization 
conditions will be applied. However, in the results that we collect in the Appendix, the counterterms are kept in a generic form, which allows for other choices of 
renormalization scheme. We apply the conventions where the sneutrino fields do not take vacuum expectation values.\footnote{For the general rotation to this basis 
see Ref.~\cite{Dreiner:2003hw}. See also Ref.~\cite{Allanach:2003eb} for a discussion of this in terms of physics at the unification scale.} Moreover, the $\lambda'$ couplings of Eq.(\ref{eq:RpVSuperpotential}) are defined in the basis of down-type mass-states, \textit{i.e.}\ 
a CKM matrix appears when the second index of $\lambda'$ connects with an up-type field, but not when it connects to a down-type field \cite{Agashe:1995qm}. 
Mixing among fields are considered to their full extent, including left/right and flavor squark mixings, charged-Higgs/slepton mixing, neutral-Higgs/sneutrino mixing, 
chargino/lepton mixing 
and neutralino/neutrino mixing. The details of our notation and the Feynman rules employed can be found in Appendix \ref{Appendix:Notations}. As a crosscheck, we 
performed the calculation using two different approaches for the fermions: the usual four-component spinor description and the two-component description 
\cite{Dreiner:2008tw}.

On the side of the EFT, the operators of Eq.(\ref{eqn:EffOps}) each contribute four tree-level Feynman diagrams to the $d_i\bar{d}_j\to d_j\bar{d}_i$ amplitude. Half
of these contributions are obtained from the other two by an exchange of the particles in the initial and final states: as the dimension $6$ operators are 
symmetrical over the simultaneous exchange of both $d_i$'s and both $d_j$'s, we may simply consider two diagrams and double the amplitude. The two remaining diagrams
correspond to an $(s\leftrightarrow t)$-channel exchange. We exploit these considerations to reduce the number of diagrams that we consider on the side of the 
RpV-MSSM to only one of the $s/t$-channels.

\begin{figure}[]
\begin{center}
\begin{subfigure}{.45\textwidth}
  \centering
  \includegraphics[width=.8\linewidth]{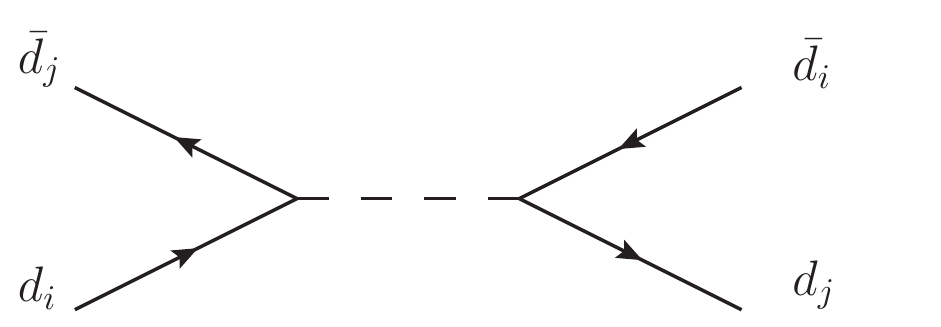}
  \caption{Tree-level Feynman diagram\\ (Appendix \ref{appendix:tree_level})}
  \label{dia:Tree}
\end{subfigure}%
\begin{subfigure}{.45\textwidth}
  \centering
  \includegraphics[width=.8\linewidth]{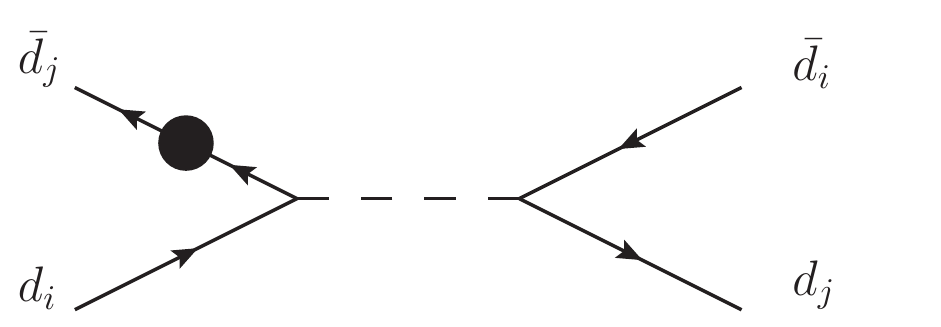}
  \caption{Tree-level Feynman diagram\\ with quark self-energies (Appendix \ref{appendix:quarkSE})}
  \label{dia:Tree_quarkSE}
\end{subfigure}
\end{center}
\begin{center}
\begin{subfigure}{.45\textwidth}
  \centering
  \includegraphics[width=.8\linewidth]{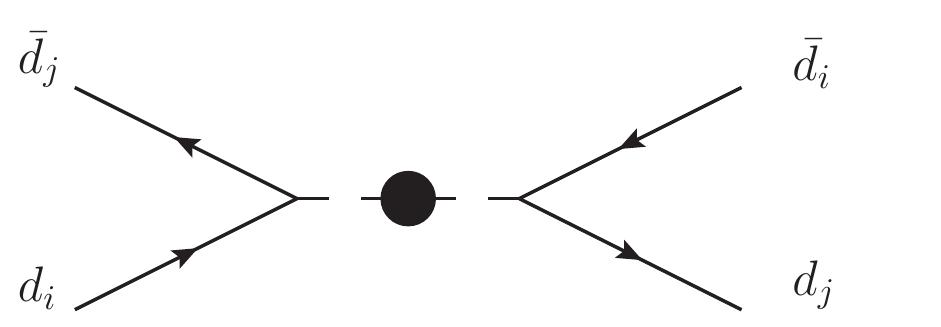}
  \caption{Tree-level Feynman diagram\\ with scalar self-energies (Appendix \ref{appendix:scalarSE})}
  \label{dia:Tree_scalarSE}
\end{subfigure}
\begin{subfigure}{.45\textwidth}
  \centering
  \includegraphics[width=.8\linewidth]{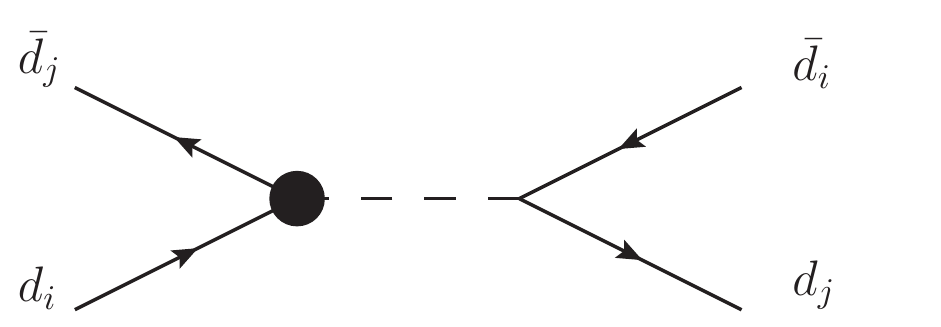}
  \caption{Tree-level Feynman diagram\\ with vertex corrections (Appendix \ref{appendix:VertexCorrection})}
  \label{dia:Tree_VC}
\end{subfigure}
\end{center}
\caption{The tree level diagram and its one-loop corrections.}
\label{fig:tree_corrections}
\end{figure}

The tree-level contribution to the $d_i\bar{d}_j\to d_j\bar{d}_i$ amplitudes is due to the $\lambda'$ couplings of Eq.(\ref{eq:RpVSuperpotential}). It involves a sneutrino exchange where, however, sneutrino-flavor and sneutrino-Higgs mixing could occur. The appearance of RpV contributions at tree-level
complicates somewhat a full one-loop analysis: one-loop contributions indeed depend on the renormalization of the $d_i\bar{d}_j$-sneutrino vertex (and of its 
external legs). In principle, one could define this vertex `on-shell', \textit{i.e.}\  impose that one-loop corrections vanish for on-shell $d_i$, $d_j$
external legs --- while the counterterm for the sneutrino field is set at momentum $p^2=M^2_{K,B}\simeq0$. In such a case, one could restrict oneself to calculating the 
box-diagram contributions to $d_i\bar{d}_j\to d_j\bar{d}_i$. However, in any other renormalization
scheme, self-energy and vertex-correction diagrams should be considered. Yet, if the $\lambda'$ couplings contributing at tree-level are small, the impact of the vertex and
self-energy corrections is expected to be limited, since these contributions retain a (at least) linear dependence on the tree-level $\lambda'$. These contributions 
are symbolically depicted in Fig.\ref{fig:tree_corrections}.
\begin{figure}[tbh]
\begin{center}
\begin{subfigure}{.33\textwidth}
  \centering
  \includegraphics[width=.8\linewidth]{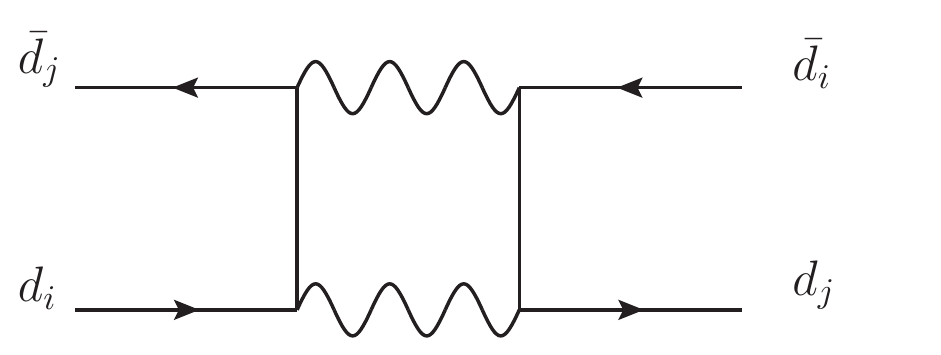}
  \caption{\\Vector/fermion/vector/fermion ``straight" box (Appendix \ref{subsec:vfvf-stright})}
  \label{dia:Box_VFVF_Straight}
\end{subfigure}
\begin{subfigure}{.33\textwidth}
  \centering
  \includegraphics[width=.8\linewidth]{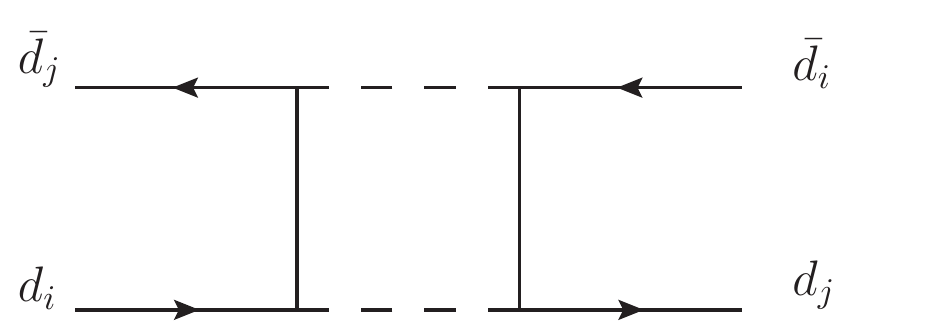}
  \caption{Scalar/fermion/scalar/fermion  ``straight" box (Appendix \ref{subsec:vfvf-stright})} 
  \label{dia:Box_SFSF_Straight}
\end{subfigure}%
\begin{subfigure}{.33\textwidth}
  \centering
  \includegraphics[width=.8\linewidth]{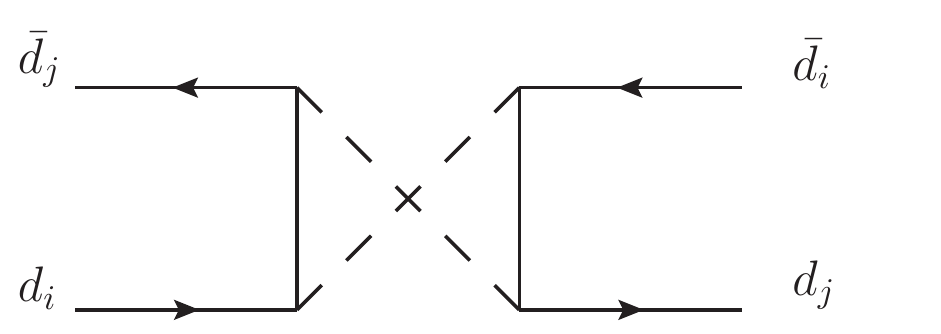}
  \caption{Scalar/fermion/scalar/fermion ``scalar-cross" box (Appendix \ref{subsec:sfsf-scross})} 
  \label{dia:Box_SFSF_ScalarX}
\end{subfigure}%
\end{center}
\begin{center}
\begin{subfigure}{.33\textwidth}
  \centering
  \includegraphics[width=.8\linewidth]{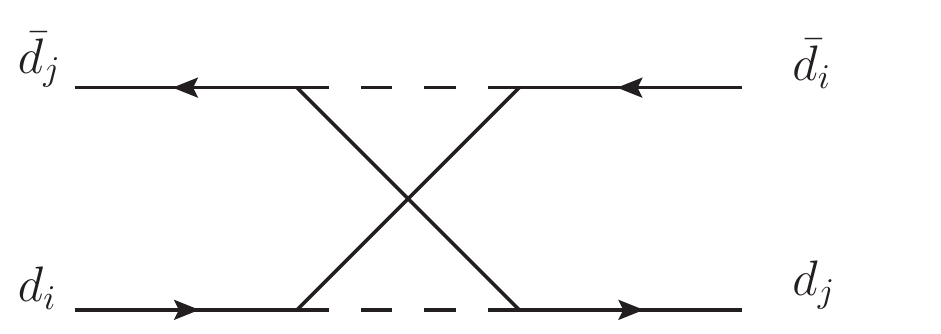}
  \caption{Scalar/fermion/scalar/fermion ``fermion-cross" box (Appendix \ref{subsec:sfsf-fcross})} 
  \label{dia:Box_SFSF_FermionX}
\end{subfigure}
\begin{subfigure}{.33\textwidth}
  \centering
  \includegraphics[width=.8\linewidth]{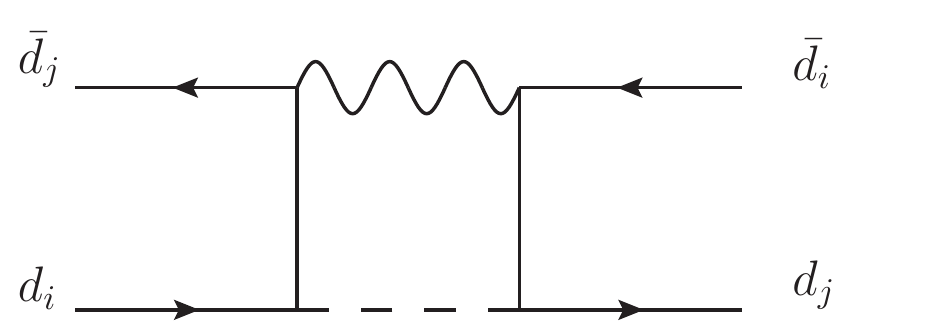}
  \caption{\\ Vector/fermion/scalar/fermion   ``straight" box (Appendix \ref{subsec:vfvf-stright})}  
  \label{dia:Box_VFSF_Straight}
\end{subfigure}%
\begin{subfigure}{.33\textwidth}
  \centering
  \includegraphics[width=.8\linewidth]{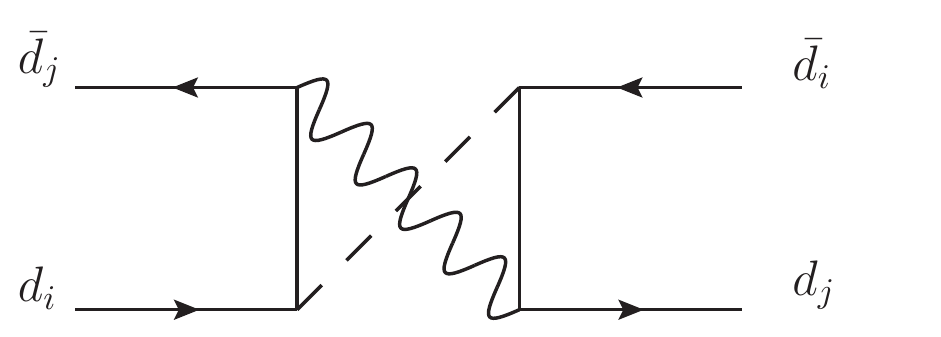}
  \caption{\\Vector/fermion/scalar/fermion ``cross" boxes (Appendix \ref{subsec:vfsf-cross})}
  \label{dia:Box_VFSF_X}
\end{subfigure}%
\end{center}
\begin{center}
\begin{subfigure}{.33\textwidth}
  \centering
  \includegraphics[width=.8\linewidth]{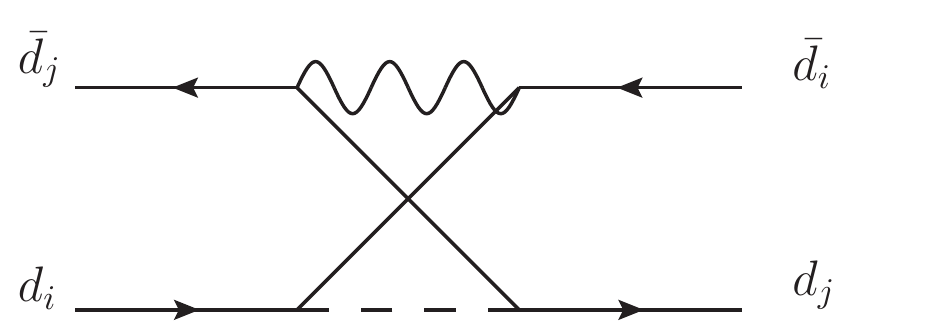}
  \caption{\\Vector/fermion/scalar/fermion ``fermion-cross" box (Appendix \ref{subsec:sfsf-fcross})}  
  \label{dia:Box_VFSF_FermionX}
\end{subfigure}
\end{center}
\caption{The topologies of box diagrams that appear in the neutral mesons mixing with the RpV-MSSM.}
\label{fig:topologies_boxes}
\end{figure}

One-loop diagrams contributing to $d_i\bar{d}_j\to d_j\bar{d}_i$ include SM-like contributions (box diagrams with internal $u$, $c$, $t$ quarks, $W$ and Goldstone bosons),
2-Higgs-doublet-model-like contributions (box diagrams with internal $u$, $c$, $t$ quarks, charged-Higgs bosons and possibly $W$ or Goldstone bosons),
${R}_p$-conserving SUSY contributions (box diagrams with chargino/scalar-up, neutralino/sdown or gluino/sdown particles in the loop) and RpV-contributions
(self-energy and vertex corrections, box diagrams with sneutrino/quark, slepton/quark, lepton/squarks, neutrino/squark or quark/squark internal lines). 
The RpV-driven mixing further intertwines these contributions, so that the distinction among \textit{e.g.}\ the ${R}_p$-conserving chargino/scalar-up and RpV
lepton/scalar-up boxes becomes largely superfluous. For all these contributions, with exception of the self-energy diagrams on the external legs, we neglect the external 
momentum, as it controls effects of order $m_{d_{i,j}}$, which are subdominant when compared to the momentum-independent pieces of order $M_W$ or 
$M_{\mbox{\tiny SUSY}}$. Yet, when a SM-fermion $f$ appears in the loop, some pieces that are momentum-independent still come with a suppression of order 
$m_f/M_{W,\mbox{\tiny SUSY}}$. We keep such pieces even though they could be discarded in view of the previous argument.

The diagrams of Fig.\ref{fig:tree_corrections} are calculated in Appendix~\ref{appendix:tree_level} (tree-level contribution), Appendix~\ref{appendix:quarkSE} 
($d_i$-quark self-energies), Appendix~\ref{appendix:scalarSE} (scalar self-energy) and Appendix~\ref{appendix:VertexCorrection} (vertex corrections). 
Fig.\ref{fig:topologies_boxes} lists the various relevant topologies involved in box diagrams. The corresponding contributions are presented in Appendix~\ref{Appendix:Box_diagrams}.
The relevant loop functions are provided in Appendix~\ref{subsec:LoopFunctions}.

While we go beyond the usual assumptions employed to study the $\Delta F=2$ Wilson coefficients in the RpV-MSSM, it is possible to compare the 
outcome of our calculation to partial results available in the literature. First, in the limit of vanishing RpV-parameters, we recover the well-known results 
in the ${R}_p$-conserving MSSM, which are summarized in \textit{e.g.}\ the appendix of Ref.~\cite{Altmannshofer:2007cs}. Then, RpV-contributions from the tree-level
and box-diagram topologies have been presented in Ref.~\cite{deCarlos:1996yh} in the no-mixing approximation. Taking this limit and neglecting further terms that are not
considered by this reference, we checked that our results coincided, with the exception of the coefficient $c'^{\lambda'}_{LR}$ of Ref.~\cite{deCarlos:1996yh} 
(a piece of the contribution to $C_5$). Transcripted to our notations, the result of Ref.~\cite{deCarlos:1996yh} reads:
\begin{eqnarray}
c'^{\lambda'}_{LR}&=&-\frac{1}{64\pi^2}\lambda'^*_{i1k}\lambda'_{j2k}\lambda'_{im1}\lambda'^*_{jm2}D_2(m^2_{N_i},m^2_{N_j},m^2_{d_k},m^2_{d_m})\nonumber\\
& &-\frac{1}{64\pi^2}\lambda'^*_{i1k}\lambda'_{j2k}\lambda'_{im1}\lambda'^*_{jm2}D_2(m^2_{\nu_i},m^2_{\nu_j},m^2_{\bar{D}^k_R},m^2_{\bar{D}_R^m}),
\end{eqnarray}
while we obtain:
\begin{eqnarray}
c'^{\lambda'}_{LR}&=&\frac{1}{32\pi^2}\lambda'^*_{i1k}\lambda'_{j2k}\lambda'_{im1}\lambda'^*_{jm2}D_2(m^2_{N_i},m^2_{N_j},m^2_{d_k},m^2_{d_m})\nonumber\\
& +&\frac{1}{32\pi^2}\lambda'^*_{i1k}\lambda'_{j2k}\lambda'_{im1}\lambda'^*_{jm2}D_2(m^2_{\nu_i},m^2_{\nu_j},m^2_{D^k_R},m^2_{D_L^m}).
\end{eqnarray}
The mismatch lies in the prefactor and the sfermion chiralities. Another class of $\lambda'$ boxes involving an electroweak charged current has been considered
in the no-mixing limit in Ref.~\cite{Bhattacharyya:1998be}. There, we find agreement with our results. As self-energy and vertex corrections have not been considered before, 
the opportunities for comparison are more limited. Still, we checked that the scalar self-energies were consistent with the results of
Ref.~\cite{Martin:2003it}. Finally, our results can be controlled in another fashion, using the automatically generated results of public tools: we detail
this in the following section. 

\section{Numerical implementation and tools}\label{sec:numerical}

In order to determine limits from the meson oscillation measurements on the parameter space of the RpV-MSSM, 
we establish a numerical tool implementing the one-loop contributions to the $\Delta F=2$ Wilson coefficients and deriving the corresponding
theoretical predictions for $\Delta M_{K,d,s}$. To this end, we make use of the {\tt Mathematica}
 package {\tt SARAH} \cite{Staub:2008uz,Staub:2009bi,Staub:2010jh,Staub:2012pb,Staub:2013tta,Staub:2015kfa} to produce a customized spectrum generator based
  on {\tt SPheno}  \cite{Porod:2003um,Porod:2011nf,Staub:2017jnp}. {\tt SPheno} calculates the complete supersymmetric particle spectrum 
  at the one-loop order and includes all important two-loop corrections to the neutral scalar masses \cite{Goodsell:2015ira}.

The routines performing the calculation of flavor 
observables are generated through the link to {\tt FlavorKit} \cite{Porod:2014xia}. {\tt FlavorKit} makes use of 
{\tt FeynArts}/{\tt FormCalc} \cite{Hahn:1998yk,Hahn:2000kx,Nejad:2013ina} to calculate the leading diagrams to quark and lepton flavor violating
 observables. For the meson mass differences, the tree-level and box diagrams as well as the double-penguin contributions are included per default. 
However, as parameters within \texttt{SPheno} are defined in the $\overline{DR}$ scheme, it is in principle necessary to implement the self-energy and 
vertex corrections. We added the vertex corrections via \texttt{PreSARAH} \cite{Porod:2014xia}, which enables the implementation of new operators
 into {\tt FlavorKit} within  certain limits.  As the scalar self-energies cannot be generated in this fashion, we incorporated these by hand.

The Wilson coefficients computed by {\tt FlavorKit} and \texttt{PreSARAH} at the electroweak matching scale are stored in analytical form 
in the {\tt Fortran} output of {\tt FlavorKit}.
We compared these expressions with our results of the previous section; we found explicit agreement in almost all cases --- and adapted the code to match our results in 
the few cases where it proved necessary.\footnote{In rare cases, we identified seemingly minor --- but numerically important --- differences between our computation and the 
{\tt FlavorKit} code, namely in a few tree-level contributions to  $C_5$ (which should be absent),
 as well as in $\tilde{C}_{2,3}$ and $C_{2,3}$ for a few one-loop box diagrams. We fixed those 
appearances in the code as well as the relative sign between tree and one-loop contributions after correspondence and cross-checking with the {\tt FlavorKit} authors.}

After the Wilson coefficients at the electroweak matching scale are computed, further steps are necessary in order to relate them to the 
observables $\Delta M_{K,d,s}$. 
The \texttt{FlavorKit} output includes a theoretical prediction for these observables, however the hadronic input parameters are more up-to-date
in the more recently-developed code \texttt{Flavio} \cite{david_straub_2017_897989}, which shares an interface with \texttt{FlavorKit} using the FLHA 
standards \cite{Mahmoudi:2010iz}. We hence use \texttt{Flavio} to process the Wilson coefficients as calculated by \texttt{FlavorKit}. First, the Wilson 
coefficients must be run to a low-energy scale using the QCD RGE's of the EFT \cite{Buras:2001ra}. 
In the case of the $K^0-\bar{K}^0$ system, the impact of the 
charm loop is sizable \cite{Brod:2011ty}: we upgraded the NLO coefficient $\eta_{cc}$ coded within \texttt{Flavio} to the NNLO value $1.87(76)$ \cite{Brod:2011ty}
and $\eta_{ct}=0.496(47)$ \cite{Brod:2010mj}. 
For consistency, the charm mass in the loop functions is set to the $\overline{\text{MS}}$ value $m_c(m_c)\simeq1.28$~GeV. 
Then, the hadronic dynamics encoded in the dimension $6$ operators must be interpreted at low-energy in the form of hadronic mixing elements: this step gives rise to 
``bag-parameters'', which are evaluated in lattice QCD. Here, \texttt{Flavio} employs the bag parameters of Ref.~\cite{Carrasco:2015pra} for the $K^0-\bar{K}^0$ system and 
of Ref.~\cite{Bazavov:2016nty} for the $B_d^0-\bar{B}_d^0$ and $B_s^0-\bar{B}_s^0$ systems. In addition, the CKM matrix elements within \texttt{Flavio} are derived from 
the four inputs $|V_{us}|$, $|V_{ub}|$, $|V_{cb}|$ and $\gamma$. We set these to the fit-results of Ref.~\cite{Olive:2016xmw}: $|V_{us}|\simeq0.22506$, 
$|V_{ub}|\simeq3.485\cdot10^{-3}$, $|V_{cb}|\simeq4.108\cdot10^{-2}$ and $\gamma\simeq1.236$. Moreover, we changed the 
$B_d^0$ decay constant to 
a numerical value of $186$~MeV \cite{Dowdall:2013tga}. Finally, we added 
the observable $\Delta M_K$ to 
Flavio 
(based on pre-included material) and made sure that the predicted
SM short-distance prediction 
was consistent 
with the theoretical SM estimate given by Ref.~\cite{Brod:2011ty}.

A quantitative comparison of the predicted $\Delta M_{K,d,s}$ with the experimental results of Eqs.(\ref{DMBexp}) and (\ref{DMKexp}) requires
an estimate of the theoretical uncertainties. The Wilson coefficients have been obtained at leading order, which implies higher-order corrections of QCD-size. In the 
case of the SM-contributions, large QCD logarithms are resummed in the evolution of the RGEs between the matching electroweak scale and the low-energy scale. However,
for the new-physics contributions, further logarithms between the new-physics and the electroweak scale could intervene --- {\tt FlavorKit} computes the new-physics
contributions to the Wilson coefficients at the electroweak scale, hence missing such logarithms. Therefore, the higher-order uncertainty is larger for contributions
beyond the SM and can be loosely estimated as $O\left(\frac{\alpha_S}{\pi}\log\frac{\mu^2_{NP}}{\mu^2_{EW}}\right)$, where $\mu_{NP}$ and $\mu_{EW}$ represent the 
new-physics and electroweak scales, respectively. Further sources of uncertainty are the RGE evolution in the EFT and the evaluation of hadronic matrix elements. 
For the SM matrix elements, the uncertainties on $\eta_{cc}$, $\eta_{ct}$ and $\eta_{tt}$ are of order $30\%$ \cite{Brod:2011ty}, $10\%$ \cite{Brod:2010mj} and $1\%$
\cite{Buras:1990fn}, respectively, leading to a large SM uncertainty in $\Delta M_K$ and a smaller one in $\Delta M_{d,s}$. For the $K^0-\bar{K}^0$ system, the 
bag-parameters are known with a precision of $\sim3\%$ in the case of $B_K^{(1)}$ and $\sim7\%$ for the other operators \cite{Carrasco:2015pra}. For the 
$B_d^0-\bar{B}_d^0$ system, the uncertainty is of order $10\%$ \cite{Bazavov:2016nty} --- and even $20\%$ for $B_{B_d}^{(3)}$. For the $B_s^0-\bar{B}_s^0$, the bag
parameters are known at about $7\%$ accuracy \cite{Bazavov:2016nty} --- $14\%$ for $B_{B_s}^{(3)}$. Finally, CKM matrix elements contribute to the uncertainty at
the level of a few percent. To summarize, we decided to estimate the theoretical uncertainties of our predictions for the meson oscillation parameters in the 
RpV-MSSM as follows:
\begin{itemize}
 \item $40\%\times\left[|\Delta M_K^{\mbox{\tiny SM, Short. Dist.}}|+|\Delta M_K^{\mbox{\tiny RpV-MSSM, Short. Dist.}}-\Delta M_K^{\mbox{\tiny SM, Short. Dist.}}|\right]$
 for the short-distance contribution to $\Delta M_K$. As explained above, we will employ the estimate of Ref.~\cite{Buras:2013raa} for the long-distance contribution:
 $\Delta M_K^{\mbox{\tiny SM, Long Dist.}}\simeq(20\pm10)\%\times\Delta M_K^{\mbox{\tiny\em exp}}$.
 \item $15\%\times|\Delta M_{d,s}^{\mbox{\tiny SM}}|+30\%\times|\Delta M_{d,s}^{\mbox{\tiny RpV-MSSM}}-\Delta M_{d,s}^{\mbox{\tiny SM}}|$ for the 
 evaluation of $\Delta M_{d,s}$.
\end{itemize}
These uncertainty estimates restore the magnitude of the SM uncertainties \cite{Lenz:2010gu,Lenz:2011ti,Artuso:2015swg,Brod:2011ty}. Concerning the new-physics part, we stress that the calculation employs a (QCD/QED) LO matching and misses running effects between the SUSY and the matching scales, which motivates conservative estimates.

Finally, we note that our calculation of the Wilson coefficients for the $\Delta F=2$ transition also provides access to CP-violating observables such as $\epsilon_K$. These would grant complementary constraints on the parameter space, in particular when the RpV-parameters of Eq.(\ref{eq:RpVSuperpotential}) are considered as complex degrees of freedom.
Obviously, in the presence of e.g.\ a large RpV tree-level contribution to the $d_i\bar{d}_j\to d_j\bar{d}_i$ amplitude, it is always possible to choose the phases of the $\lambda'$-parameters such that, amongst others, $\epsilon_K$ is in agreement with the experimental measurement (within uncertainties that are dominated by the theoretical evaluation \cite{Brod:2011ty}). On the other hand, it is less trivial whether such an adjustment would be possible within the magnitude of the NP contributions that is compatible with $\Delta M$'s.
For simplicity --- keeping in mind that our numerical studies are strictly illustrative in purpose and do not aim at conveying an exhaustive picture of possible RpV-effects associated to the meson-oscillation parameters ---, we restrict ourselves to real values of the RpV-parameters and do not consider the CP-violating observables below.
In practice, the $R_p$-conserving contributions beyond the SM in the scenarios that we consider in the following section are always subleading to RpV effects, so that any deviation of the CP-violating observables from the SM predictions (caused by the CKM phase) is proportional to the RpV parameters and could be compensated via the corresponding RpV phases. Of course, if one chooses not to exploit this degree of freedom, the scenario with real RpV parameters itself would be subject to stronger limits when the CP-violating observables are also taken into account.

\section{Numerical results}
We are now in a position to study the limits on RpV-parameters that are set by the meson-oscillation parameters. However, it makes limited sense to 
scan blindly over the RpV-MSSM parameter space imposing only constraints from the $\Delta M$'s. Comparable analyses of all the relevant observables 
for which experimental data is available would be necessary. We will thus restrict ourselves to a discussion of the bounds over a restricted number of parameters 
and in a few scenarios. The input parameters that we mention below correspond to the \texttt{SPheno} input defined at the $M_Z$ scale.

We first consider the case where no explicit source of flavor violation appears in the $R_p$-conserving parameters. The flavor transition is thus 
strictly associated to the CKM matrix or to the RpV-effects. The latter can intervene in several fashions:
\begin{itemize}
 \item Flavor violation in the $\lambda'$ couplings could lead to tree-level contributions to the $\Delta M$'s. The relevant combinations --- in the absence of sneutrino mixing --- are of the form $\lambda'_{fIJ}\lambda'^*_{fJI}$, where $(I,J)$ are the indices of the valence quarks of the considered meson --- \textit{i.e.}\ 
 $(1,2)$, $(1,3)$ and $(2,3)$ for $\Delta M_K$, $\Delta M_d$ and $\Delta M_s$ respectively --- and $f$ is the flavor of the sneutrino mediator.
 \item Flavor violation in the $\lambda'$ couplings could also intervene at the loop-level only. This happens when, for instance, 
one product of the form $\lambda'_{mnI}\lambda'^*_{mnJ}$ or $\lambda'_{mIn}\lambda'^*_{mJn}$ is non-zero --- again, $(I,J)$ corresponds
to the valence quarks of the meson; $m$ and $n$ are internal to the loop.
\item Finally, the flavor transition can be conveyed by the $\lambda''$ couplings, in which case it appears only at the loop level in the $\Delta M$'s.
Possible coupling combinations include $\lambda''_{m12}\lambda''_{m23}$, $\lambda''_{m12}\lambda''_{m13}$ or $\lambda''_{m13}
\lambda''_{m23}$.
\end{itemize}
Below, we first consider these three cases separately, before we investigate possible interferences between tree- and loop-level generated diagrams for several non-zero 
$\lambda'$ couplings. 
However, we avoid considering simultaneously non-zero $LQ\bar D$ and $\bar U \bar D \bar D$ couplings: then, discrete symmetries no longer protect the 
proton from decay, so that the phenomenology would rapidly come into conflict with associated bounds. 
Still, we note that some diagrams 
contributing to the meson mixing parameters would combine both types of couplings: these are also provided in the appendix.

Then, flavor transitions can also be mediated by $R_p$-conserving effects. In this case, flavor violation could originate either in the CKM 
matrix, as in the Minimal Flavor Violation scenario \cite{DAmbrosio:2002vsn}, or in new-physics parameters, such as the soft squark bilinear and trilinear terms. We briefly 
discuss possible interferences with RpV-contributions.

For simplicity, we consider only the case of real $\lambda^{\prime (\prime)}$ and disregard the bilinear $R$-parity violating terms (though they are included in our 
analytical results in the appendix).

\subsection{Bounds on a pair of simultaneously non-zero  \boldmath $LQ\bar D$ couplings}
\label{subsec:two_nonzeroLQD}

\begin{table}[htb]
\centering
\begin{tabular}{l|cccccc}
Scenario          & $M_A$/TeV & $\mu$/TeV & $\tan{\beta}$ & $m_{\tilde{q}}$/TeV                    & $M_{1,2}$/TeV & $M_3$/TeV \\\hline
SM-like           & $3.5$     & $2$       & $10$      & $2$                                    & $2$           & $2$       \\
2HDM              & $0.8$     & $2$       & $10$      & $2$                                    & $2$           & $2$       \\
SUSY-RpV(a)       & $1.2$     & $0.6$     & $10$      & $\simeq 2$                            & $0.5$         & $2$       \\
SUSY-RpV(b)       & $1.2$     & $0.3$     & $10$      & $\simeq 2 \& 1_{\tilde{t},\tilde{b}}$ & $0.5$         & $2$                                                    
\end{tabular}
\caption{Input parameters for various scenarios under consideration. With $2 \& 1_{\tilde{t},\tilde{b}}$ we imply $m_{\tilde q_{1,2}}= 2\,$TeV while keeping a lighter 
third generation, $m_{\tilde q_{3}}= 1\,$TeV. }
\label{Tbl:parameters_scenarios}
\end{table}

\afterpage{\FloatBarrier}
\begin{figure}[tbh]
  \centering
  \includegraphics[width=0.45\textwidth]{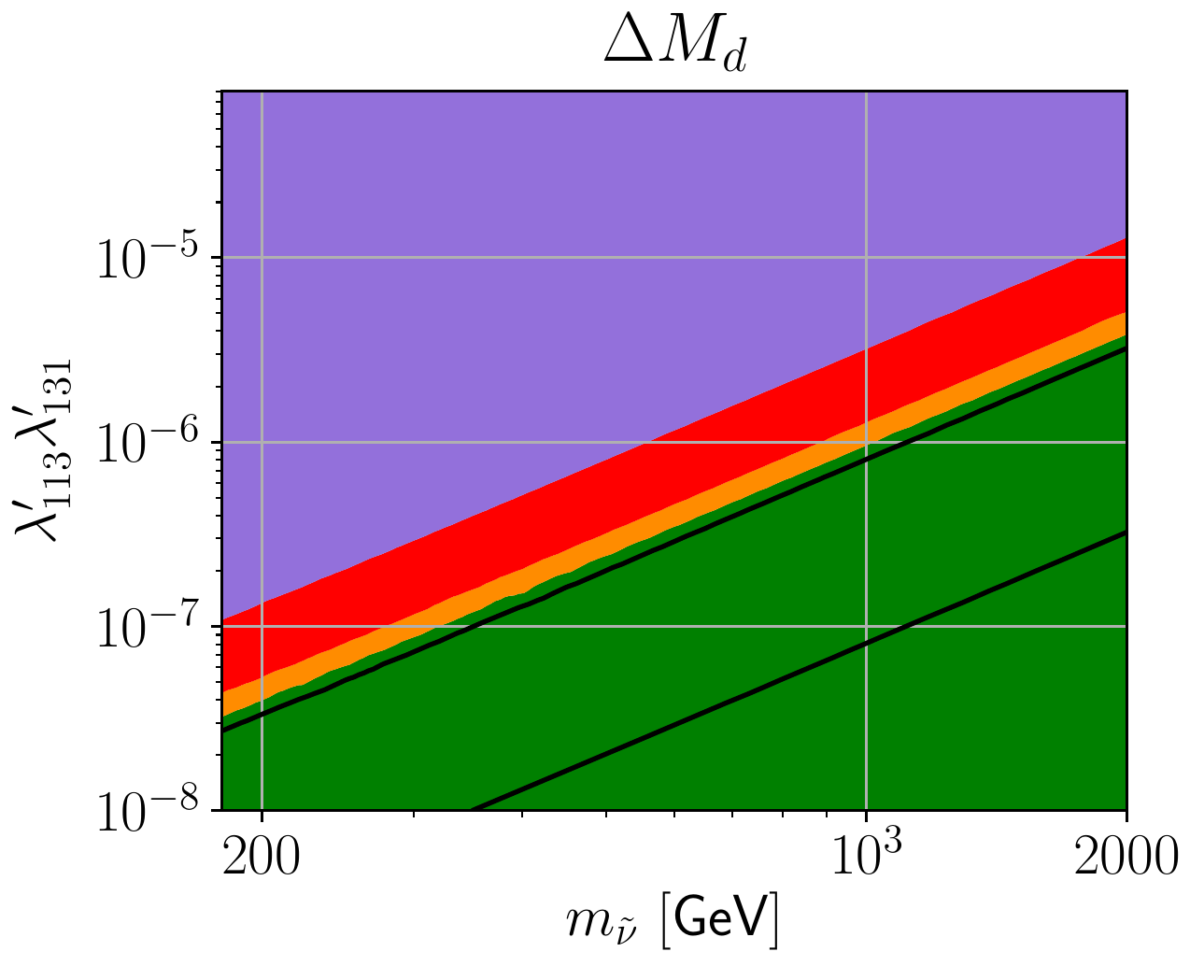}
  \includegraphics[width=0.45\textwidth]{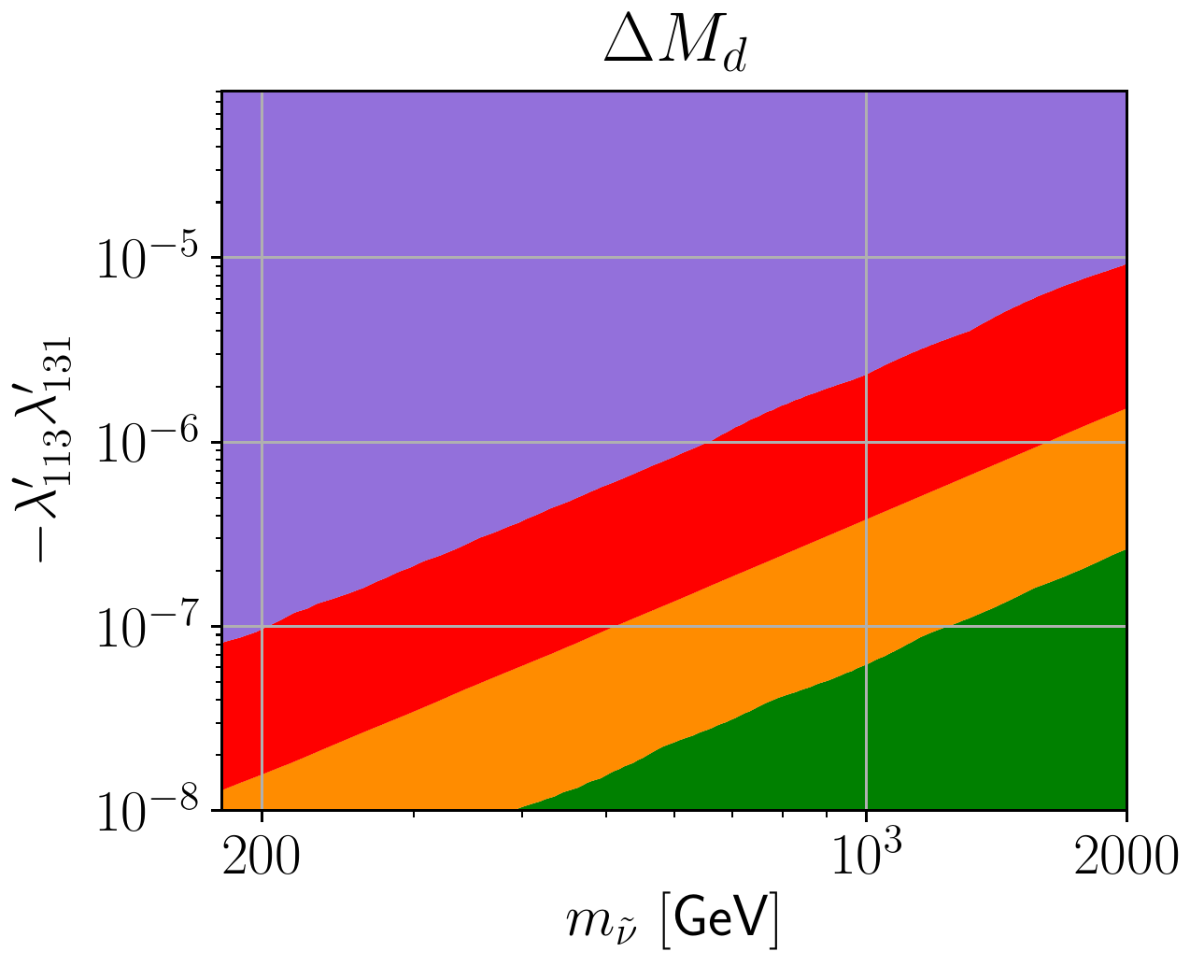}
  \includegraphics[width=0.45\textwidth]{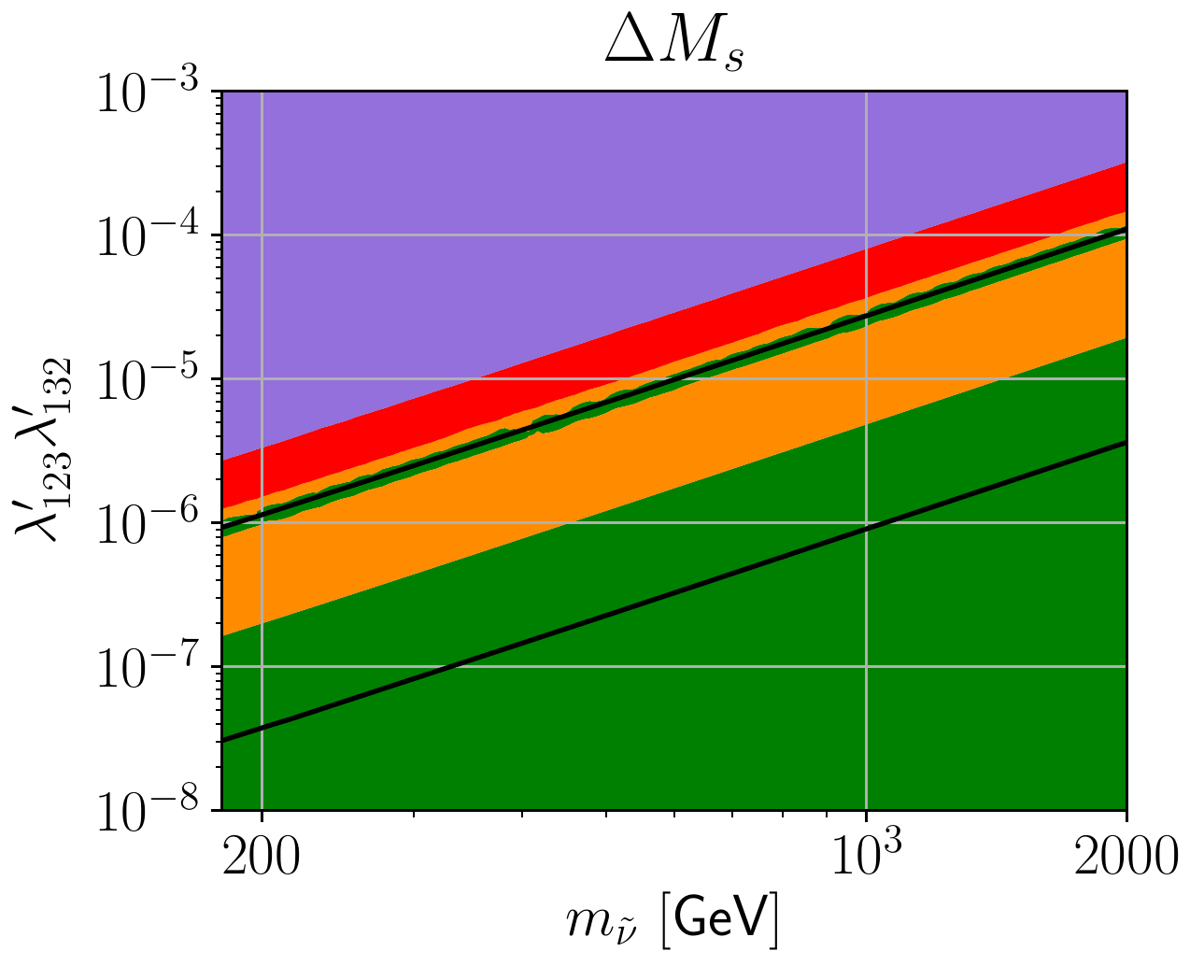}
  \includegraphics[width=0.45\textwidth]{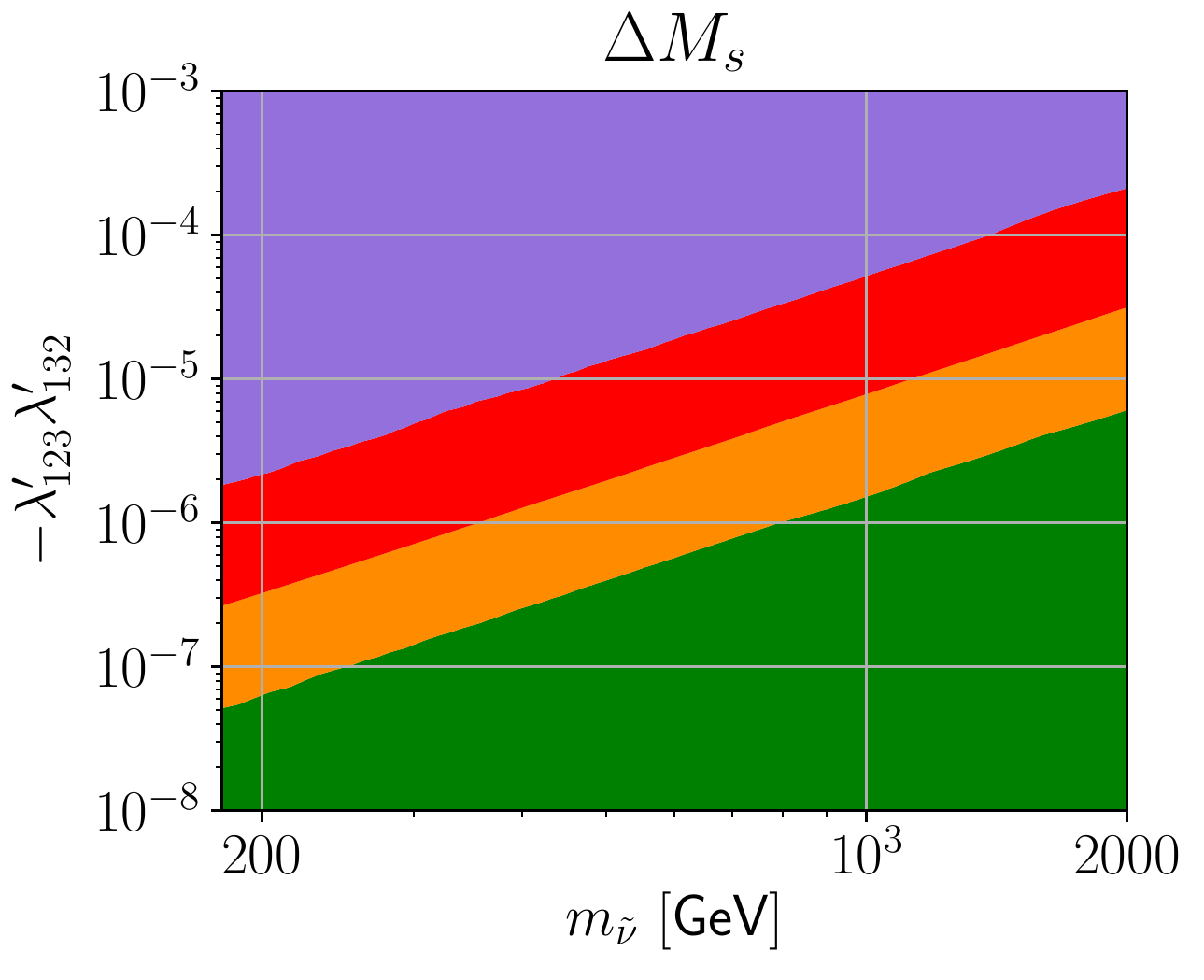}
  \includegraphics[width=0.45\textwidth]{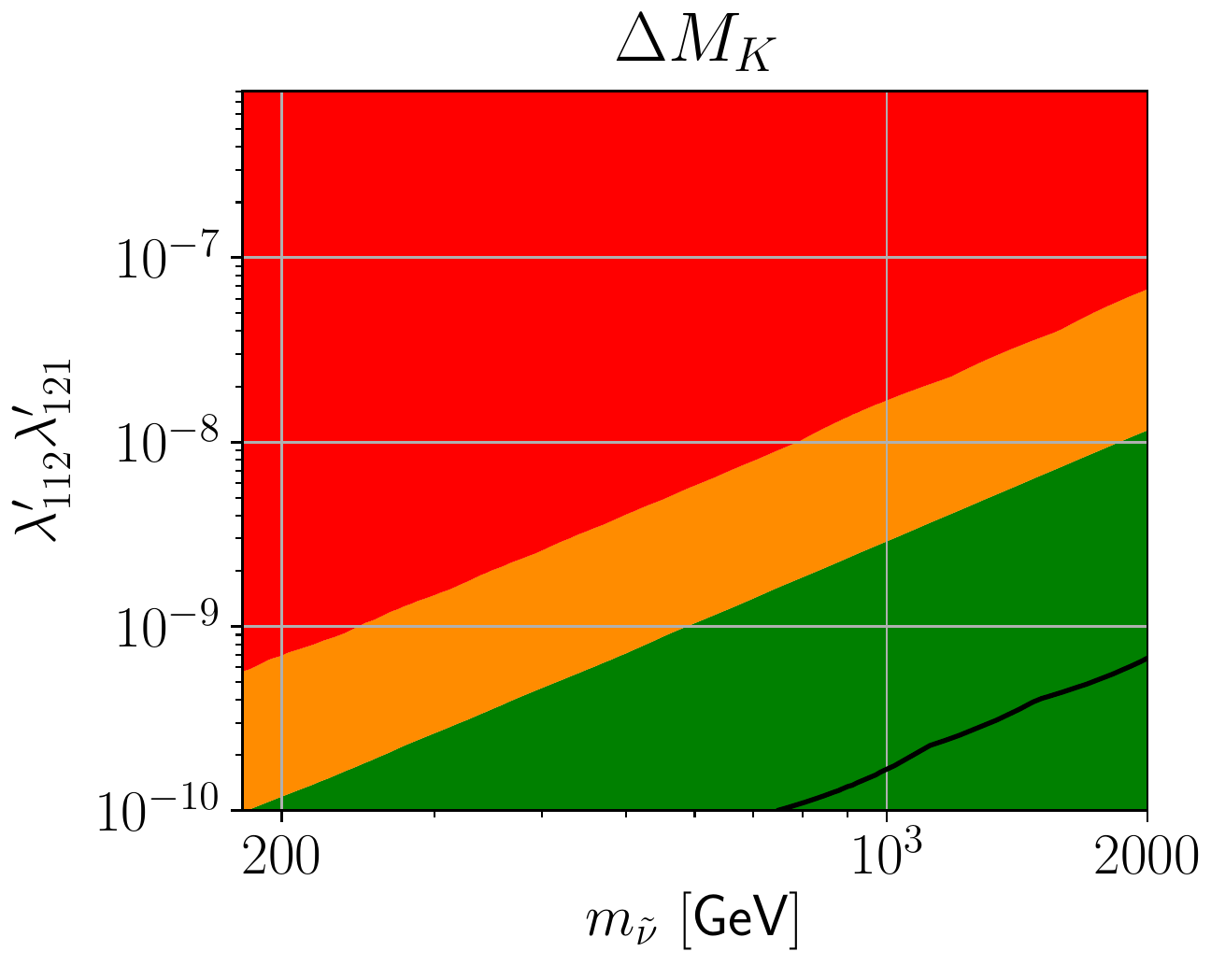}
  \includegraphics[width=0.45\textwidth]{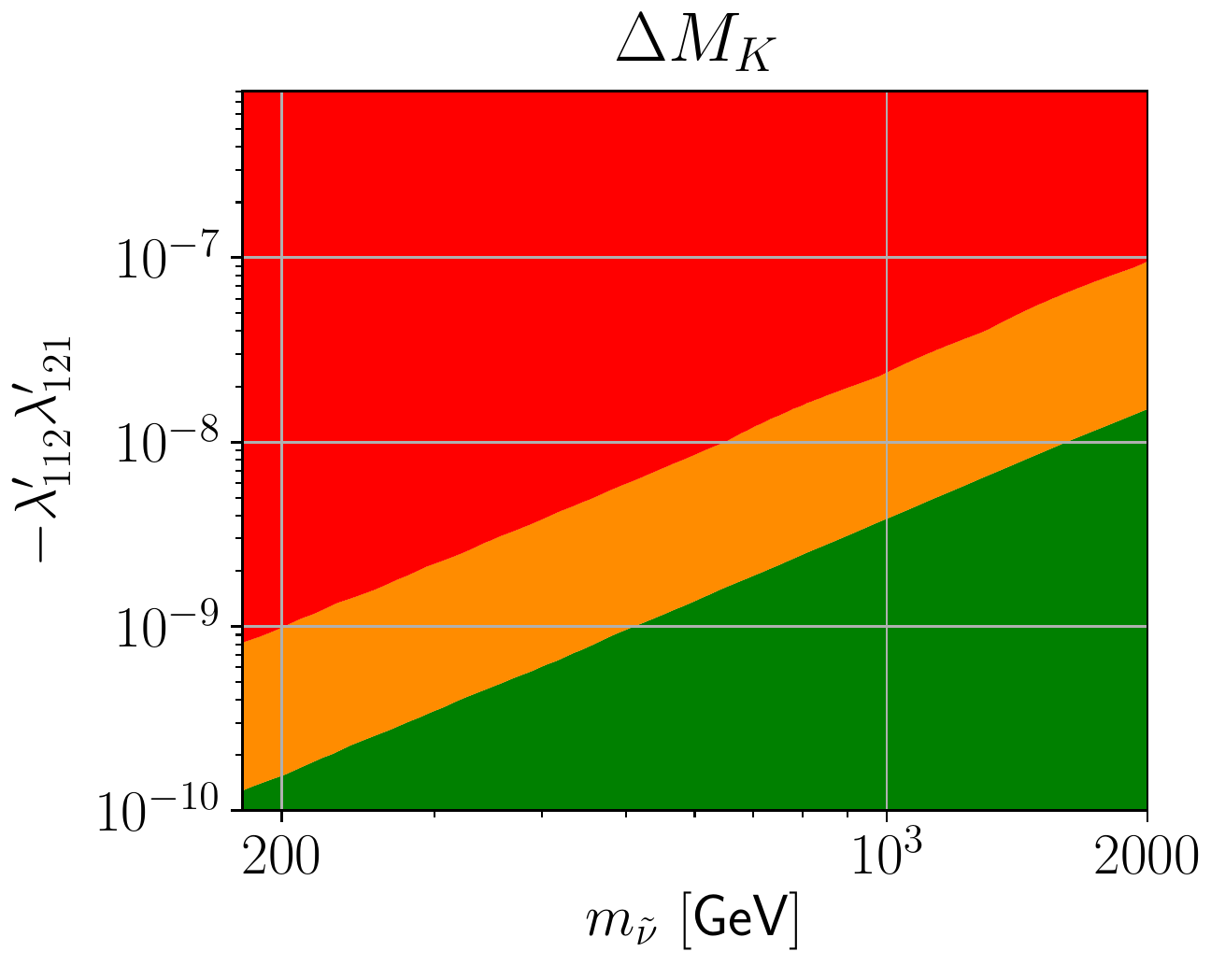}
  \caption{Constraints from the $\Delta M$'s on scenarios with RpV-mediated flavor violation contributing at tree-level, 
  as a function of the sneutrino mass.  The plots on the left correspond to the upper limit on positive $\lambda'\cdot\lambda'$; those on the right to lower limits on 
  negative $\lambda'\cdot\lambda'$ combinations. The green, orange, red and purple colors represent regions within $[0,1\sigma]$, $[1\sigma,2\sigma]$, $[2\sigma,3\sigma]$ 
  and $>3\sigma$ bounds, respectively. The experimental central value is exactly recovered on the black lines. For these plots, the parameter set of the scenario SM-like of Table \ref{Tbl:parameters_scenarios} has been employed.
  \label{treelambda}}
\end{figure}

\subsubsection{Tree Level Contributions}\label{subsec:tlc}
Let us begin with the case where only two $LQ\bar D$ couplings are simultaneously non-vanishing and contribute to the $\Delta M$'s at tree-level.
For doing so, we choose a   spectrum of the form of an effective SM at low mass, 
where we have fixed the squark, higgsino and  gaugino masses to 2\,TeV, while varying 
all the slepton masses simultaneously in the range $0.2-2$~TeV. The important parameter values are listed in the first line of Table~\ref{Tbl:parameters_scenarios}.
In addition, the stop trilinear coupling $A_t$, of order $3$~TeV (without endangering (meta)stability of the potential however\footnote{The stability of the electroweak minimum was tested for individual points. To this end, 
		we generated a model file allowing for non-vanishing squark 
		VEVs with {\tt SARAH} and tested it through the numerical code {\tt Vevacious} \cite{Camargo-Molina:2013qva}, interfaced with {\tt CosmoTransitions} \cite{Wainwright:2011kj}. A parameter point is deemed unstable on cosmological time-scales, and therefore ruled out, if the mean tunnelling time is smaller than 21.7\% of the age of the Universe.}), 
is adjusted so that the lighter Higgs mass satisfies $m_h\approx 125$~GeV (within $3$~GeV). We also considered several other scenarios, listed in 
Table~\ref{Tbl:parameters_scenarios}, 
\textit{e.g.}\ involving lighter charged Higgs or lighter squarks of the third generation, but the general properties of the constraints remained 
qualitatively unchanged. In fact, the predicted values of $\Delta M$'s in the $R_p$-conserving limit only differ at the percent level (a barely noticeable variation in view of the uncertainties) between these four scenarios, which can be placed into the perspective of the systematic suppression of the SUSY $R_p$-conserving loops due to the high squark masses. As the $R_p$-conserving contributions do not depend on the parameters that we vary in this subsection, the $n\,\sigma$-boundaries ($n=0,\cdots,3$) are only shifted by an imperceptible amount in parameter space when comparing the various scenarios of Table~\ref{Tbl:parameters_scenarios}. Therefore, we only present the results in the SM-like scenario here. 
All the input is defined at the electroweak scale, so that we can discuss the various classes of RpV-contributions to the $\Delta M$'s
without the blurring effect due to the propagation of flavor-violation via RGE's between a high-energy scale and the electroweak scale.

In Fig.~\ref{treelambda}, we present the limits set by $\Delta M_d$, $\Delta M_s$ and $\Delta M_K$ on the tree-level flavor violating contributions.
The plots in the first column are obtained for a positive product $\lambda'\cdot\lambda'$, while those in 
the second column correspond to 
negative $\lambda'\cdot\lambda'$. 
For each observable, the most relevant $\lambda'\cdot\lambda'$ 
combination, leading to a tree-level contribution, was selected. The individual sub-figures depict the extension 
of the $0,1,2,3\,\sigma$ regions in the plane defined by the corresponding flavor-violating $\lambda'\cdot\lambda'$ product and the slepton mass. 
The colors in Fig.~\ref{treelambda} are chosen such that purple regions are excluded at three standard deviations or more; red regions are excluded at 
$\geq 2\,\sigma$ --- which is the limit that we apply later on, in order to decide 
whether a point in parameter space
is excluded or allowed experimentally; the orange regions correspond to a prediction of the $\Delta M$ within $1$ and $2\,\sigma$; finally, the green areas are 
consistent with the experimental measurement within $1\,\sigma$, while the black curves reproduce the central values exactly. 
Experimental and theoretical uncertainties are added in quadrature to define the total uncertainty 
$U_{tot}=\sqrt{U_{theo}^2+U_{exp}^2}$. 
In the case of $\Delta M_K$, the theoretical uncertainties from long-distance and short-distance contributions are also combined 
quadratically. Since experimentally one cannot tell apart the 
two mass eigenstates of $B^0_{d/s}$, we simply consider 
the absolute value of $\Delta M_{d/s}$ in our evaluation. When we plot $\Delta M_{d,s}$, this feature may result in a doubling of the solutions for the 
central value or of the $1\,\sigma$-allowed regions, such as in the upper-left and middle-left plots of Fig.~\ref{treelambda}. For $K^0$, instead, the mass ordering, 
and hence the sign of $\Delta M_K$ is known. 

The limits that we obtain on the $\lambda'$ couplings contributing at tree-level are relatively tight. In the scenarios of Fig.\ref{treelambda}, 
the $2\sigma$ bounds read approximately:
\begin{equation}\begin{cases}\label{TLbounds}
\lambda'_{i13}\lambda'_{i31}\lesssim 1.6\times 10^{-6}\left(\frac{m_{\tilde{\nu}_i}}{1\,\text{TeV}}\right)^2,\qquad -\lambda'_{i13}\lambda'_{i31}\lesssim 4\times 
10^{-7}\left(\frac{m_{\tilde{\nu}_i}}{1\,\text{TeV}}\right)^2,\\
\lambda'_{i23}\lambda'_{i32}\lesssim 3.6\times 10^{-5}\left(\frac{m_{\tilde{\nu}_i}}{1\,\text{TeV}}\right)^2,\qquad -\lambda'_{i23}\lambda'_{i32}\lesssim 8\times 
10^{-6}\left(\frac{m_{\tilde{\nu}_i}}{1\,\text{TeV}}\right)^2,\\
|\lambda'_{i12}\lambda'_{i21}|\lesssim 2.2\times 10^{-8}\left(\frac{m_{\tilde{\nu}_i}}{1\,\text{TeV}}\right)^2,
\end{cases}\end{equation}
where we assume that only one lepton flavor, namely $i$, has non-vanishing RpV-couplings --- therefore the bounds only depend on the 
mass of the corresponding sneutrino $\tilde \nu_i$. Alternatively, with degenerate sneutrinos, we could sum over the index $i$ on the left-hand side of Eq.~(\ref{TLbounds}).
Limits on these products of couplings have been presented in Ref.~\cite{Allanach:1999ic} for a SUSY mass of $100$~GeV and in \cite{Wang:2010vv} for a mass of $500$~GeV
-- as explained above, our limits can be confronted to the bounds applying on $\sum_i\lambda'_{i13}\lambda'_{i31}$, \textit{etc.}, in these references.
In comparison, the bounds that we obtain in Fig.\ref{treelambda} are somewhat stronger, at least by a factor $\sim3$.
This result should be put mainly in the perspective of the reduction of the experimental uncertainty in the recent years.

\subsubsection{1-Loop Contributions to Flavor Transition}

\begin{figure}[htb]
  \centering
  \includegraphics[width=0.45\textwidth]{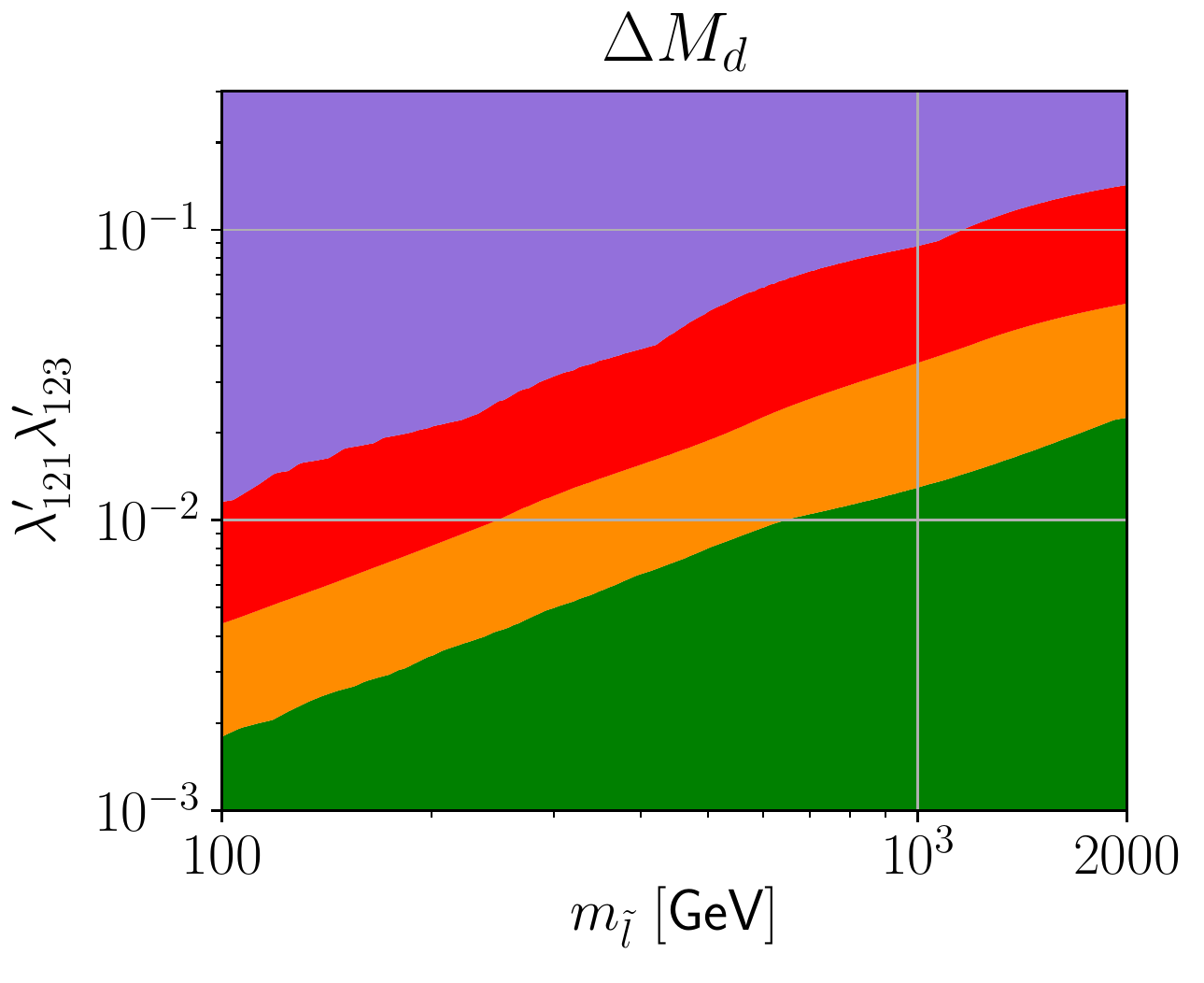}
  \includegraphics[width=0.45\textwidth]{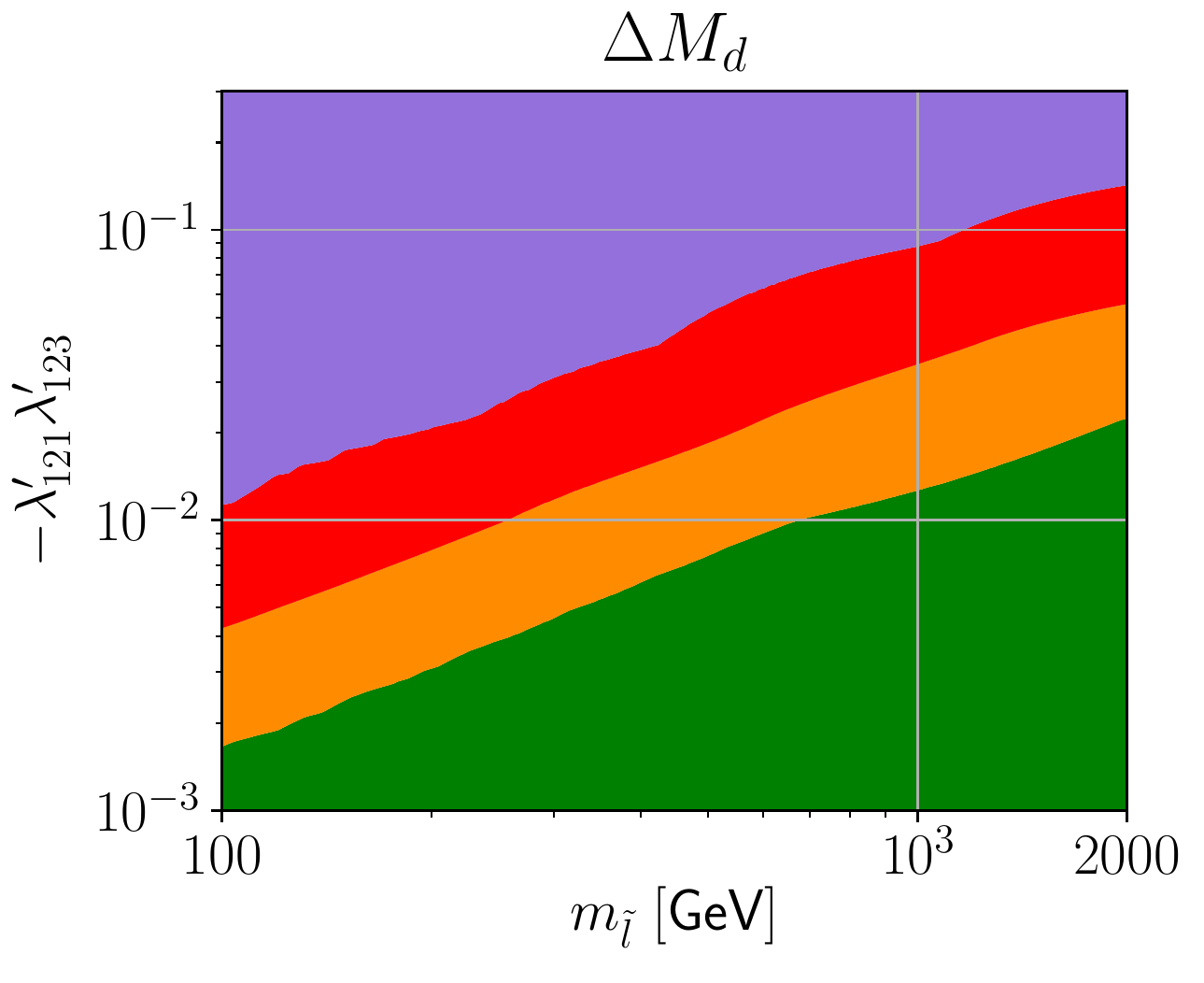}
  \includegraphics[width=0.45\textwidth]{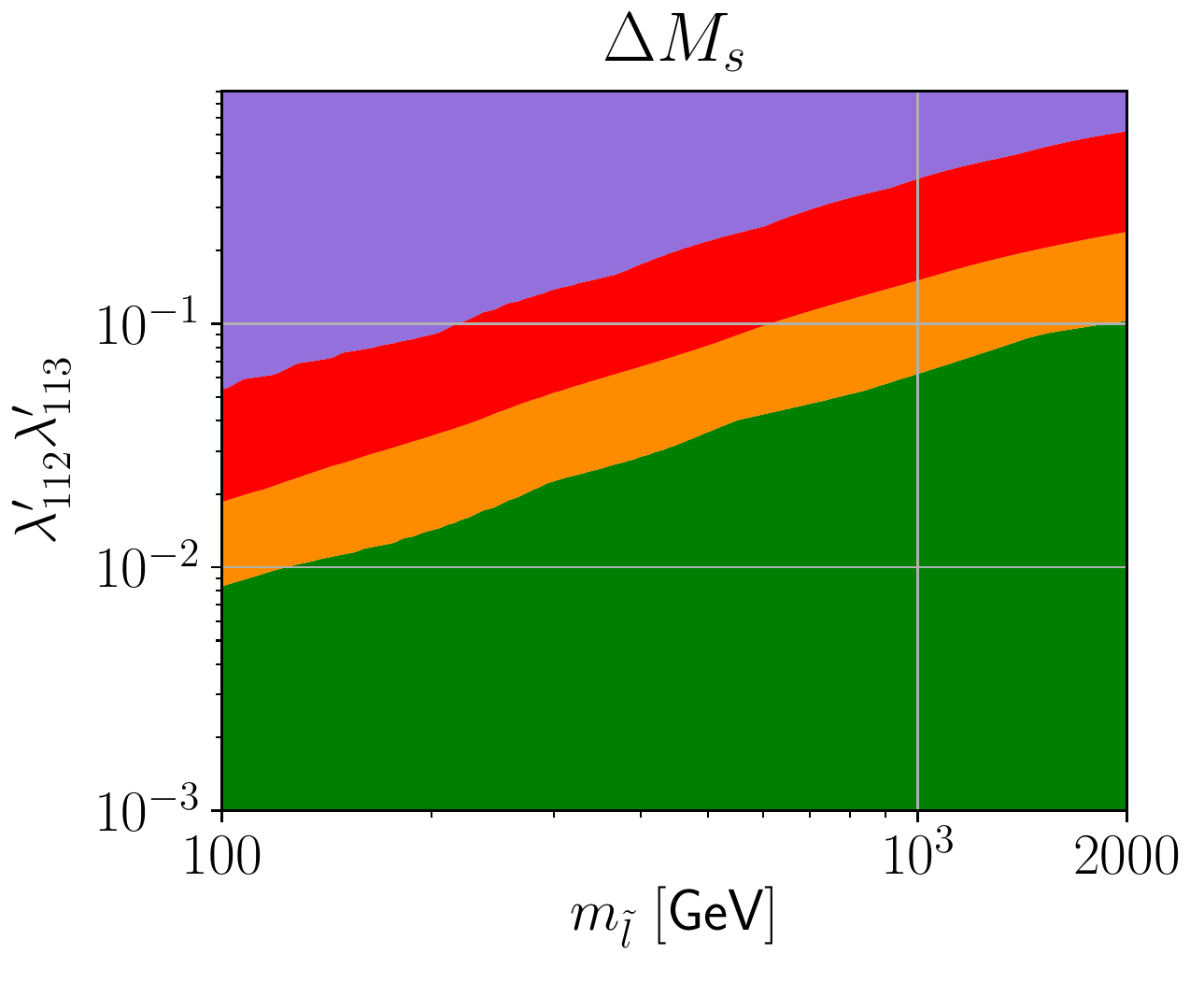}
  \includegraphics[width=0.45\textwidth]{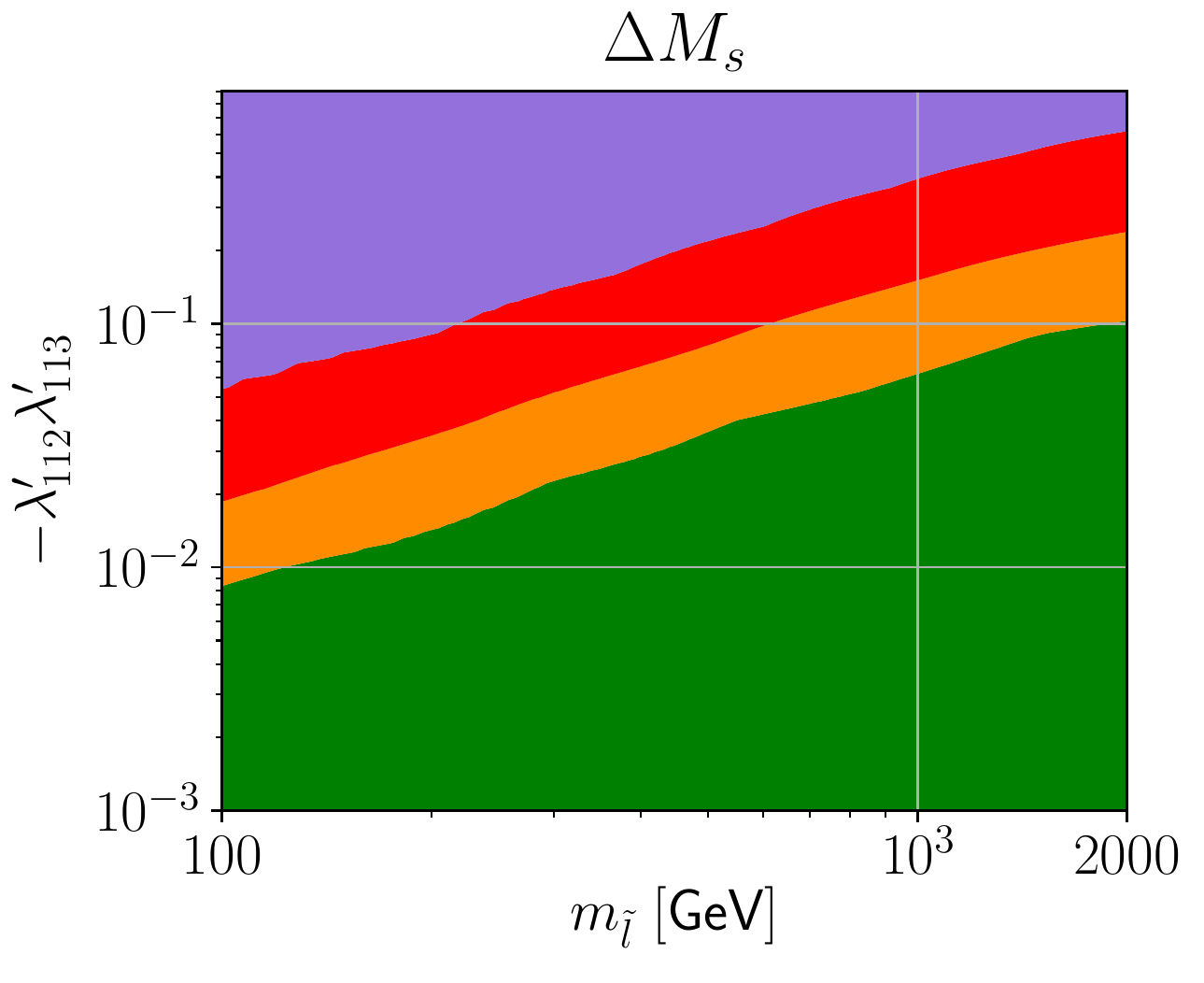}
  \includegraphics[width=0.45\textwidth]{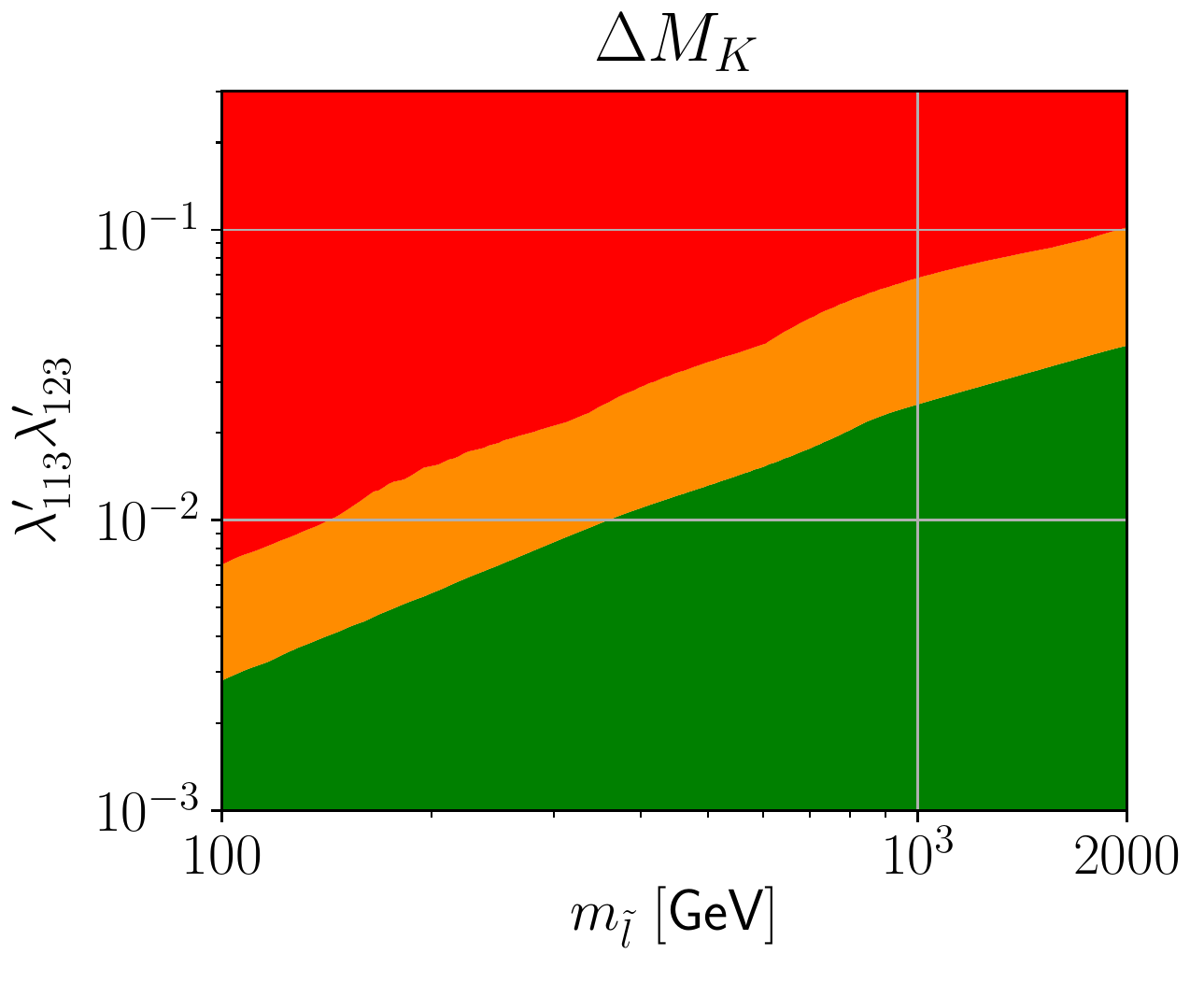}
  \includegraphics[width=0.45\textwidth]{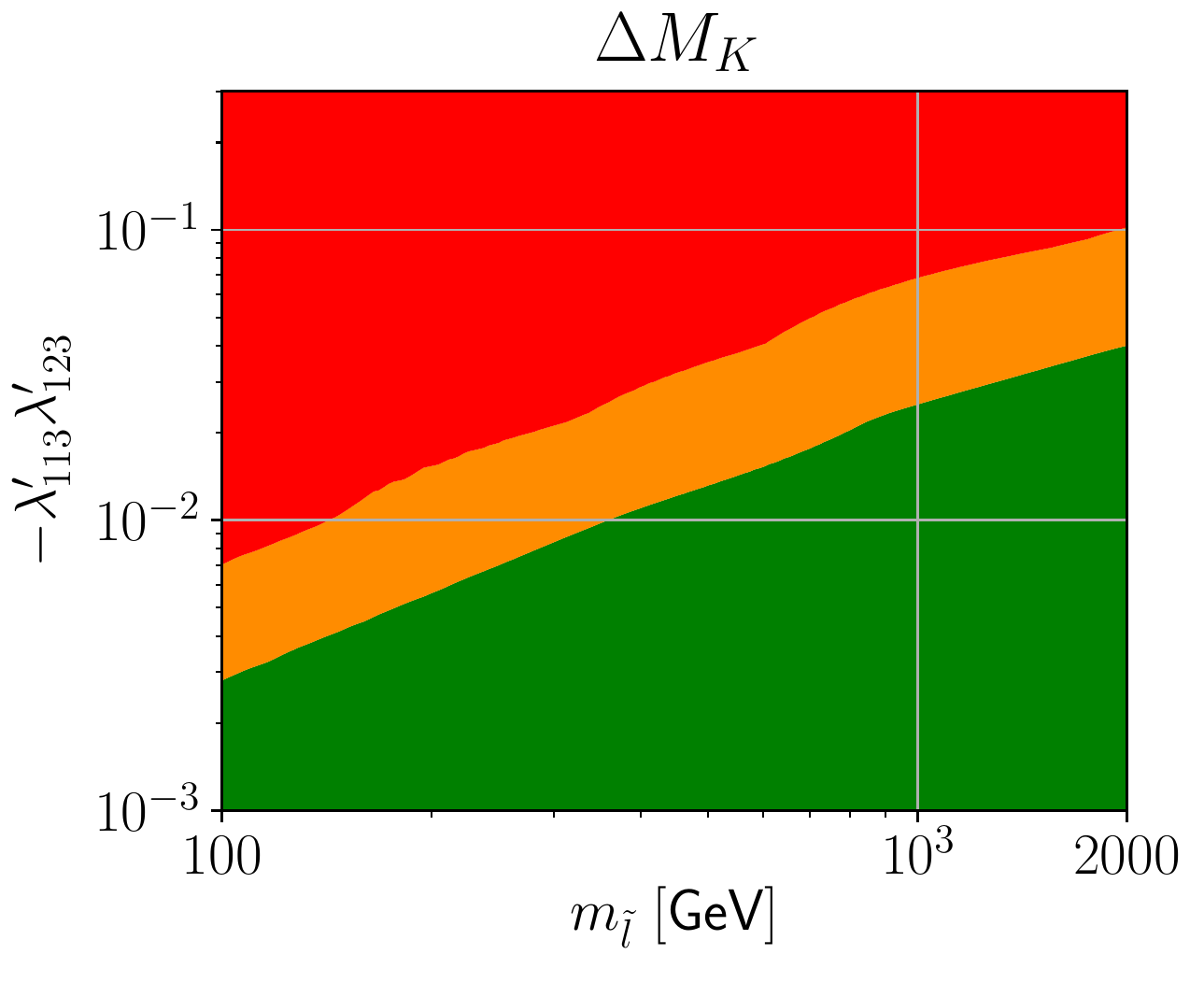}
  \caption{Constraints from the $\Delta M$'s on scenarios with RpV-mediated flavor violation of $LQ\bar D$-type,
  where the RpV-violating contribution is dominated by a box diagram. The limits are plotted against the slepton mass and follow the same
  color-code as Fig.\ref{treelambda}. For these plots, the parameter set of the scenario SUSY-RpV(a) of Table \ref{Tbl:parameters_scenarios} has been employed.
  \label{LQD_LOOP_BoxD}}
\end{figure}

Next, we turn to the case where a pair of $LQ\bar D$ couplings mediate the flavor transition only at the loop-level and we focus on coupling
combinations of the form $\lambda'_{mnI} \lambda'^*_{mnJ}$ or $\lambda'_{mIn} \lambda'^*_{mJn}$ (with $I,J$ the valence quarks of the meson). In principle we 
could consider other combinations, such as $\lambda'_{mnI} \lambda'^*_{\tilde{m}nJ}$, $\lambda'_{mnI} \lambda'^*_{m\tilde{n}J}$, $\lambda'_{mIn} \lambda'^*_{\tilde{m}Jn}$ 
or $\lambda'_{mIn} \lambda'_{mJ\tilde{n}}$ (with $m\neq\tilde{m}$, $n\neq\tilde{n}$). However, either the associated contributions are CKM suppressed
or they would require several $\lambda'\cdot\lambda'$ products to be simultaneously non-zero or non-degenerate scalar / pseudoscalar sneutrino fields.
We thus restrict ourselves to the two types mentioned above. For these, we note that the limits are independent of the flavor $m$ of the slepton field. In this context, RpV-effects in $\Delta M$'s are dominated by diagrams 
involving the comparatively light (charged or neutral) sleptons. We thus concentrate on these below. We can distinguish two types of contributions:
\begin{itemize}
 \item If one of the pair of non-vanishing $LQ\bar D$ couplings is one of those involved for the tree-level exchange diagram --- \textit{i.e.}\ if it contains
 the two flavor indices of the valence quarks of the meson --- 
 we find that quark self-energy corrections on the tree-level diagram can be comparable to or even dominant over box contributions.
 \item If neither of the non-vanishing $LQ\bar D$ couplings participates in the tree-level diagrams, box diagrams are the main contributions.
\end{itemize}
This difference impacts both the magnitude of the resulting bounds and their dependence on the slepton mass, as we shall see below. 

The spectrum that we focus on in this subsection (and later on) is described in the third row of Table~\ref{Tbl:parameters_scenarios}. The choice of the scenario SUSY-RpV(a) instead of SM-like is motivated by the wish not to systematically suppress the loop diagrams associated with charginos/neu\-tralinos. We will also comment on the mild differences that we obtain in the other scenarios of Table~\ref{Tbl:parameters_scenarios}.

In Fig.\ref{LQD_LOOP_BoxD}, we consider non-vanishing 
$\lambda'_{121}\lambda'_{123}$, $\lambda'_{112}\lambda'_{113}$ and, finally, $\lambda'_{113}\lambda'_{123}$. In these cases, 
the box diagrams dominate over the fermionic self-energy corrections. 
For each scenario, the limits from the $\Delta M$'s essentially originate in one of the three observables
$\Delta M_d$, $\Delta M_s$ or $\Delta M_K$. 
The corresponding limits approximately read:
\begin{equation}\begin{cases}
|\lambda'_{i21}\lambda'_{i23}|\lesssim 3.4\times 10^{-2}\left(\frac{m_{\tilde{\l}_i}}{1\,\text{TeV}}\right),\\
|\lambda'_{i12}\lambda'_{i13}|\lesssim 1.6\times 10^{-1}\left(\frac{m_{\tilde{\l}_i}}{1\,\text{TeV}}\right),\\
|\lambda'_{i13}\lambda'_{i23}|\lesssim 6.3\times 10^{-2}\left(\frac{m_{\tilde{\l}_i}}{1\,\text{TeV}}\right),
\end{cases}\end{equation}
where $m_{\tilde{\l}_i}$denotes the mass of the degenerate sneutrinos and charged sleptons. Here, we note that the mass dependence of the form $(\lambda'\cdot \lambda')^2 < c\cdot m_{\tilde \ell}^2$ differs from that appearing 
when the RpV-contribution intervenes at tree-level. 
It is characteristic of the leading RpV-diagrams in the considered setup, corresponding to the box formed out of two charged sleptons and
two up-type quarks in the internal lines and to the box consisting of two sneutrinos and two down-type quarks: these diagrams roughly scale as 
$(\lambda'\cdot \lambda')^2/m^2_{\tilde \ell}$. 
As a consequence, the limits for positive and negative $\lambda'\cdot \lambda'$ products are comparable.
In addition, the bounds on $\lambda'\cdot \lambda'$ now scale about linearly with the sparticle mass.

Expectedly, the limits are much weaker in these box-dominated scenarios than in the case where the flavor transition appears at tree-level.
Refs.~\cite{Bhattacharyya:1998be,Saha:2003tq,Wang:2010vv} presented limits on the corresponding coupling-combinations for a sfermion mass of $100$ or 
$500$~GeV. The bounds that we derive are of the same order. Similarly to the case where the RpV-contribution to the flavor transition is mediated at tree-level, the investigation of the various scenarios of Table~\ref{Tbl:parameters_scenarios} results in very little variations.

\begin{figure}[htb]
  \centering
  \includegraphics[width=0.45\textwidth]{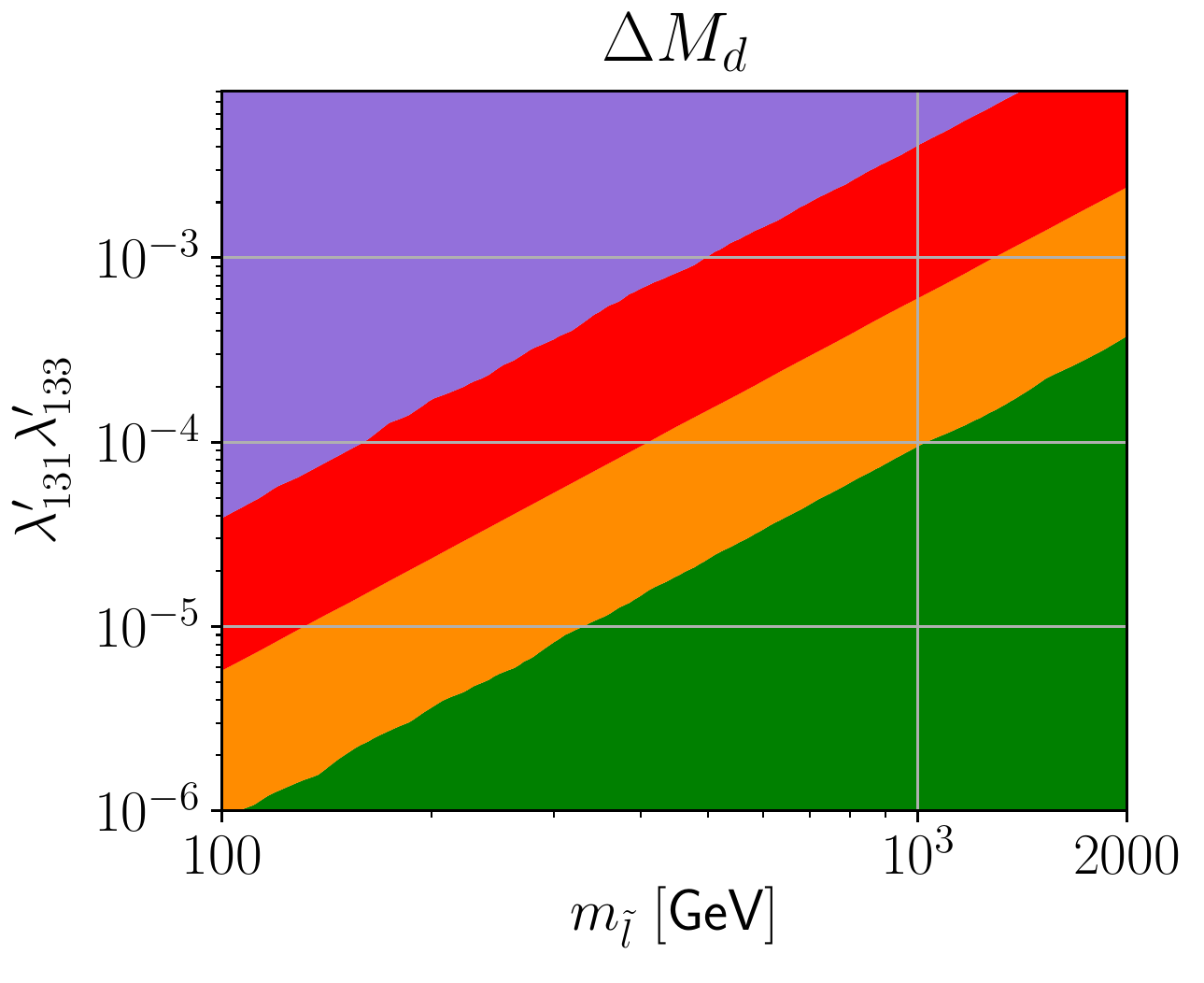}
  \includegraphics[width=0.45\textwidth]{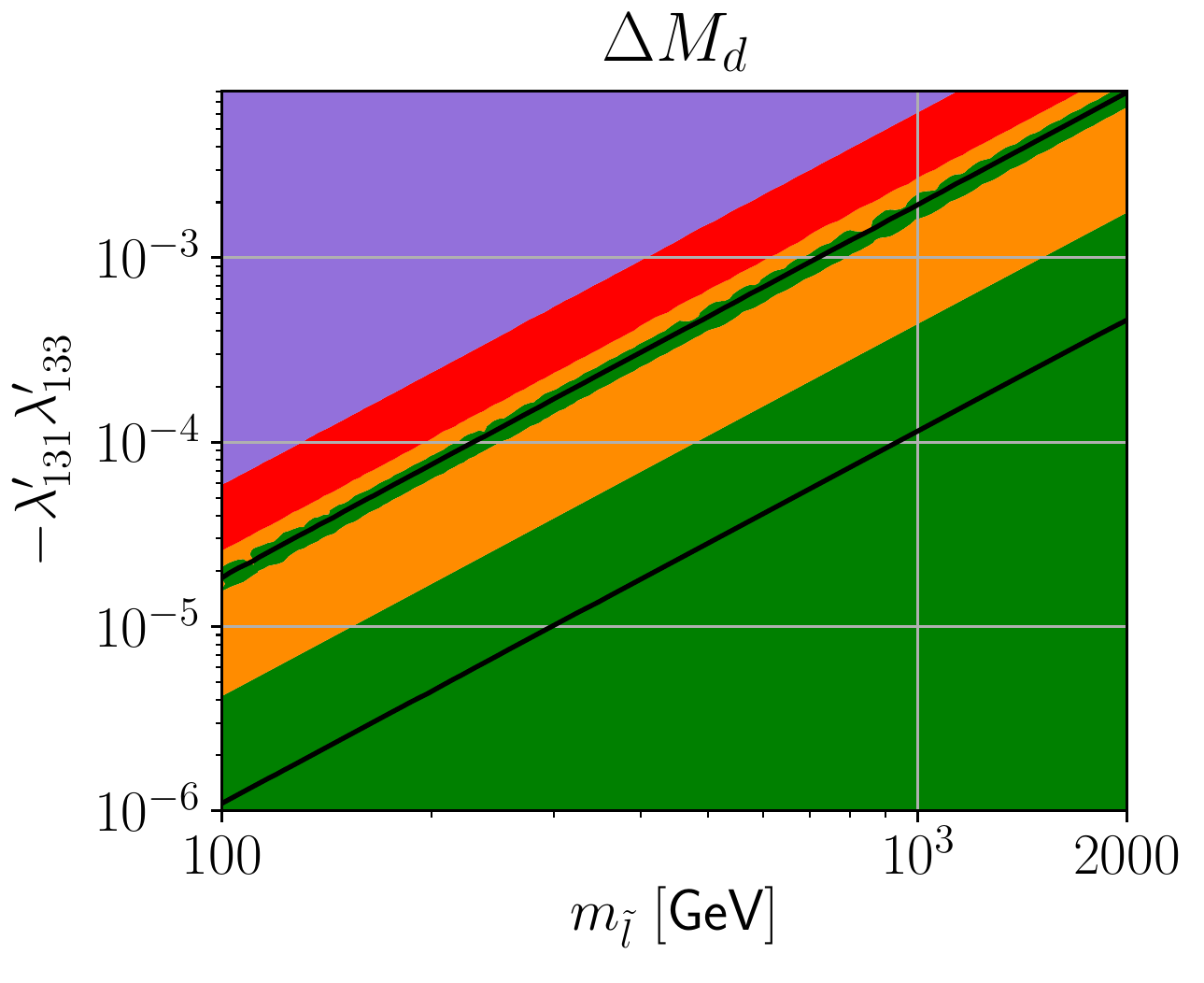}
  \includegraphics[width=0.45\textwidth]{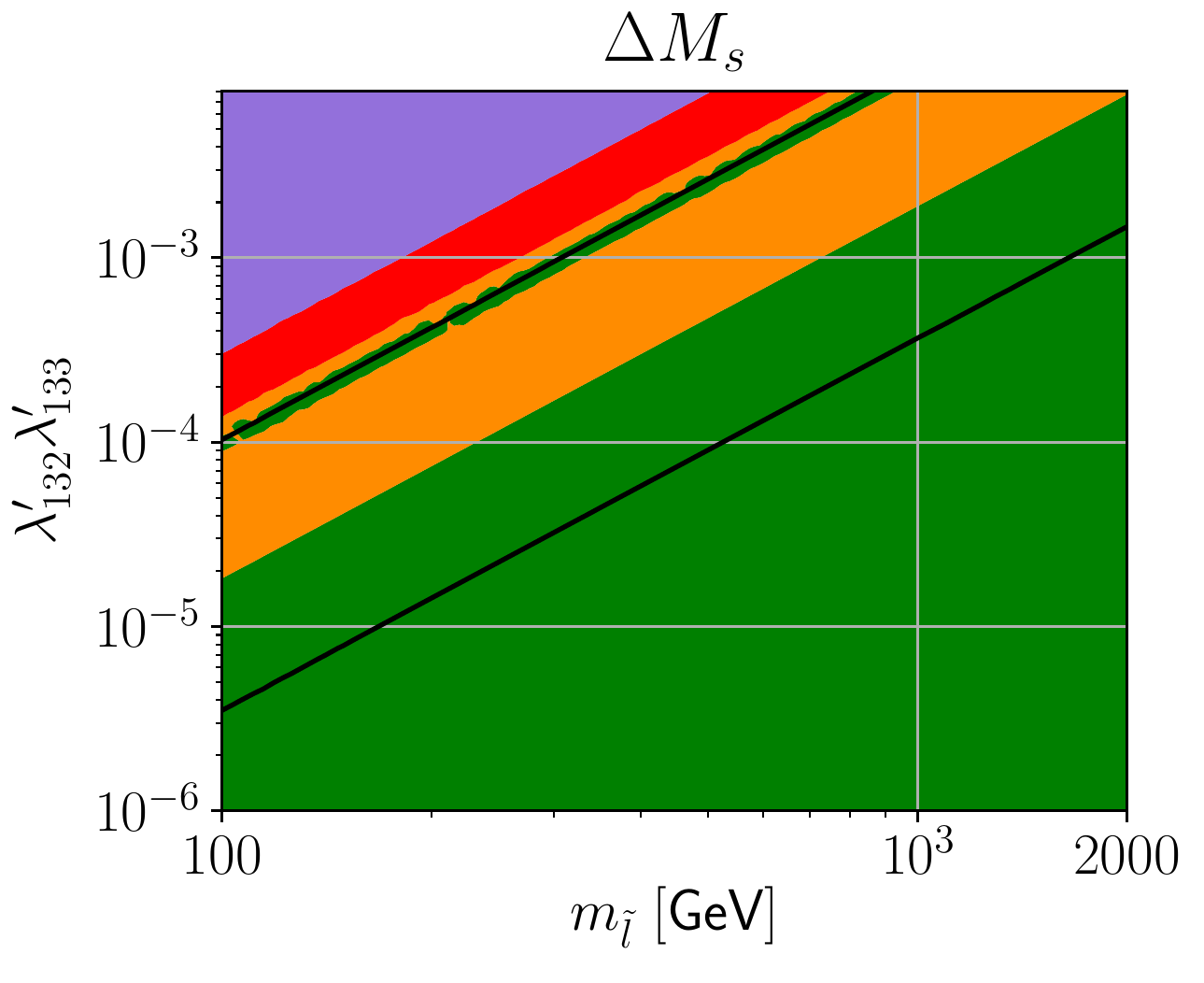}
  \includegraphics[width=0.45\textwidth]{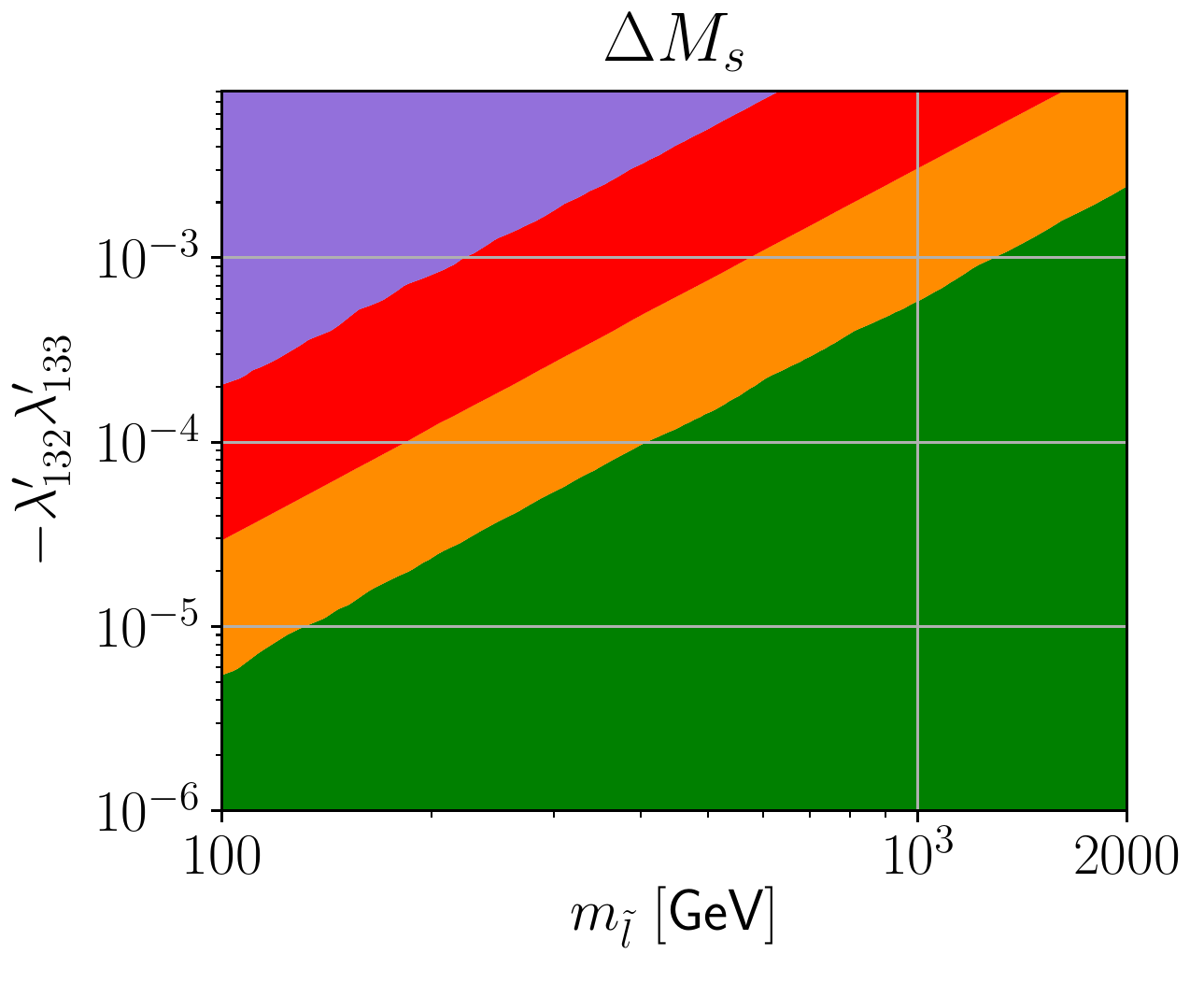}
  \includegraphics[width=0.45\textwidth]{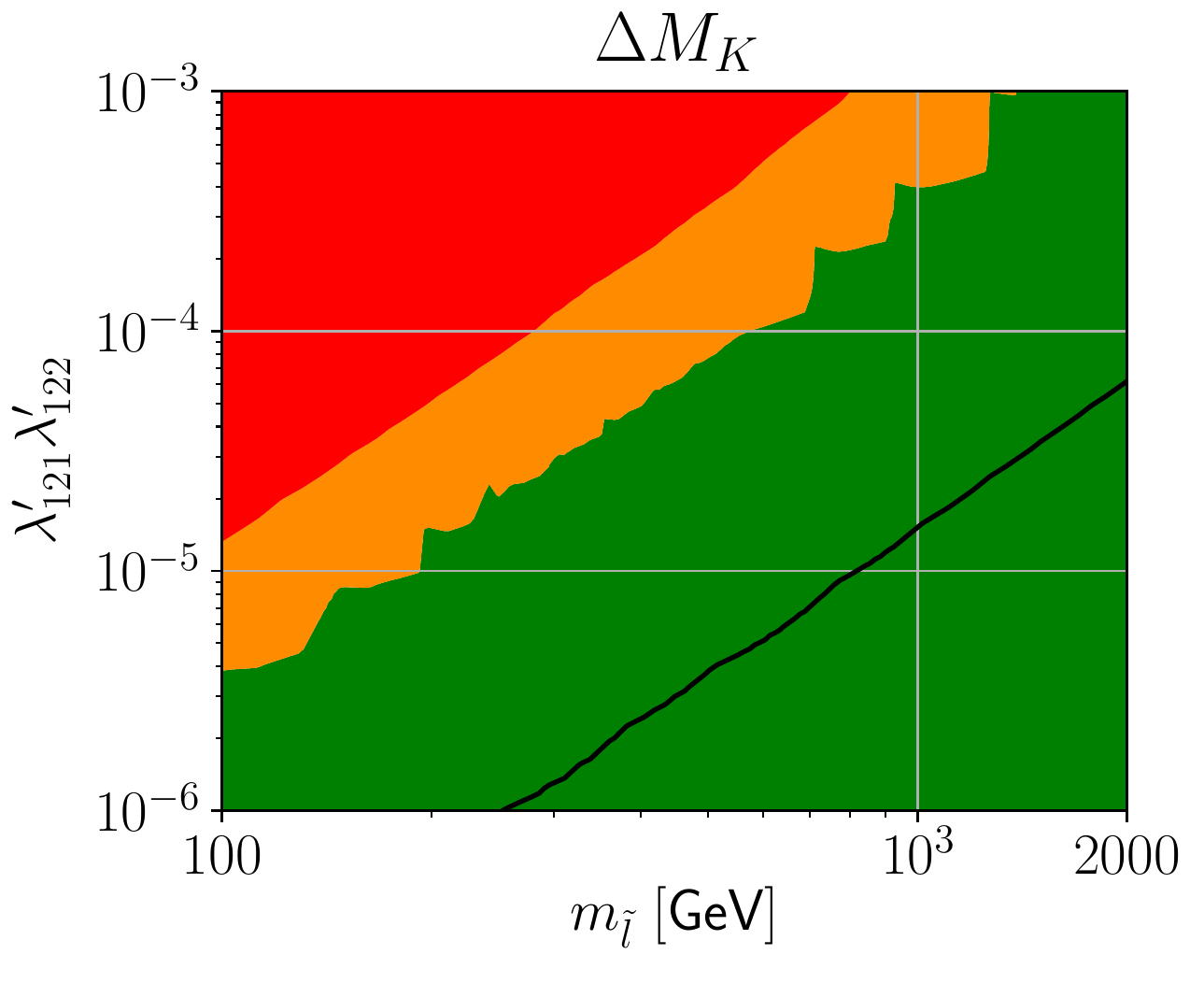}
  \includegraphics[width=0.45\textwidth]{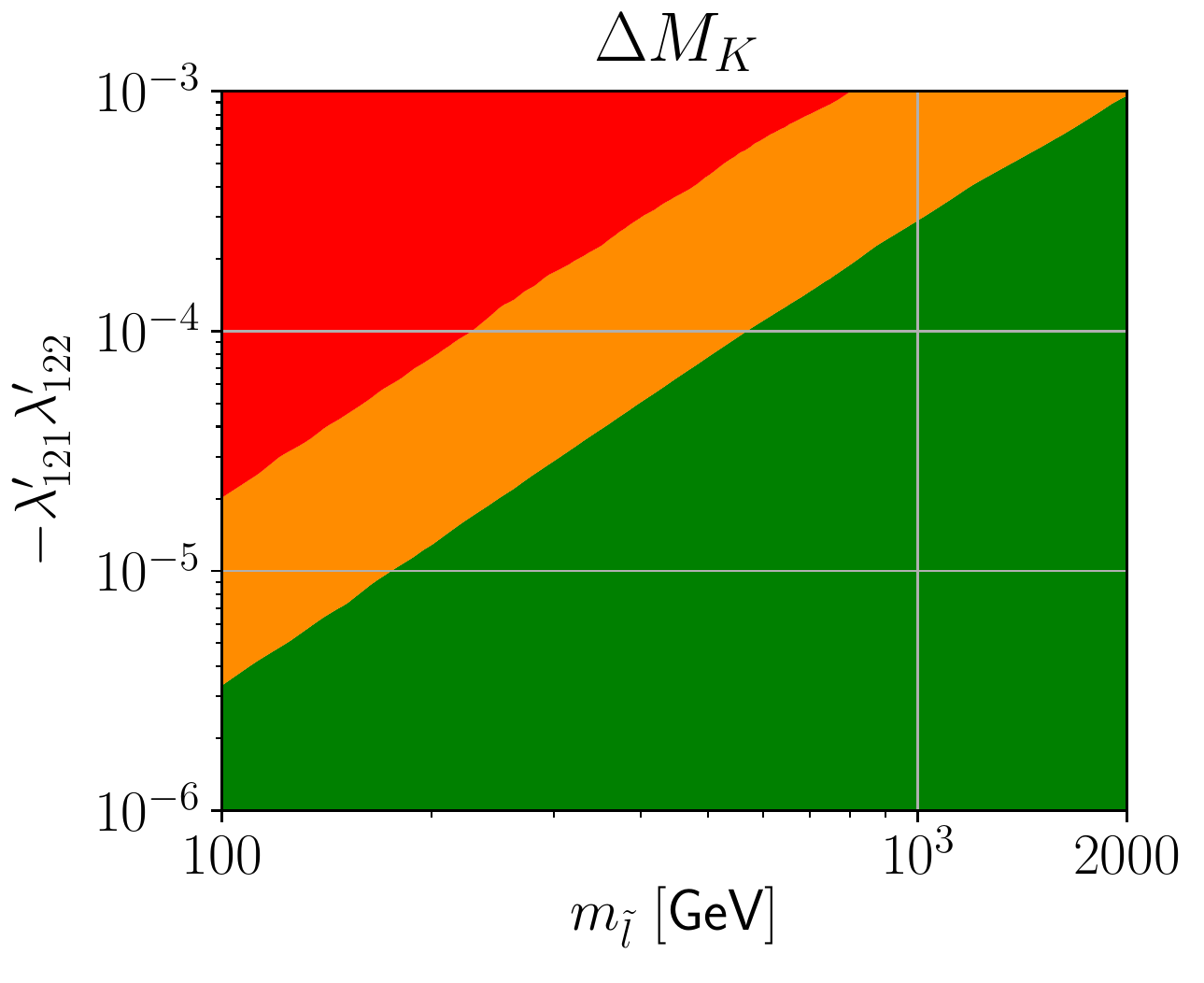}
  \caption{Constraints from the $\Delta M$'s on scenarios with RpV-mediated flavor violation of $LQ\bar D$-type, where the dominant
  RpV-diagram involves a one-loop quark self-energy. The limits are plotted against the sneutrino mass and follow the color code of Fig.~\ref{treelambda}. For these plots, the parameter set of the scenario SUSY-RpV(a) of Table \ref{Tbl:parameters_scenarios} has been employed.
 \label{LQD_LOOP_QSED}} 
\end{figure}

Finally, we turn to the case where one of the non-vanishing $\lambda'$ involves both flavors of the valence quarks of the $K^0$, $B^0_{d,s}$ meson while
the other is flavor-diagonal (and contains only one of the valence flavors). Then, the dominant diagrams are of the form of Fig.~\ref{dia:Tree_quarkSE}: one 
$\Delta F=1$ transition is mediated by the non-vanishing $\lambda'$ with both valence-flavor indices, while the second $\Delta F=1$ transition appears at the 
loop level --- typically through a SM loop ($W$/up-type quark), \textit{i.e.}\ in association with the CKM matrix. We stress that such contributions were dismissed 
in previous analyses and are considered here for the first time.

\begin{table}
\centering
\begin{tabular}{|c|c|c|c|c|c|} \hline
\multicolumn{2}{|c|}{$\Delta m_{B_d^0}$} & \multicolumn{2}{|c|}{$\Delta m_{B_s^0}$} & \multicolumn{2}{|c|}{$\Delta m_{K^0}$}\\ \hline
$|\lambda'_{ijk} \cdot \lambda'_{imn}|$ & $2\sigma$ bound & $|\lambda'_{ijk} \cdot \lambda'_{imn}|$ & $2\sigma$ bound & $|\lambda'_{ijk} \cdot \lambda'_{imn}|$ & $2\sigma$ bound \\ \hline \hline
$(i31) (i13)^{(\text{T})}$ & $1.6\times 10^{-6}$ &$(i32) (i23)^{(\text{T})}$ & $3.6\times 10^{-5}$ & $(i12) (i21)^{(\text{T})}$ &  $2.2\times 10^{-8}$   \\
$(i11) (i13)^{(\text{S})}$ & $1.8\times 10^{-3}$ & $(i22) (i23)^{(\text{S})}$ & $9.5\times 10^{-3}$ & $(i12) (i11)^{(\text{S})}$ & $1.5\times 10^{-3}$   \\
$(i21) (i13)^{(\text{S})}$ & $[2.8\times 10^{-4}]$ & $(i12) (i23)^{(\text{S})}$ & $[4.2\times 10^{-2}]$  &$(i22) (i21)^{(\text{S})}$ & $1.5\times 10^{-3}$   \\
$(i31) (i23)^{(\text{S})}$ & $0.15$  &$(i32) (i13)^{(\text{S})}$ & $0.33$ & $(i12) (i31)^{(\text{S})}$ & $9\times 10^{-6}$   \\
$(i31) (i33)^{(\text{S})}$ & $2.7\times 10^{-3}$  &$(i32) (i33)^{(\text{S})}$ & $1.4\times 10^{-2}$ &  $(i32) (i21)^{(\text{S})}$ &   $4.2\times 10^{-5}$  \\
$(i21) (i23)^{(\text{B})}$ & $3.4\times 10^{-2}$ & $(i12) (i13)^{(\text{B})}$ & $0.16$  &  $(i32) (i11)^{(\text{B})}$  & $0.64$    \\
$(i21) (i33)^{(\text{B})}$ & $0.64$ & $(i22) (i33)^{(\text{B})}$ & $0.74$ & $(i22) (i31)^{(\text{B})}$  &   $0.24$ \\
$(i11) (i33)^{(\text{B})}$ & $0.64$ &   $(i12) (i33)^{(\text{B})}$ & 4  &$(i22) (i11)^{(\text{B})}$ &  4  \\
$(i11) (i23)^{(\text{B})}$ & N/A &   $(i22) (i13)^{(\text{B})}$ & N/A  & $(i32) (i31)^{(\text{B})}$  &   $0.01$ \\ \hline
$(i12) (i31)^{(\text{S})}$ & $[0.012]$ & $(i23) (i31)^{(\text{S})}$ & N/A  & $(i21) (i11)^{(\text{S})}$ & $5\times 10^{-3}$   \\
$(i13) (i32)^{(\text{S})}$ & $[0.73]$ & $(i22) (i32)^{(\text{S})}$ & $0.23$ & $(i22) (i12)^{(\text{S})}$ & $5.8\times 10^{-3}$   \\
$(i13) (i33)^{(\text{B})}$ & $0.05$ & $(i23) (i33)^{(\text{S})}$ & $0.24$ &$(i23) (i12)^{(\text{S})}$ &   $2.2\times 10^{-2}$  \\
$(i11) (i31)^{(\text{B})}$ & $0.07$ & $(i21) (i32)^{(\text{S})}$ & $[2.25]$ & $(i21) (i13)^{(\text{S})}$ &  $2.3\times 10^{-4}$  \\
$(i12) (i32)^{(\text{B})}$ & $0.05$  &  $(i21) (i31)^{(\text{B})}$ & $0.21$  &  $(i23) (i13)^{(\text{B})}$ & $6.3\times 10^{-2}$  \\
\hline

\end{tabular}
\caption{Compilation of the latest bounds on relevant couplings of $LQ\bar D$ operators, coming from the considered meson oscillation observables. 
These limits were established with the spectrum defined in the row SUSY-RpV(a) of Table~\ref{Tbl:parameters_scenarios}, with slepton and sneutrino masses
of 1\,TeV. The precise $2\,\sigma$ boundary obviously depends on the sign of the non-vanishing $\lambda'\cdot\lambda'$ product: we always apply the most conservative (weakest) 
limit.
In the list of couplings, the comment ``(T)/(S)/(B)'' indicates that the coupling product is dominated by a tree-level/quark self-energy/box contribution. ``N/A'' means that we did not identify
upper-limits on the couplings below $4\pi$ (a rough limit from perturbativity considerations).
Above the horizontal line, the non-vanishing coupling combinations select right-handed external quarks. Below this line, the external quarks are left-handed.
The scaling with the sneutrino/slepton mass is roughly quadratic for all $\lambda'\cdot\lambda'$ products that contain both valence flavors in (at least) 
one of the non-vanishing $\lambda'$, linear otherwise: see more precise explanation in the main body of the text. 
Some combinations contribute to two observables, such as $\lambda'_{i13}\lambda'_{i32}$, relevant for both $\Delta M_d$ and $\Delta M_s$. In such a case, 
the square brackets identify the weaker limit.}

\label{tab:bounds_LQD}
\end{table}

Corresponding scenarios are displayed in Fig.\ref{LQD_LOOP_QSED}, where $\Delta M_{B_d}$, $\Delta M_{B_s}$ and $\Delta M_{K}$ are plotted against $\lambda'_{131}\cdot \lambda'_{133}$,
$\lambda'_{132}\cdot \lambda'_{133}$ and $\lambda'_{121}\cdot \lambda'_{122}$, respectively. The bounds have a comparable scaling to that appearing in the scenario with 
tree-level sneutrino exchange, but the constraints are far weaker. At $2\,\sigma$:
\begin{equation}
\begin{cases}
\lambda'_{i31}\lambda'_{i33}\lesssim  6\times 10^{-4}  \left(\frac{m_{\tilde{\nu}_i}}{1\,\text{TeV}}\right)^2, \qquad  -\lambda'_{i31}\lambda'_{i33}\lesssim  2.7\times 10^{-3}  \left(\frac{m_{\tilde{\nu}_i}}{1\,\text{TeV}}\right)^2,\\
\lambda'_{i32}\lambda'_{i33}\lesssim 1.4\times 10^{-2} \left(\frac{m_{\tilde{\nu}_i}}{1\,\text{TeV}}\right)^2,\qquad -\lambda'_{i32}\lambda'_{i33}\lesssim 3\times 10^{-3} \left(\frac{m_{\tilde{\nu}_i}}{1\,\text{TeV}}\right)^2,\\
|\lambda'_{i21}\lambda'_{i22}|\lesssim 1.5\times 10^{-3} \left(\frac{m_{\tilde{\nu}_i}}{1\,\text{TeV}}\right)^2,
\end{cases}
\end{equation}
where $-\lambda'_{i31}\lambda'_{i33}, -\lambda'_{i32}\lambda'_{i33}>0$. Due to the inclusion of the missing and obviously relevant
self-energy diagrams, the bounds that we report are accordingly tighter than in the literature\cite{Bhattacharyya:1998be,Saha:2003tq,Wang:2010vv}. If we compare the various scenarios of Table~\ref{Tbl:parameters_scenarios}, we again observe little change at the qualitative level. However, the exact position of the $n\,\sigma$ ($n=0,\cdots,3$) boundaries is shifted by a numerical prefactor of order unity, homogeneous in the whole range of scanned parameters of Fig.~\ref{LQD_LOOP_QSED}. This prefactor is characteristic of the magnitude $R_p$-conserving loop entering the off-diagonal quark self-energy. For example, the upper-bounds on $\lambda'_{131}\lambda'_{133}$ are stronger by a factor $\sim2$ in the SM-like scenario, as compared to the scenario SUSY-RPV(a) (shown in the plots), by a factor $\sim1.3$ in the scenario 2HDM and by a factor $\sim1.6$ in the scenario SUSY-RPV(b). Other numbers (of the same order) intervene for the two other considered sets of $\lambda'\cdot\lambda'$.

In Table~\ref{tab:bounds_LQD}, we compile the $2\,\sigma$ bounds on $\lambda'\cdot\lambda'$ products that we derive for $1$\,TeV sleptons in the scenario
SUSY-RpV(a) of Table~\ref{Tbl:parameters_scenarios} (the limits depend only weakly on the chosen scenario). In this list, the pairs $\lambda'\cdot\lambda'$ are taken
non-zero only one at a time and, in particular, for a unique (s)lepton flavor $i$. As explained above, the scaling with the slepton/sneutrino mass depends on the choice 
of non-vanishing $\lambda'$: essentially quadratic if at least one of the non-vanishing $\lambda'$ contains both valence-flavors of the decaying meson, linear otherwise.
One of the $\Delta M$'s is usually more sensitive to a specific $\lambda'\cdot\lambda'$ product than the other two. etc.

\subsection{Bounds on a pair of simultaneously non-zero \boldmath $\bar U \bar D\bar D$ couplings}
\label{subsubsec:UDD}

 \begin{figure}[h]
\centering
\includegraphics[width=.45\linewidth]{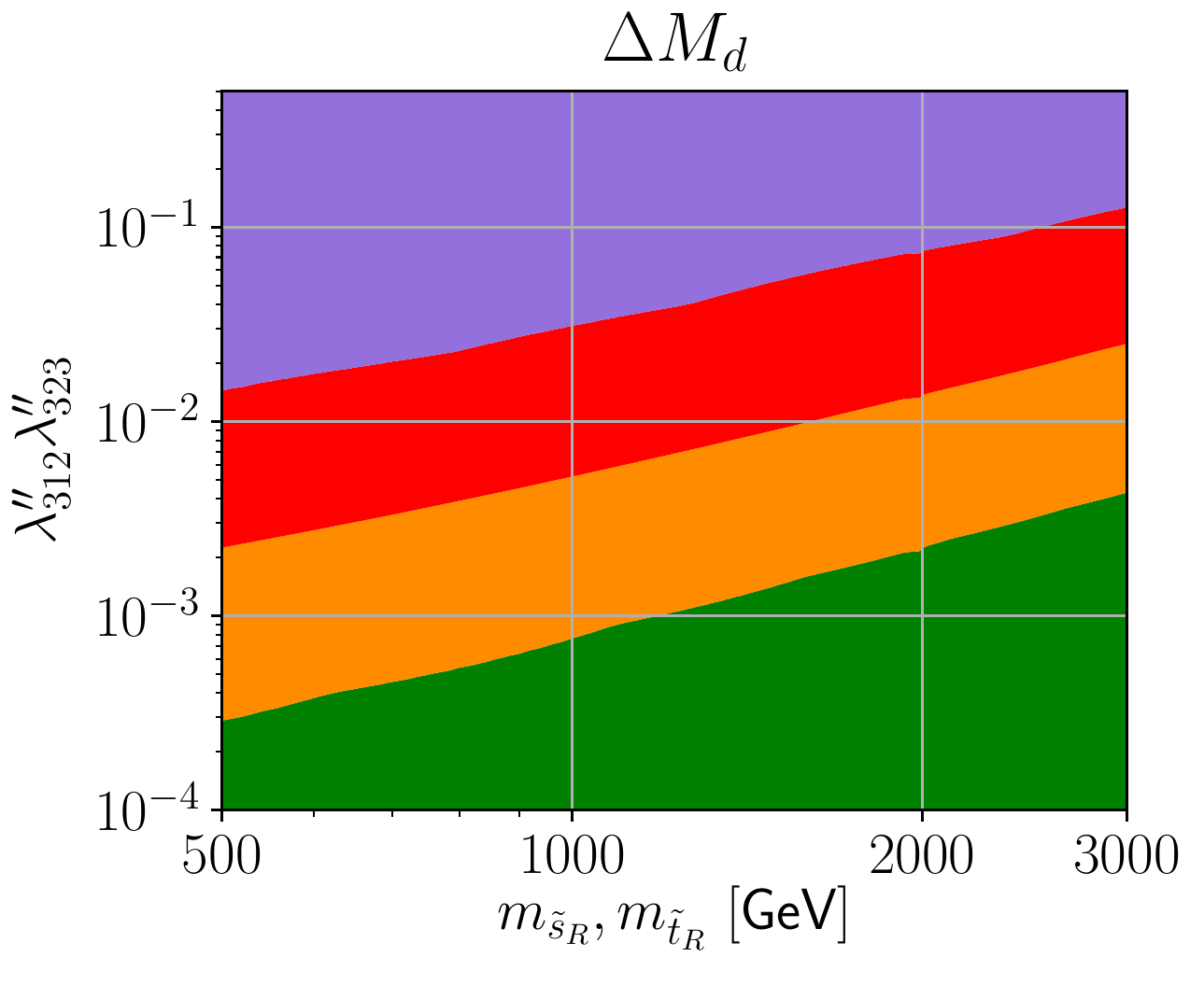}
\includegraphics[width=.45\linewidth]{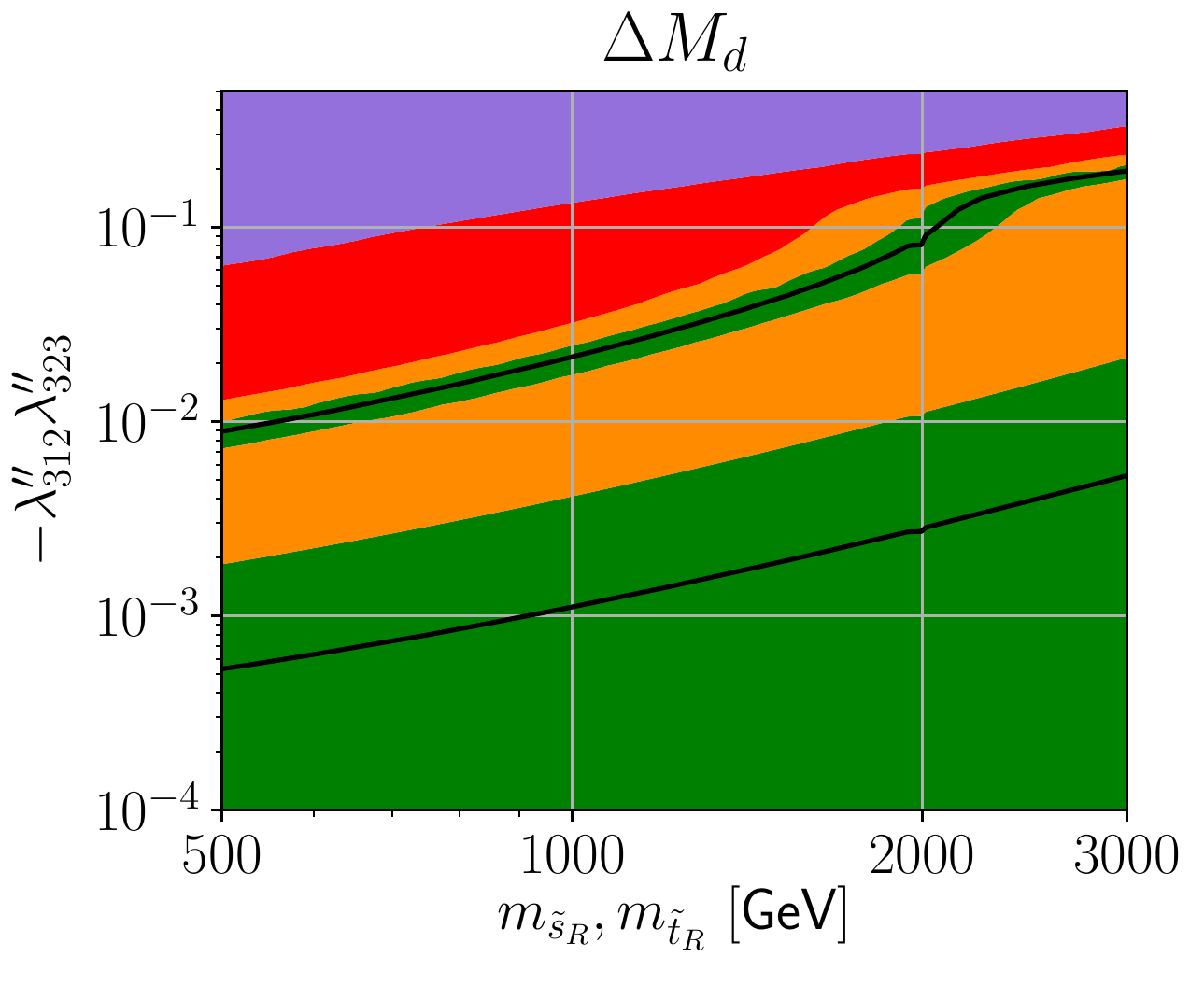}
\includegraphics[width=.45\linewidth]{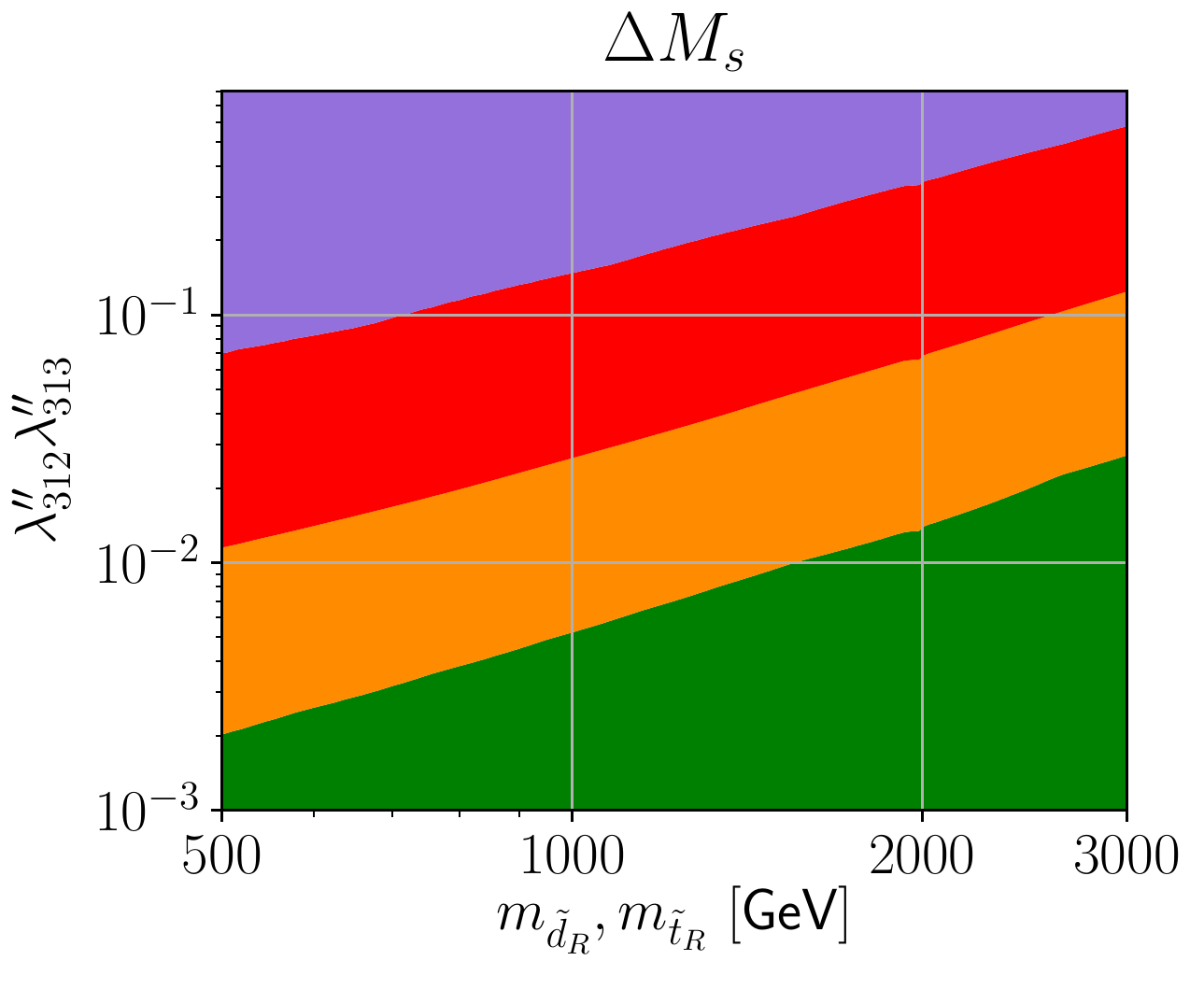}
\includegraphics[width=.45\linewidth]{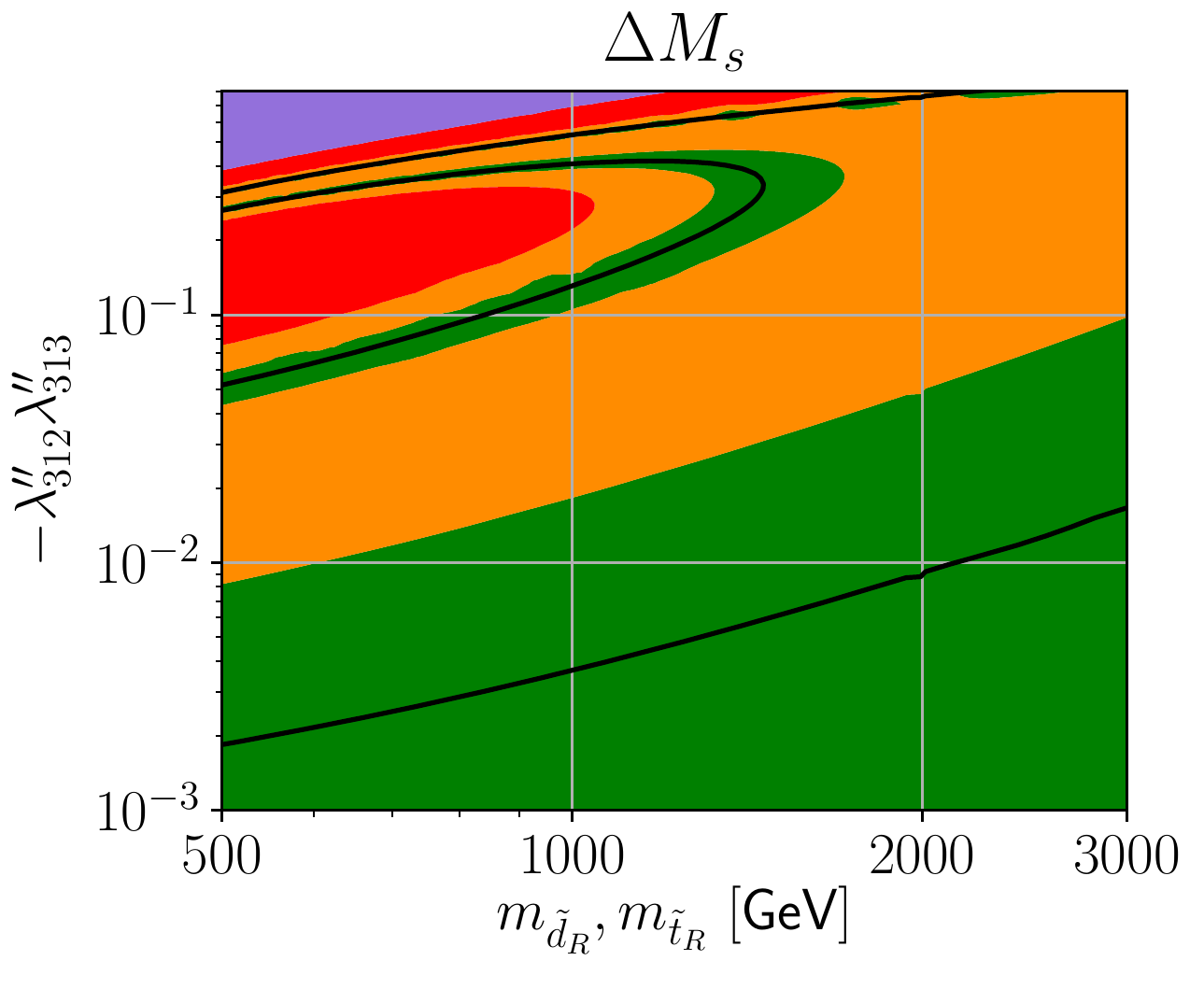}
\includegraphics[width=.45\linewidth]{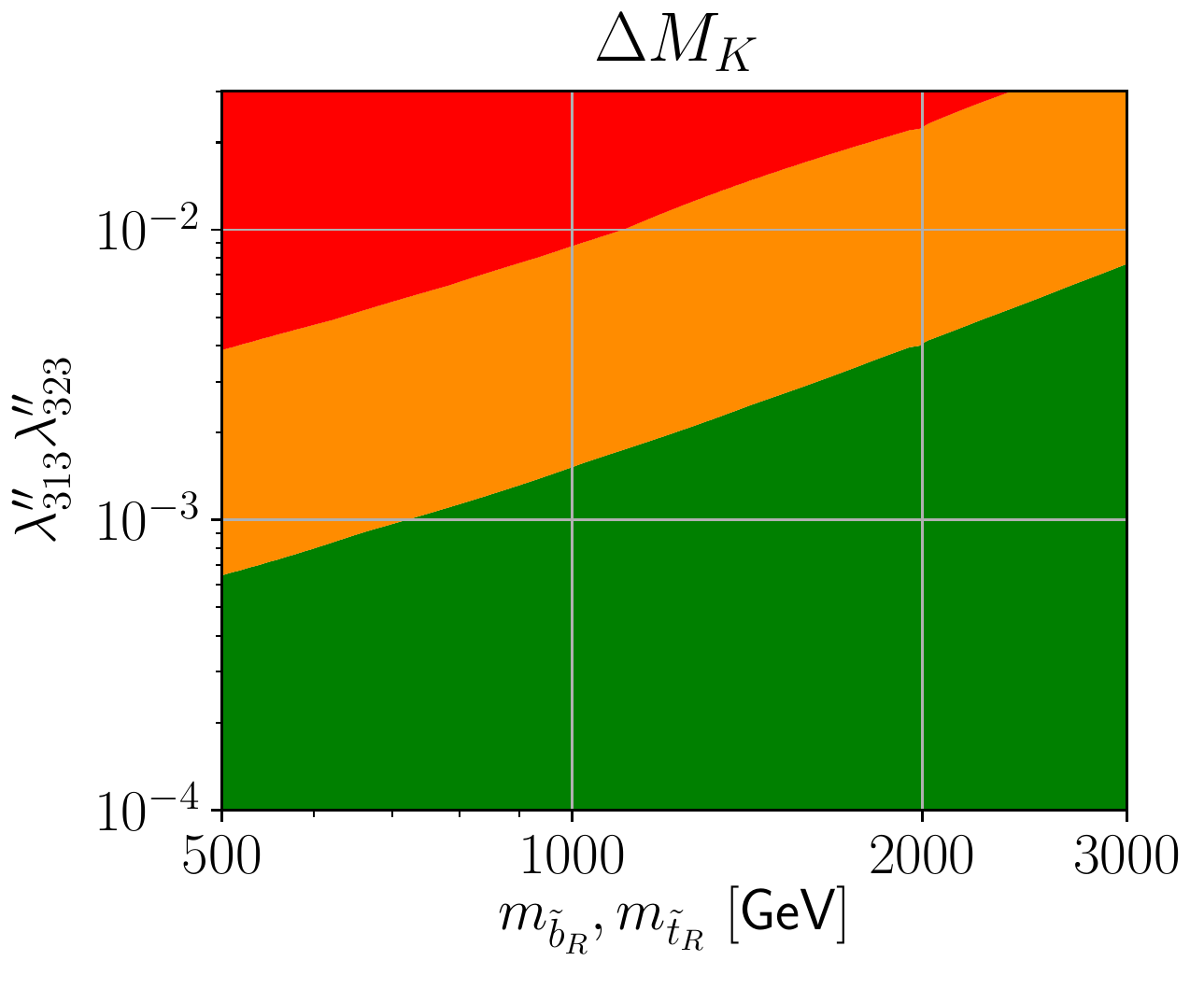}
\includegraphics[width=.45\linewidth]{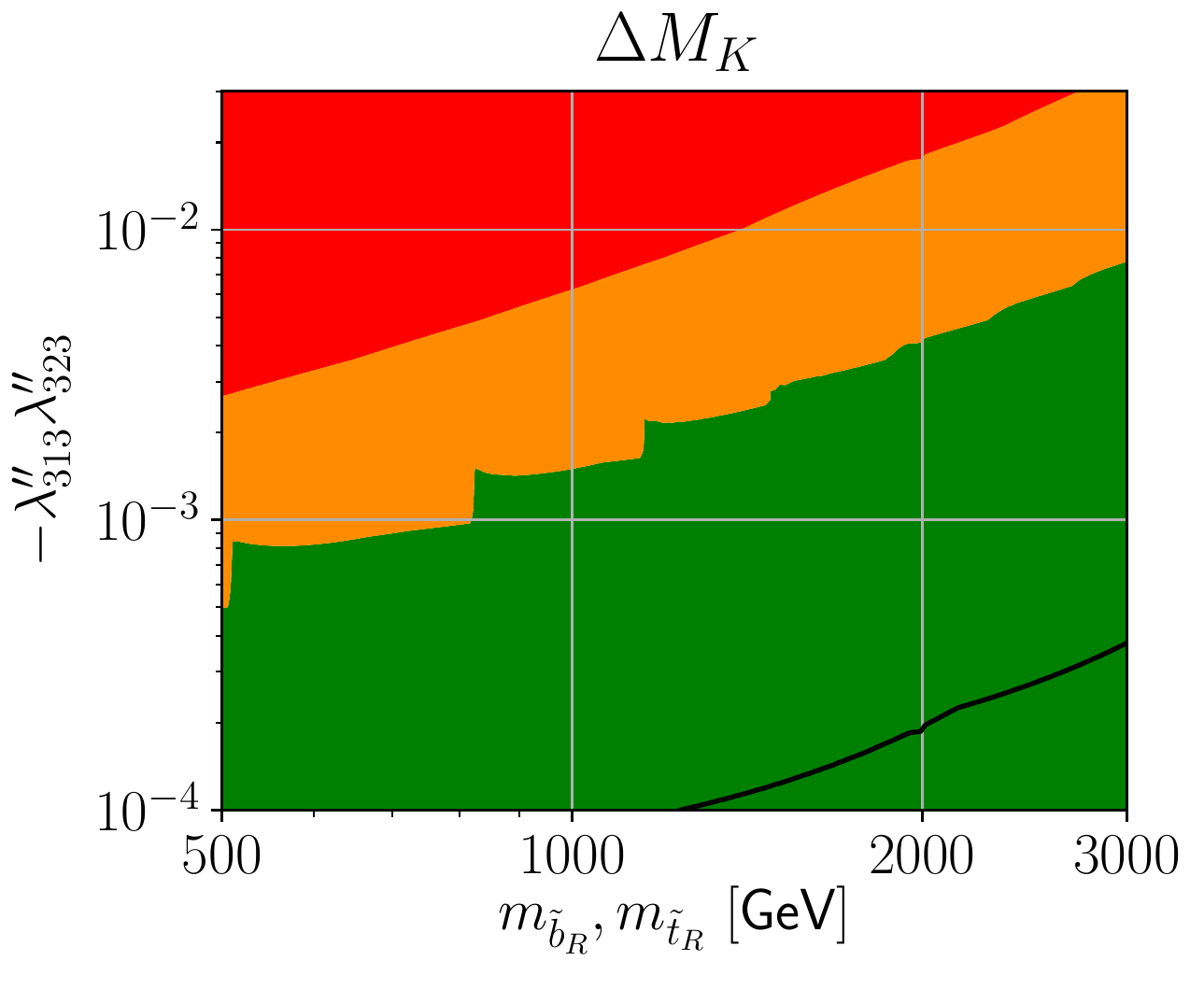}
\caption{Limits on $\bar U_3 \bar D_i \bar D_j$ couplings from the meson oscillation parameters. Internal (s)top lines are allowed by such couplings.
The color code is similar to that of the previous plots. For these plots, the parameter set of the scenario SUSY-RpV(a) of Table \ref{Tbl:parameters_scenarios} has been employed except for the squark masses that are scanned over.
\label{fig:squark_mass_vs_box_couplings_UDD}}
\end{figure}

We proceed with our analysis and now consider baryonic RpV, \textit{i.e.}\ non-zero $\bar U \bar D \bar D$ couplings. 
The corresponding RpV-effects appear only at the radiative level and are dominated by box diagrams. Contrarily to existing analyses \cite{deCarlos:1996yh}, 
we always consider heavy gluinos (as indicated by the current status of LHC searches), so that the associated diagrams generally remain subdominant.
In this setup, three classes of diagrams compete: (1) boxes including two squarks and two quarks in internal lines, which scale like $(\lambda''\cdot\lambda'')^2$, 
(2) boxes including two quarks, one squark and a $W$-boson, which scale like $\lambda''\cdot\lambda''$ but involve a CKM-suppression and a quark-chirality flip, 
and (3) similarly boxes with two squarks, one quark and a chargino, which scale like $\lambda''\cdot\lambda''$. 
The matter of the chirality flip can be easily understood as only right-handed quarks couple via $\lambda''$ but only left-handed quarks couple to a $W$.
Therefore, such diagrams with an internal $W$ line are mostly relevant when the internal quark line involves a top-quark. 
As to the boxes with an internal chargino line, we also find that such contributions are mainly relevant for an internal stop line: indeed, the 
higgsino contribution scales with the Yukawa coupling, hence is suppressed for squarks of first or second generation. In addition, the gaugino contribution relies
on left-right squark mixing, which we keep negligible for squarks of the first and second generation --- making the assumption that the trilinear
soft terms are proportional to the Yukawa couplings \cite{Nilles:1983ge}.

 \begin{figure}[tbh]
\centering
\includegraphics[width=.45\linewidth]{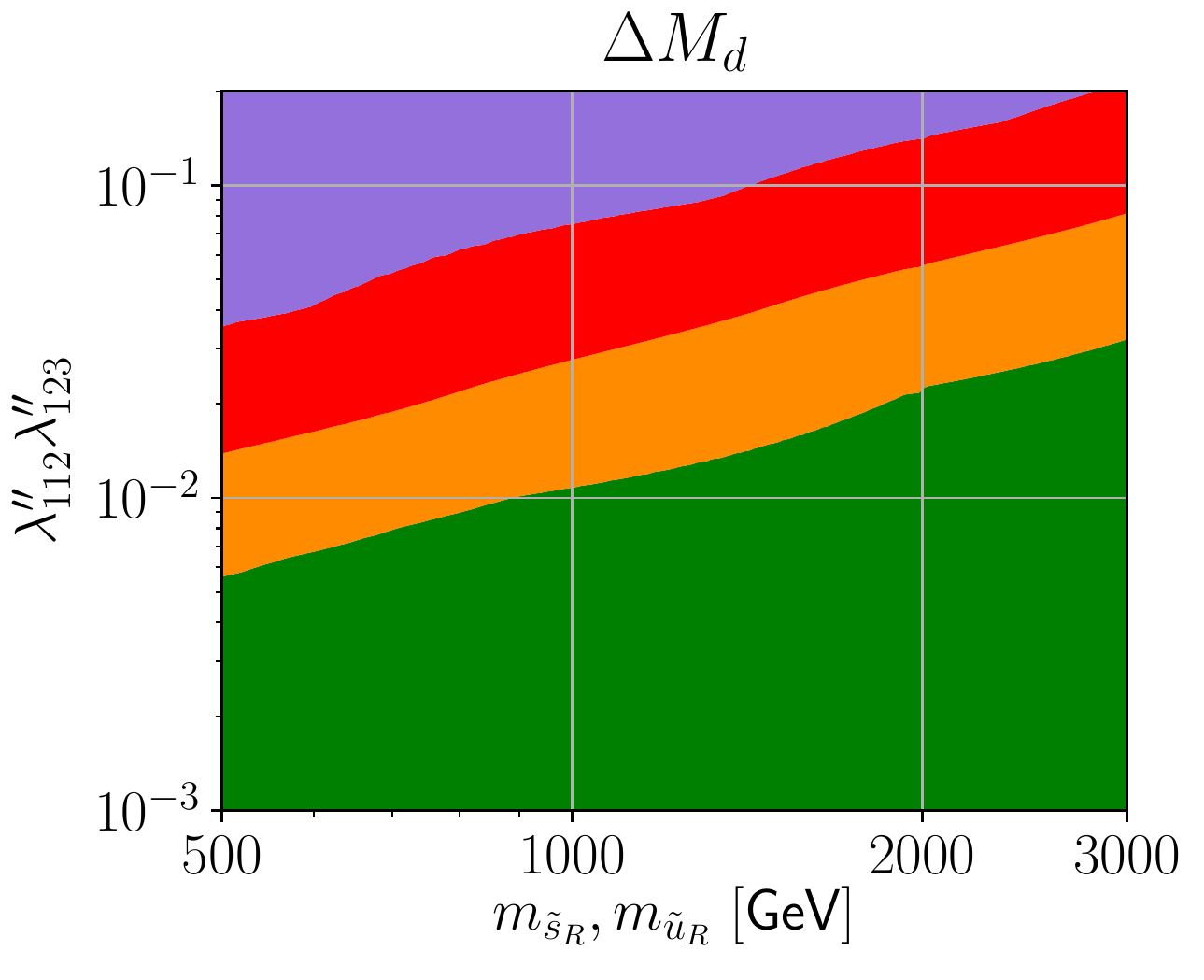}
\includegraphics[width=.45\linewidth]{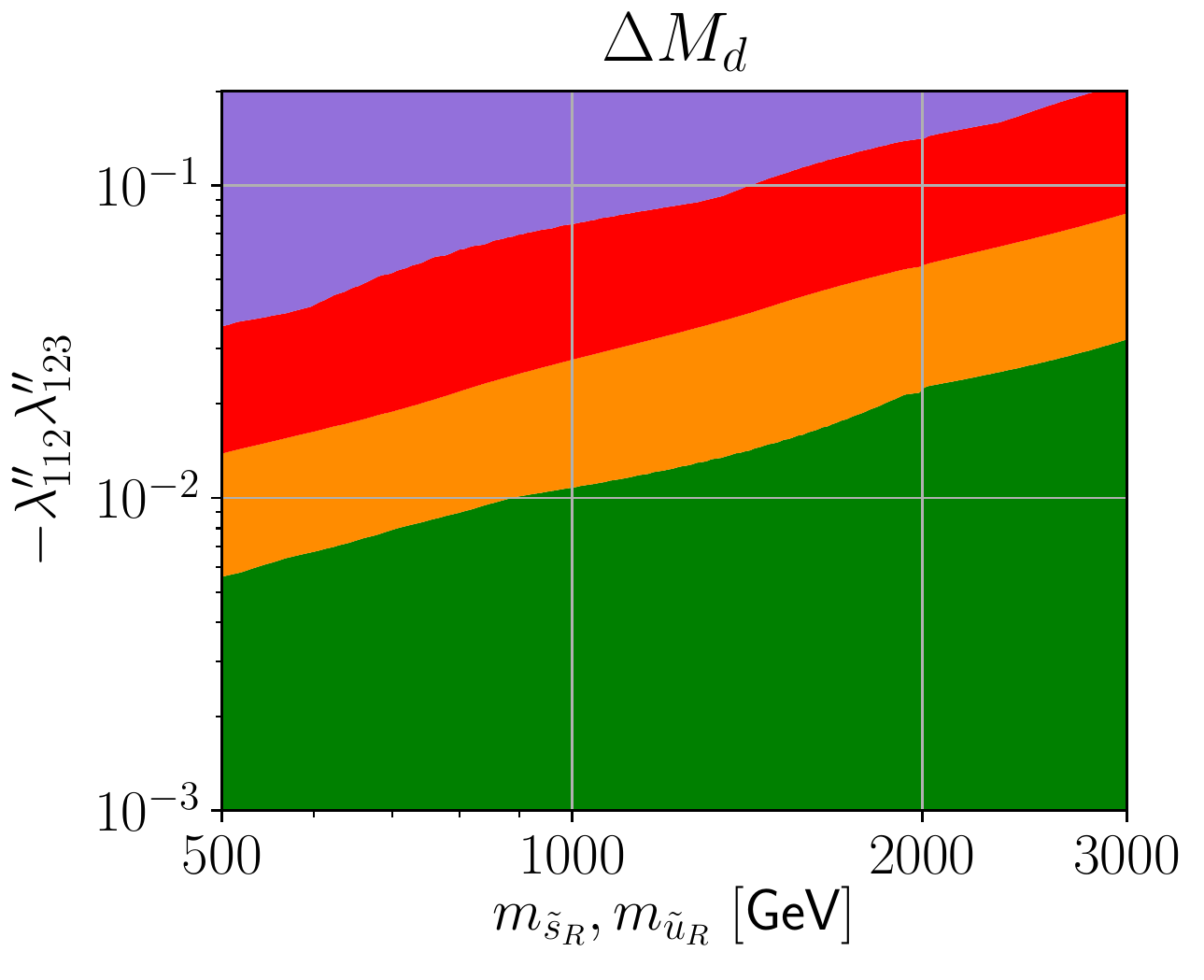}
\includegraphics[width=.45\linewidth]{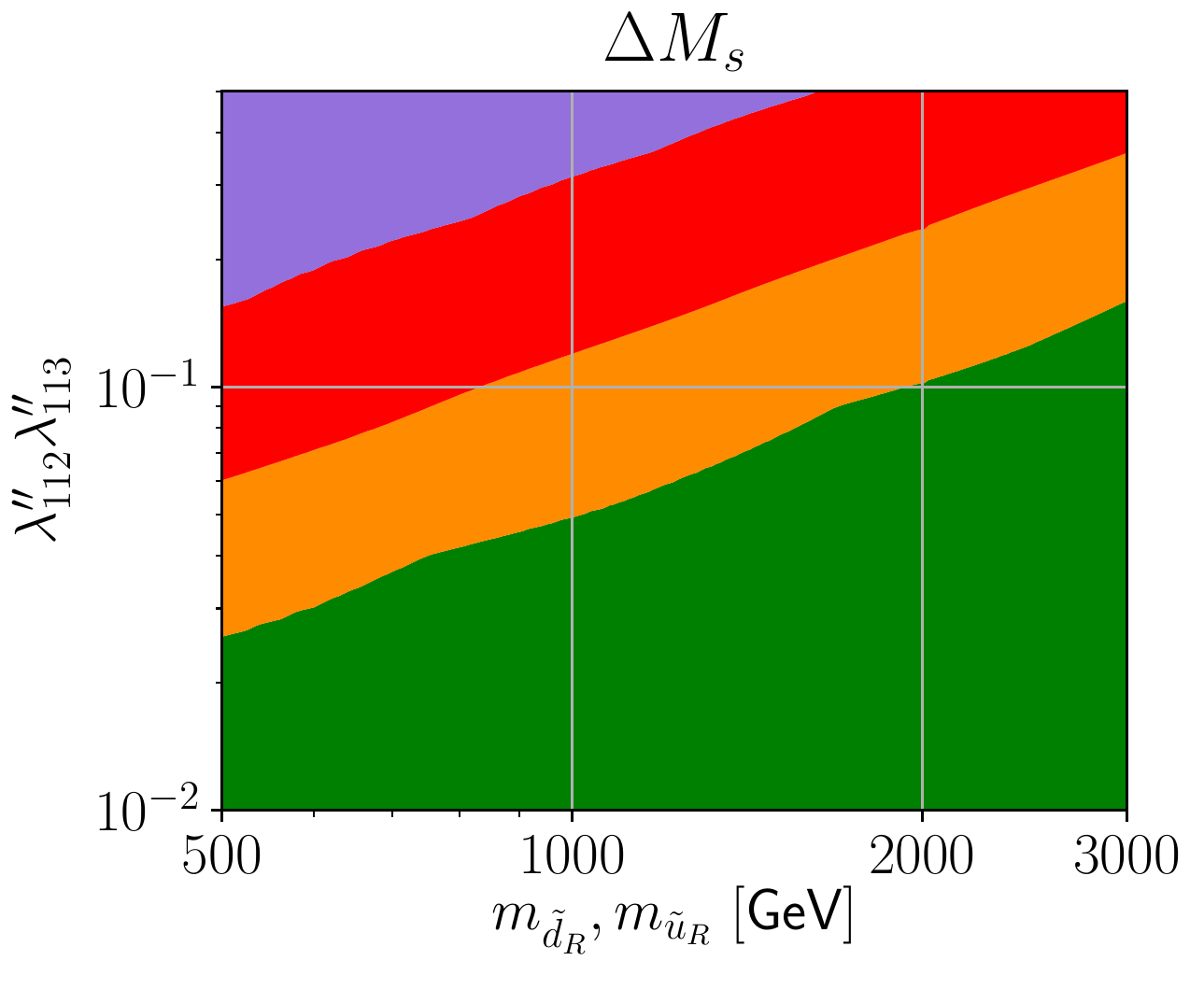}
\includegraphics[width=.45\linewidth]{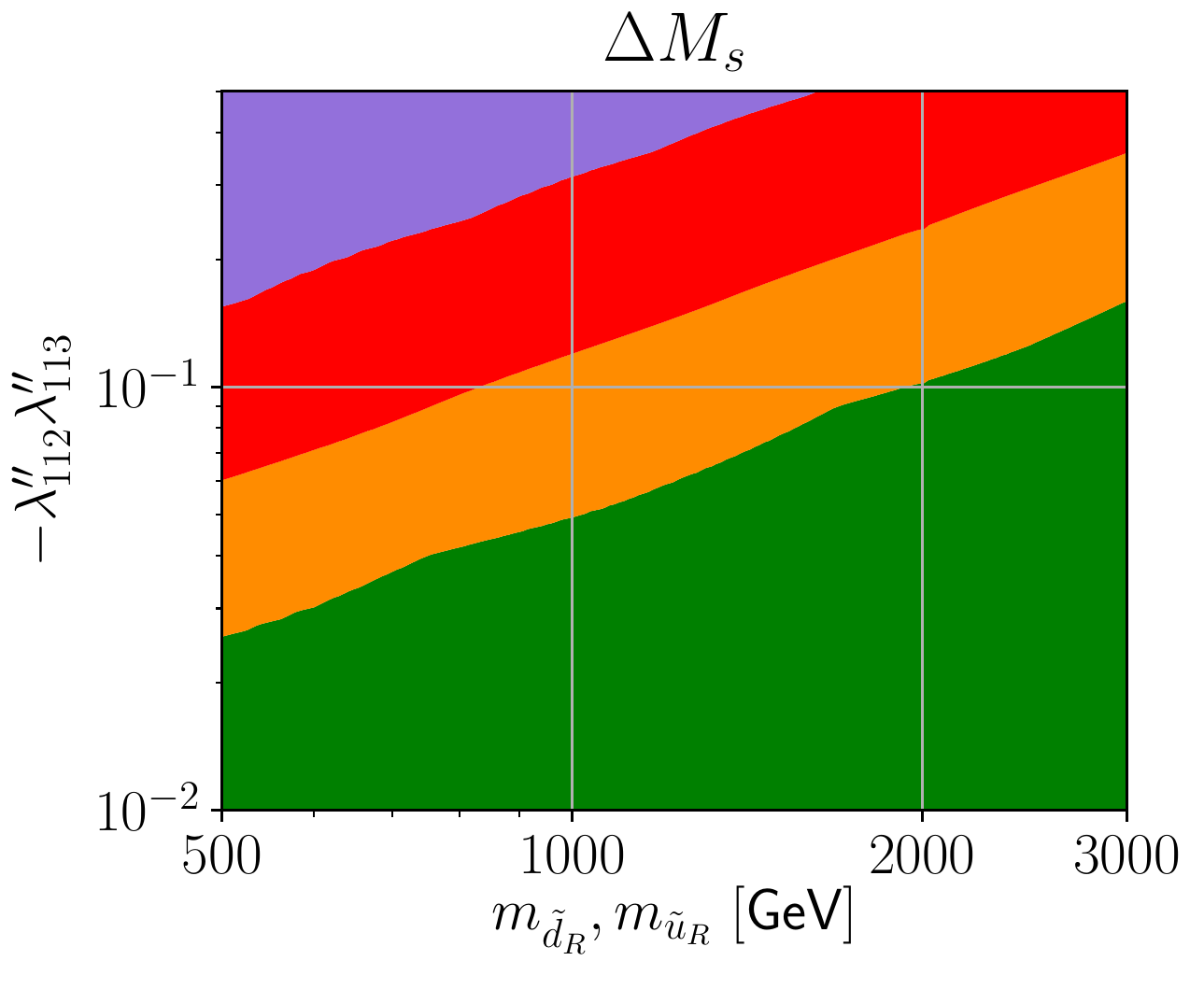}
\includegraphics[width=.45\linewidth]{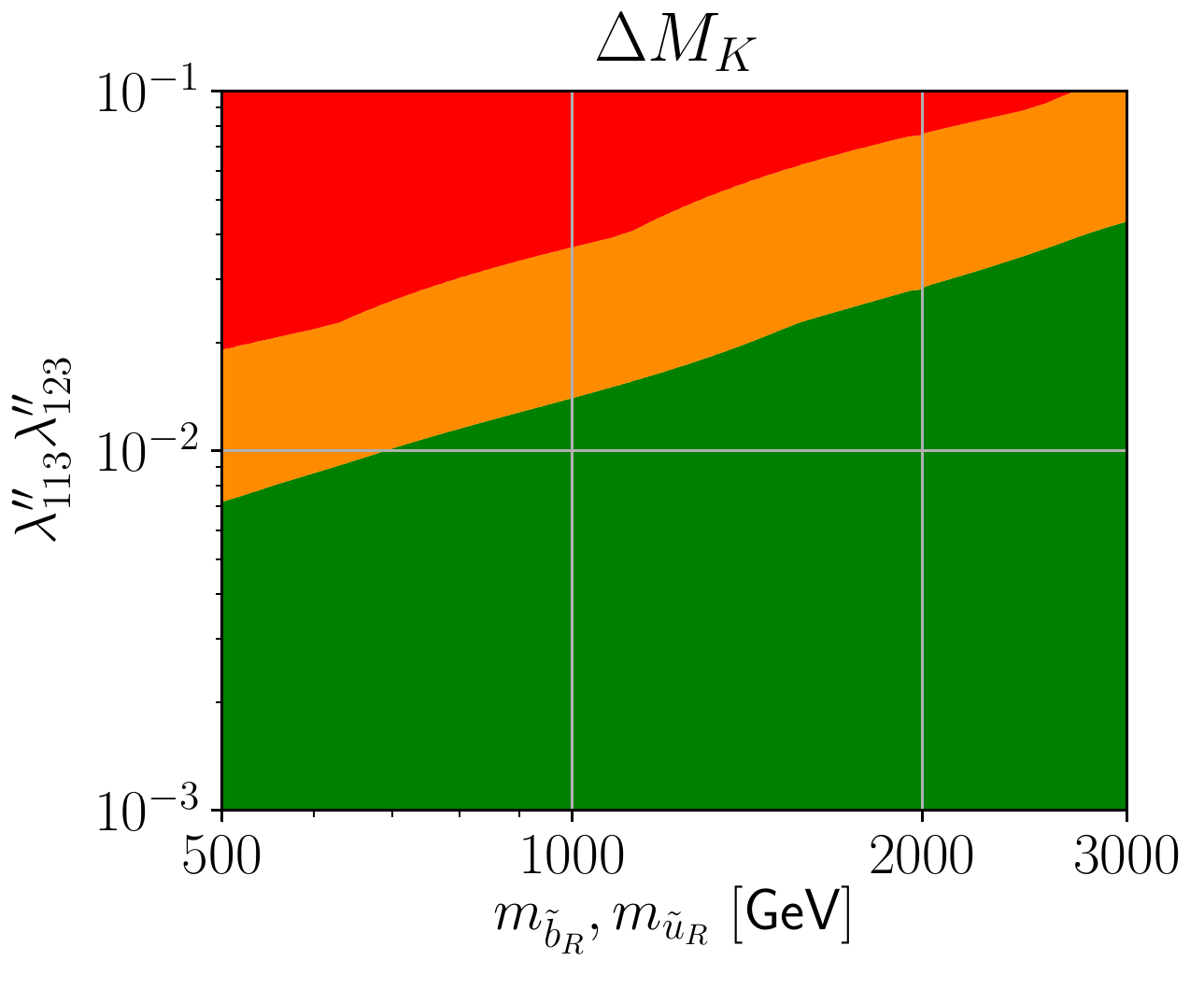}
\includegraphics[width=.45\linewidth]{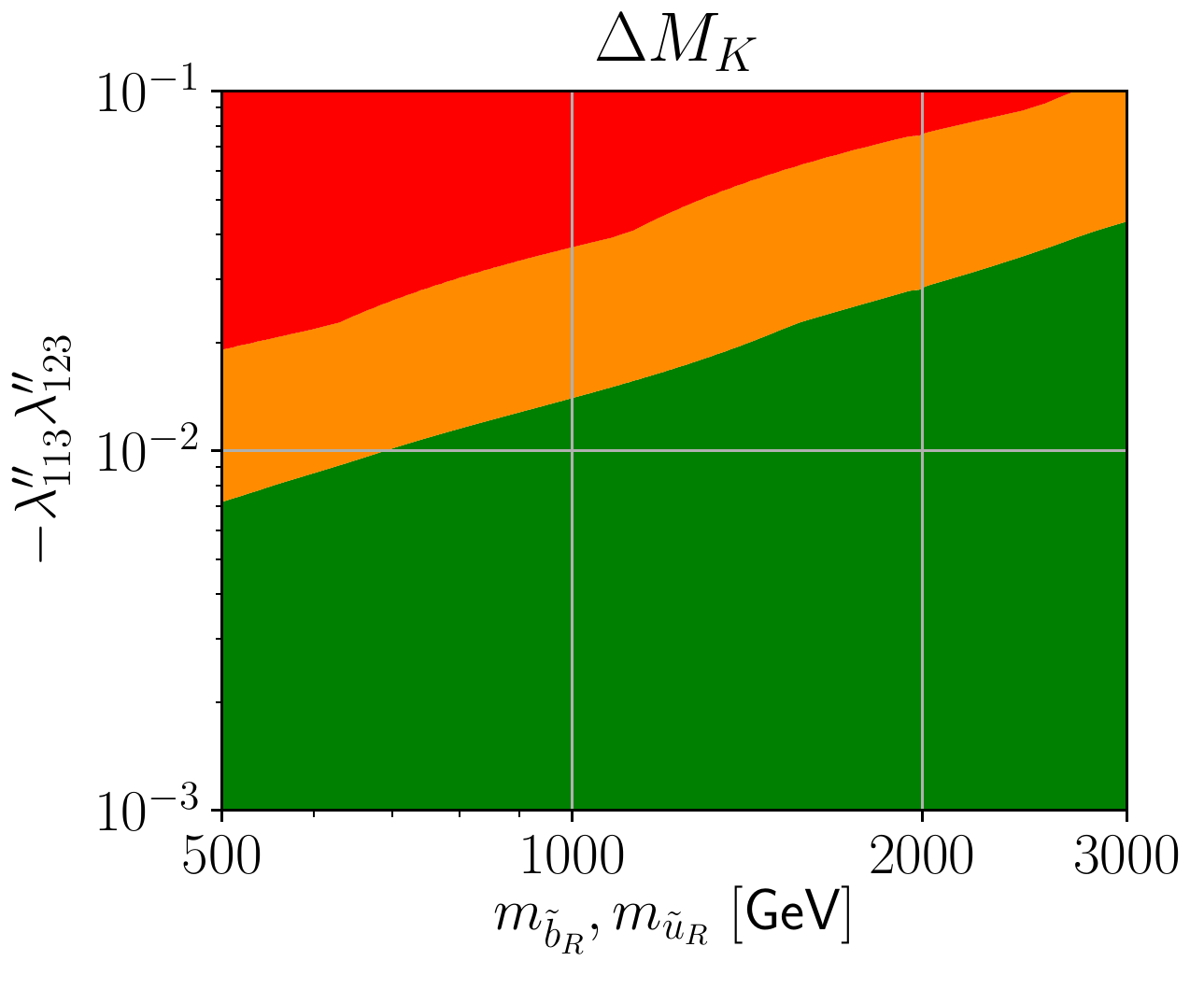}
\caption{Limits on $\bar U_1 \bar D_i \bar D_j$ couplings from the meson oscillation parameters. In this case, amplitudes with internal top lines vanish.
The color code is similar to that of the previous plots. For these plots, the parameter set of the scenario SUSY-RpV(a) of Table \ref{Tbl:parameters_scenarios} has been employed except for the squark masses that are scanned over.
\label{fig:squark_mass_vs_box_couplings_UDD2}}
\end{figure}

From now on, all the parameters are set to the values of the scenario SUSY-RpV(a) of Table~\ref{Tbl:parameters_scenarios}, except for those that are explicitly scanned over (e.g.\ the squark masses). In Fig.~\ref{fig:squark_mass_vs_box_couplings_UDD}, we present the $1,\,2$ and $3\,\sigma$ limits on coupling combinations allowing for internal (s)top lines.  
The relevant right-handed squarks are assumed to be mass-degenerate. The regime with small $\lambda''$ couplings is dominated by the box diagrams involving 
$W$ bosons and top quarks in the internal lines. We find that,
for low mass values, this contribution scales with the squark mass in an intermediate fashion between linear and quadratic, because of the finite top mass effects.
These effects largely vanish for squark masses above $\mathcal O(1\,{\rm TeV})$ and we then recover the scaling with $\frac{\lambda'' \cdot \lambda''}{m_{\tilde q}^2}$.
The supersymmetrized version of the $W$ boxes, \textit{i.e.}\ boxes with internal charginos, are also contributing with a scaling of $\frac{\lambda'' \cdot \lambda''}{m_{\tilde q}^2}$. 
However their impact w.r.t.\ the $W$ boxes is always reduced.
At large values of the couplings and for light squarks, the purely $\bar U \bar D \bar D$-mediated diagrams appear to be the most relevant, scaling with $\frac{(\lambda'' \cdot \lambda'')^2}{m_{\tilde q}^2}$ 
--- in analogy to the slepton box-diagrams with non-vanishing $LQ\bar D$ coupling --- so that the bounds on $\lambda'' \cdot \lambda''$ show a roughly linear dependence 
with the squark mass. 
Then, for both large $|\lambda'' \cdot \lambda''|$ and heavier quarks, the $W$-mediated diagrams and these purely $\bar U \bar D \bar D$ boxes can be of comparable magnitude, 
hence lead to interference structures. This interplay between various contributions brings about a non-trivial mass dependence of the bounds on the $\lambda''$ couplings, 
with both constructive as well as destructive effects between the individual amplitudes. The plots for negative $\lambda''\cdot \lambda''$ couplings perfectly illustrate this fact,
in particular in the case of $\Delta M_s$.  
Beyond this interference regime, at sufficiently large squark masses, the contribution from the UDD box with an internal W-line eventually supersedes the pure UDD amplitude.

Since the bounds on the individual coupling combinations do not scale with a simple power law in $m_{\tilde q_R}$, we refrain from showing approximate 
expressions as we did in the scenarios with flavor-violation of $LQ\bar D$-type.

In Fig.~\ref{fig:squark_mass_vs_box_couplings_UDD2}, 
by contrast, the choice of non-vanishing $\lambda''$ couplings does not allow for internal (s)top lines. Thus the RpV-diagrams 
with mixed $W$/squark or chargino/quark internal lines are suppressed, and the scaling of the limits from meson-oscillation parameters is closer to linear. In addition, the 
$2\,\sigma$ bounds are somewhat milder than in the previous case and roughly symmetrical for positive and negative $\lambda''\cdot\lambda''$ products.
Thus, in this case, we extract the approximate bounds on $\bar U_1 \bar D_i \bar D_j$ coupling pairs:
\begin{equation}
\begin{cases}
|\lambda''_{112}\lambda''_{123}|\lesssim  2.8\times 10^{-2}  \left(\frac{m_{\tilde{s}_R,\tilde u_R}}{1\,\text{TeV}}\right),\\
|\lambda''_{112}\lambda''_{113}|\lesssim 1.2\times 10^{-1} \left(\frac{m_{\tilde{d}_R,\tilde u_R}}{1\,\text{TeV}}\right),\\
|\lambda''_{113}\lambda''_{123}|\lesssim 3.6\times 10^{-2} \left(\frac{m_{\tilde{b}_R,\tilde u_R}}{1\,\text{TeV}}\right),
\end{cases}
\end{equation}

Given that the scaling of the bounds on $\lambda''\cdot\lambda''$ pairs decidedly depends on the specific choice of couplings, we refrain from showing a compilation table as Table~\ref{tab:bounds_LQD} for the $L Q\bar D$ couplings, since it would only be representative of a specific SUSY spectrum.

\subsection{Competition among \boldmath $LQ\bar D$-driven contributions}
\label{subsec:more_than_two_nonzero}

\begin{figure}[tbh]
\centering
\includegraphics[width=.45\linewidth]{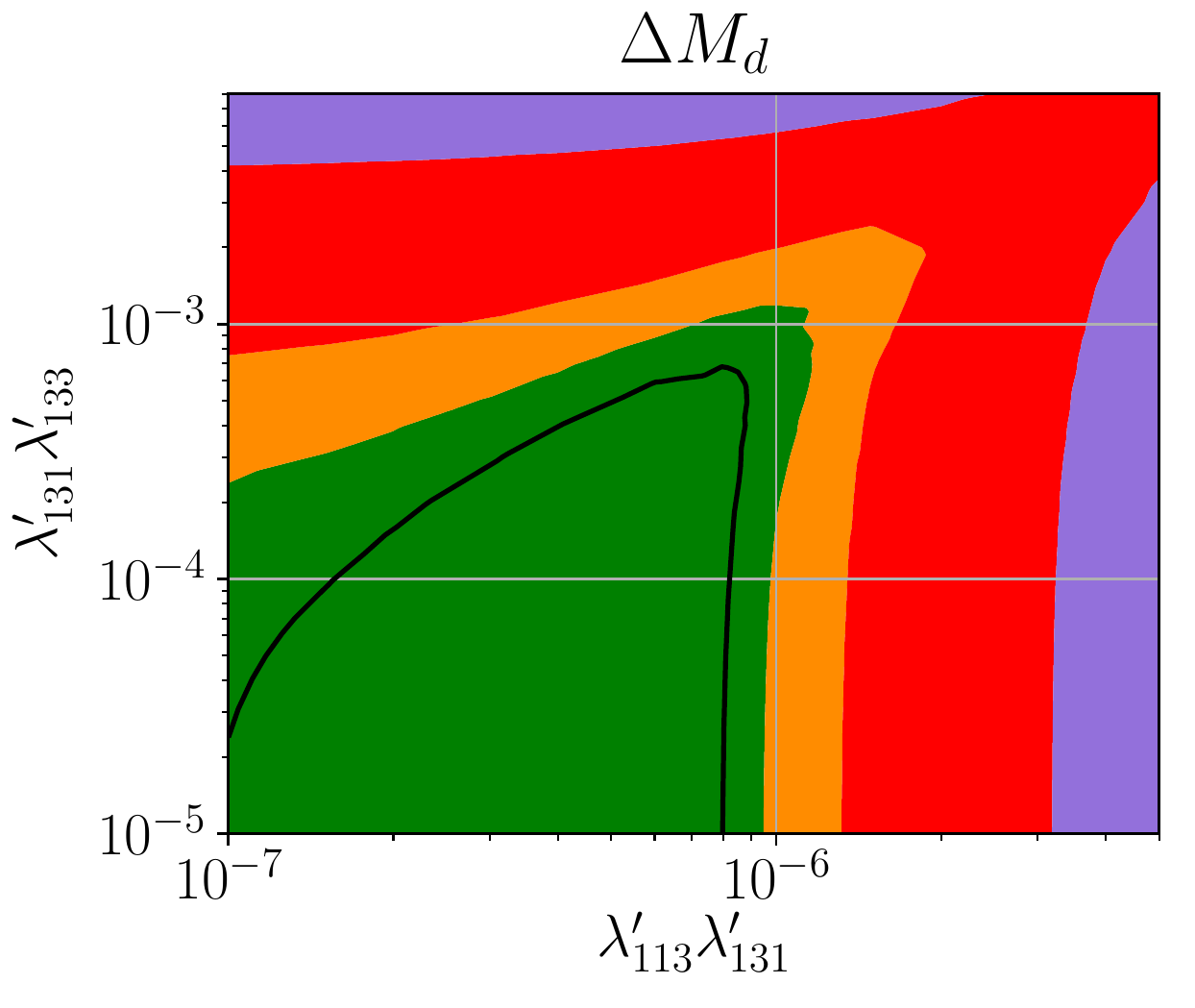}
\includegraphics[width=.45\linewidth]{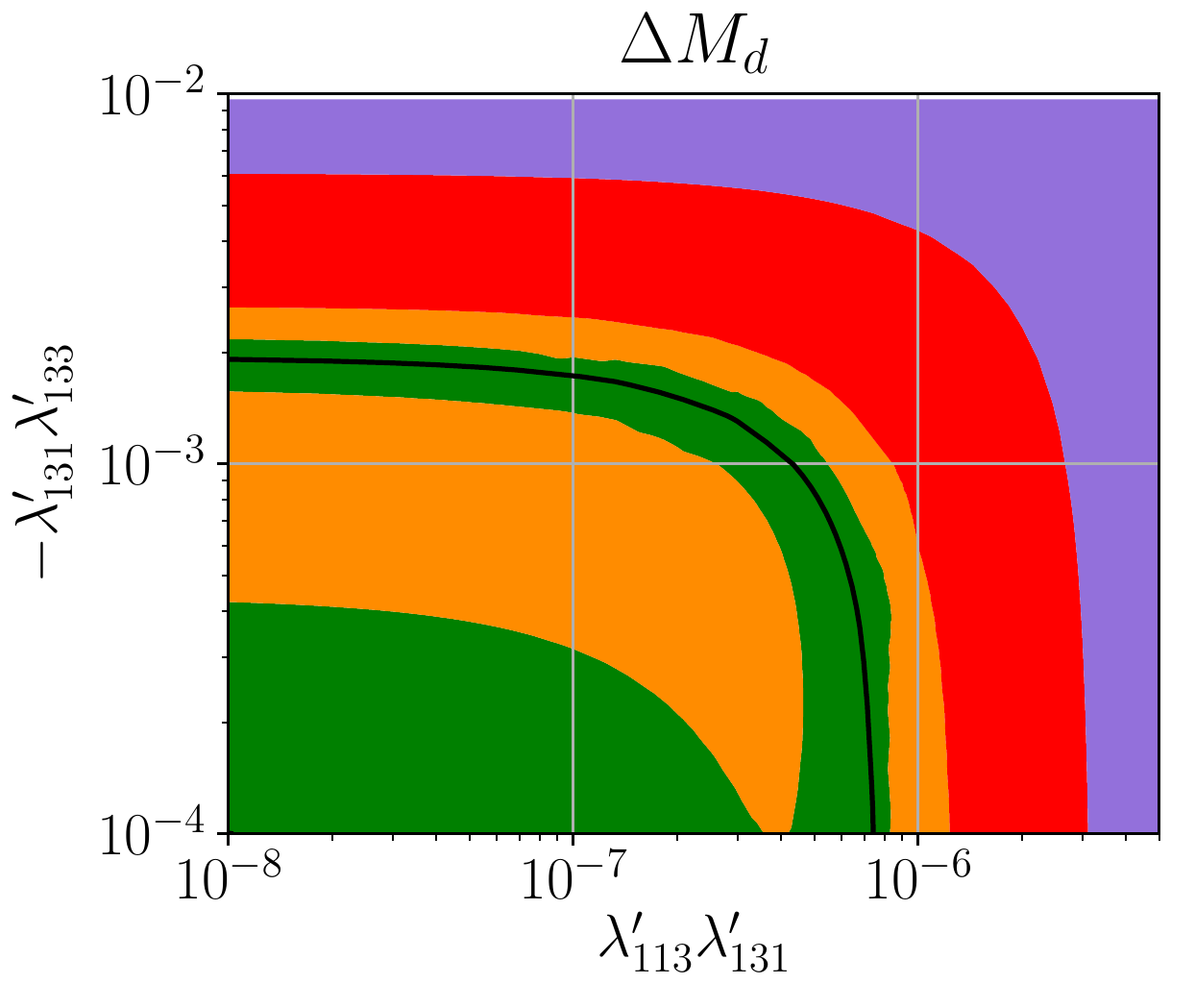}
\includegraphics[width=.45\linewidth]{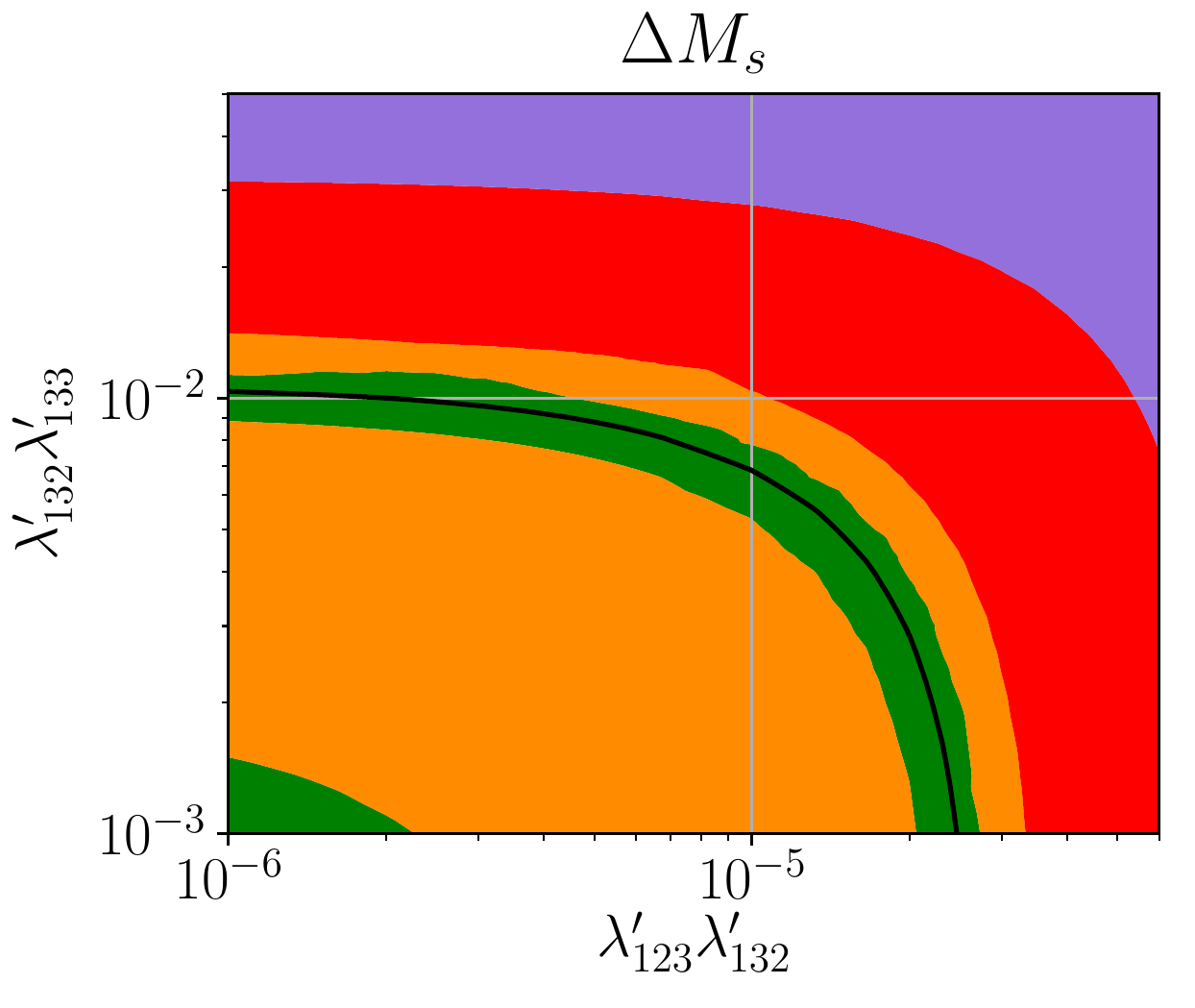}
\includegraphics[width=.45\linewidth]{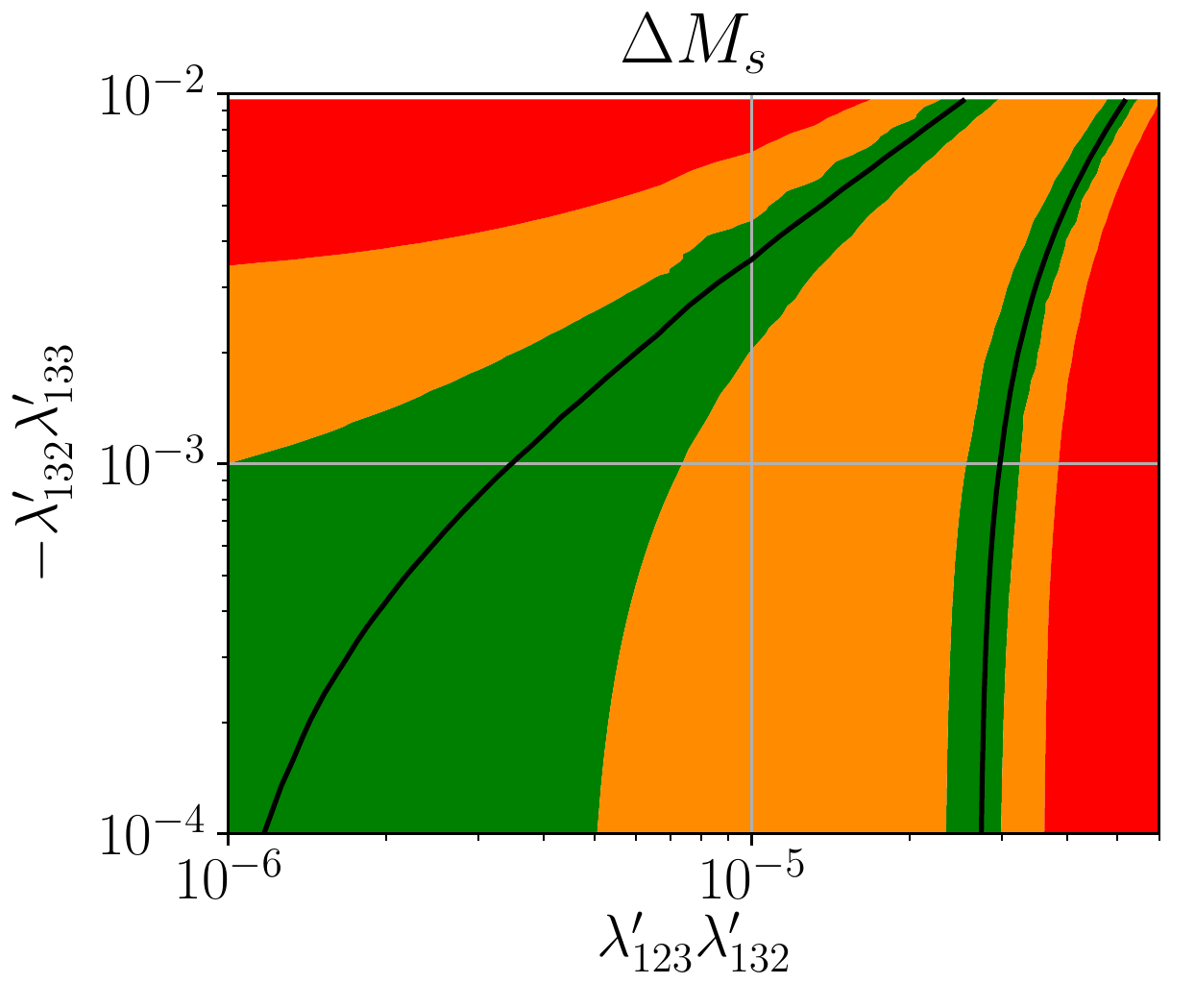}
\includegraphics[width=.45\linewidth]{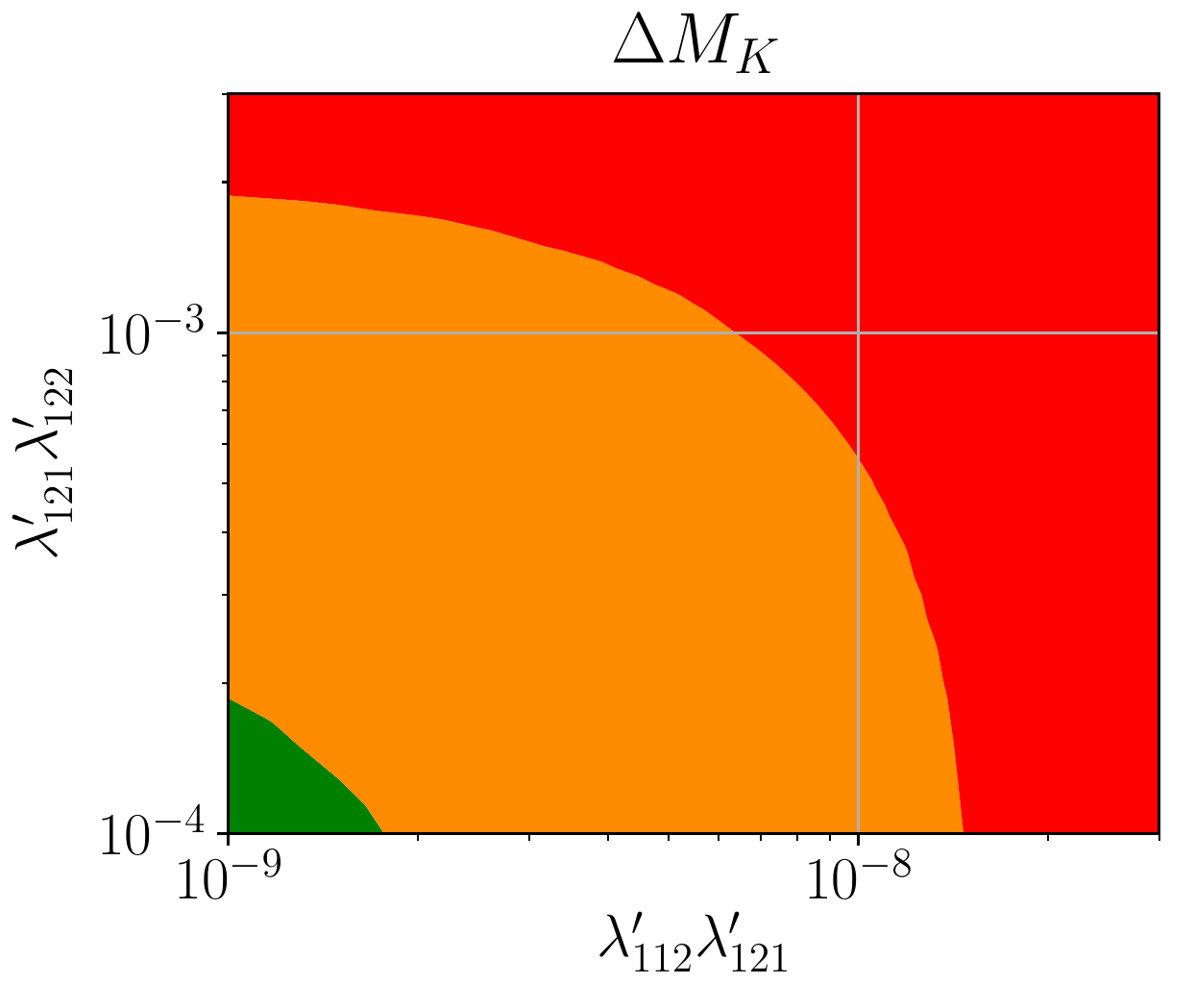}
\includegraphics[width=.45\linewidth]{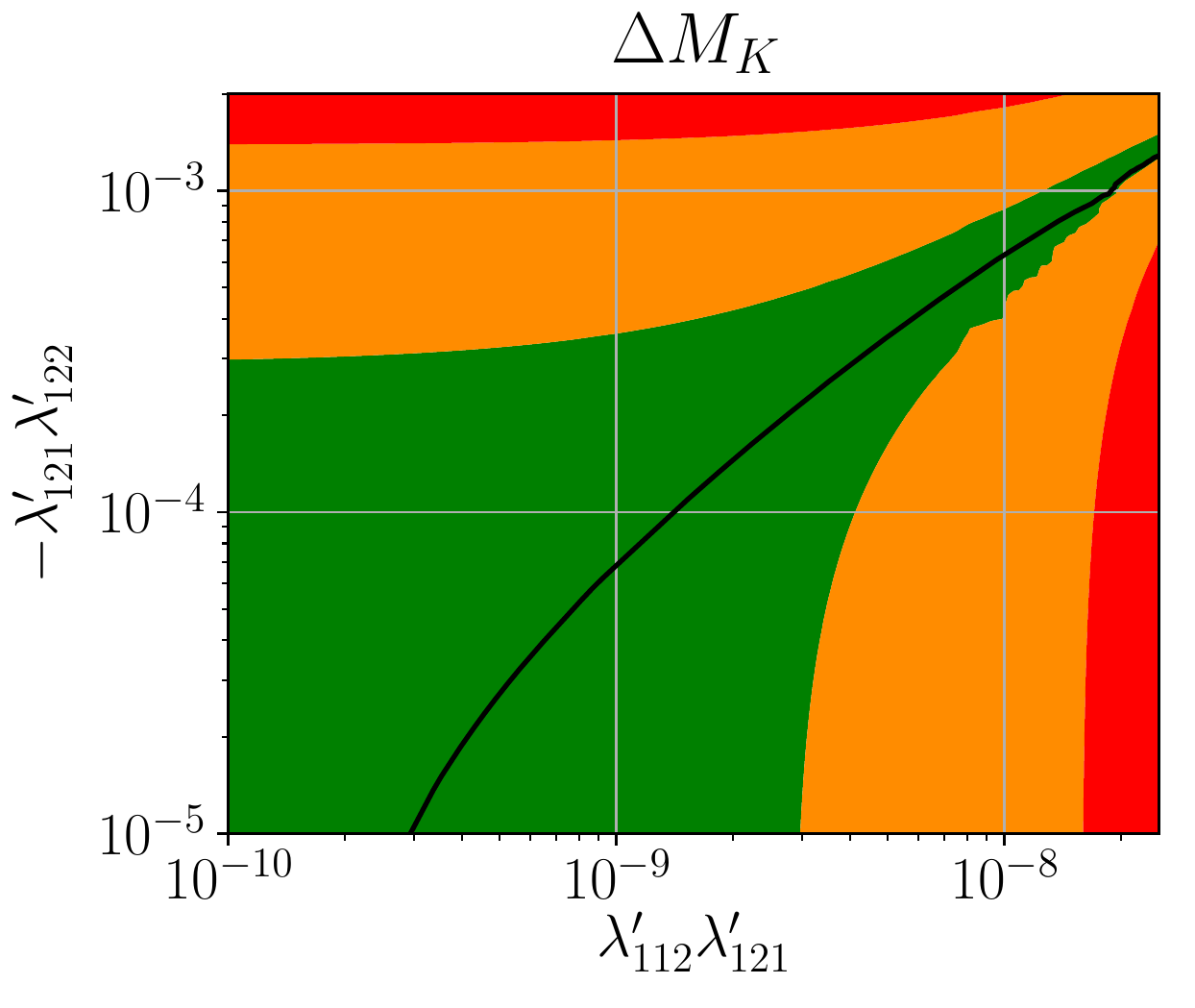}
\caption{Limits from the meson-oscillation parameters on two RpV-directions of $LQ\bar D$-type.
The parameters are set to the values in the third row of Table~\ref{Tbl:parameters_scenarios}, with slepton/sneutrinos of $1$\,TeV. 
As in the previous plots, the color code reflects the level of tension between our predictions and the experimental measurements.
\label{fig:loop_vs_tree}}
\end{figure}

Bounds on individual RpV-coupling products may be misleading, in the sense that several RpV-effects could cancel
one another. In fact, the decomposition along the line of the low-energy flavors provides likely-undue attention to these specific directions of RpV,
while the latter have no deep specificity from the high-energy perspective. In particular, RGE's are expected to mix the various flavor-directions of non-vanishing 
RpV-couplings, while the boundary condition at, say, the GUT scale, has no particular reason for alignment with the low-energy flavor directions
\cite{Allanach:1999mh,Allanach:2003eb}.
Obviously however, the relevant directions in flavor space are highly model-dependent and we have no particular suggestion to make from the low-energy perspective
of this work. Instead, we simply wish to illustrate the possibility of allowed directions with large RpV-couplings. To this end, we allow for two 
non-vanishing $\lambda'\cdot\lambda'$ coupling products and investigate the limits originating in the $\Delta M$ measurements.

If we consider Figs.~\ref{treelambda} and \ref{LQD_LOOP_QSED}, the tree-level diagram for $\lambda'_{i31}\cdot \lambda'_{i13}=\mathcal O(10^{-6})$
and the RpV-box for $\lambda'_{i31}\cdot \lambda'_{i33}=\mathcal O(10^{-4})$ --- implying a hierarchy $\lambda'_{i13}/\lambda'_{i33}=\mathcal O(10^{-2})$ 
-- naively contribute to $\Delta M_d$ by amplitudes of comparable magnitude. Whether these contributions can interfere destructively clearly depends on the form of
the amplitudes but also on the sign of the non-vanishing couplings.
In Fig.~\ref{fig:loop_vs_tree}, we complete the results from Figs.~\ref{treelambda} and \ref{LQD_LOOP_QSED} 
by now allowing for three non-vanishing couplings. In practice, we set the slepton/sneutrino mass to $1$\,TeV and keep one $LQ\bar D$ coupling 
to a constant value: $\lambda'_{131}=0.01$, $\lambda'_{132}=0.1$, or $\lambda'_{121} = 0.1$. Then, we vary two independent $\lambda'$, our choice depending again on the 
valence quarks of the considered $\Delta M$. However, we stress that this procedure in fact opens three non-trivial $\lambda'\cdot\lambda'$ directions, so that the game
is somewhat more complex than just playing one contribution versus the other.

As expected, in the plots of Fig.~\ref{fig:loop_vs_tree}, the interplay of various RpV-contributions opens funnel-shaped allowed regions for 
comparatively large values of the $LQ\bar D$ couplings, highlighting the possibility of destructive interferences. We note that, considering that the tree-level and 
radiative contributions do not necessarily have the same scaling with respect to the slepton/sneutrino mass, the `allowed angle' depends on the sfermion spectrum. Of 
course, the choice of parameters falling within the allowed funnels appears to be fine-tuned from the perspective of this work, but might be justified from a high-energy 
approach. On the other hand, constructive interferences lead to the `rounded edges' observed in some of the plots.

As mentioned earlier, we will not consider the interplay of $LQ\bar D$- and $\bar U \bar D \bar D$-couplings, since such scenarios are of limited 
relevance without a quantitative analysis of the proton decay rate. On the other hand, our discussion in this subsection points to the relevance of 
considering a full evaluation of the $\Delta M$'s (and other observables), when considering RpV-scenarios beyond the simplistic one-coupling-dominance approach.

\subsection{Competition between flavor violation in the R-parity conserving and R-parity violating sectors}
\label{subsec:different_sectors}

 \begin{figure}
\centering
\includegraphics[width=.45\linewidth]{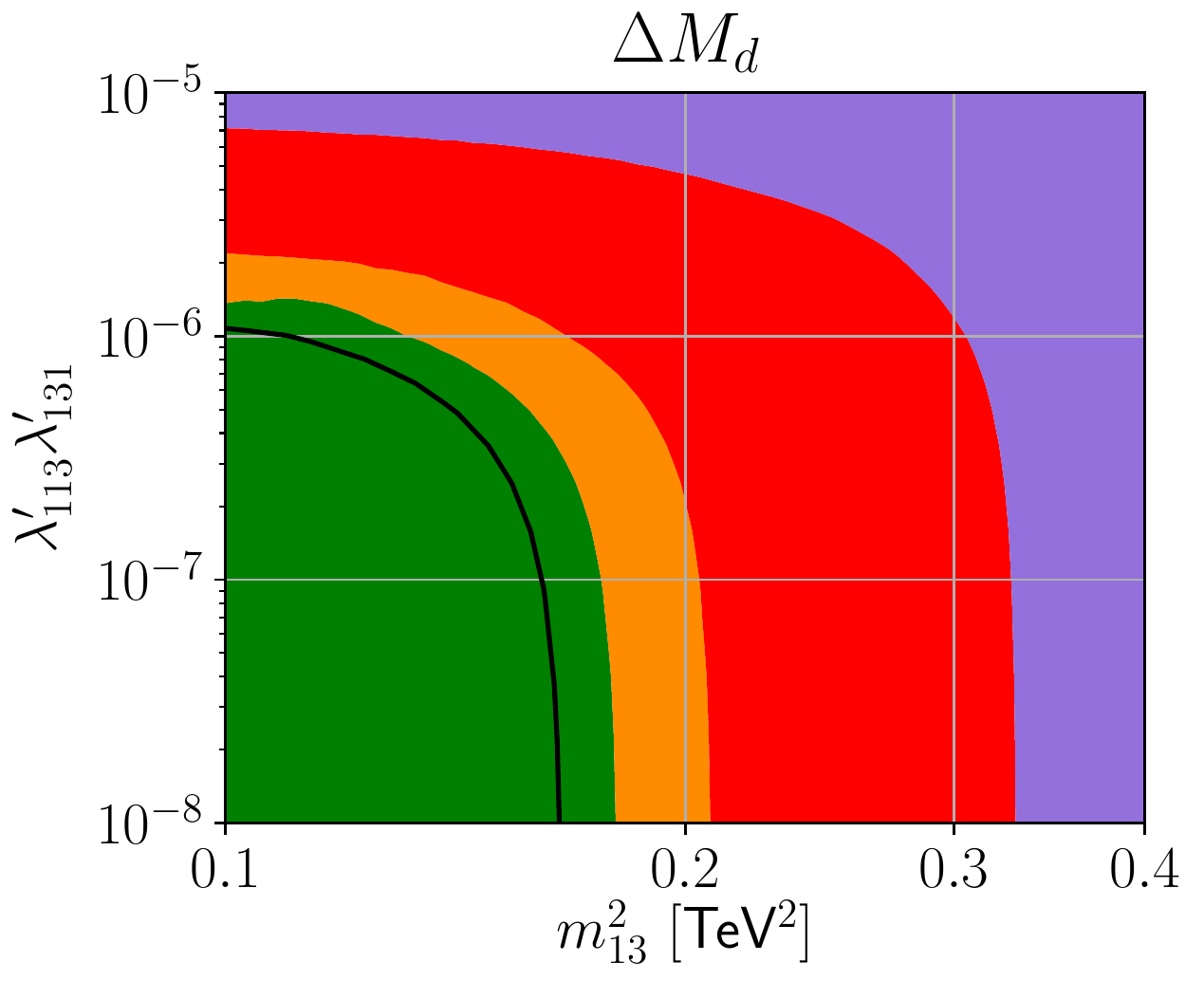}
\includegraphics[width=.45\linewidth]{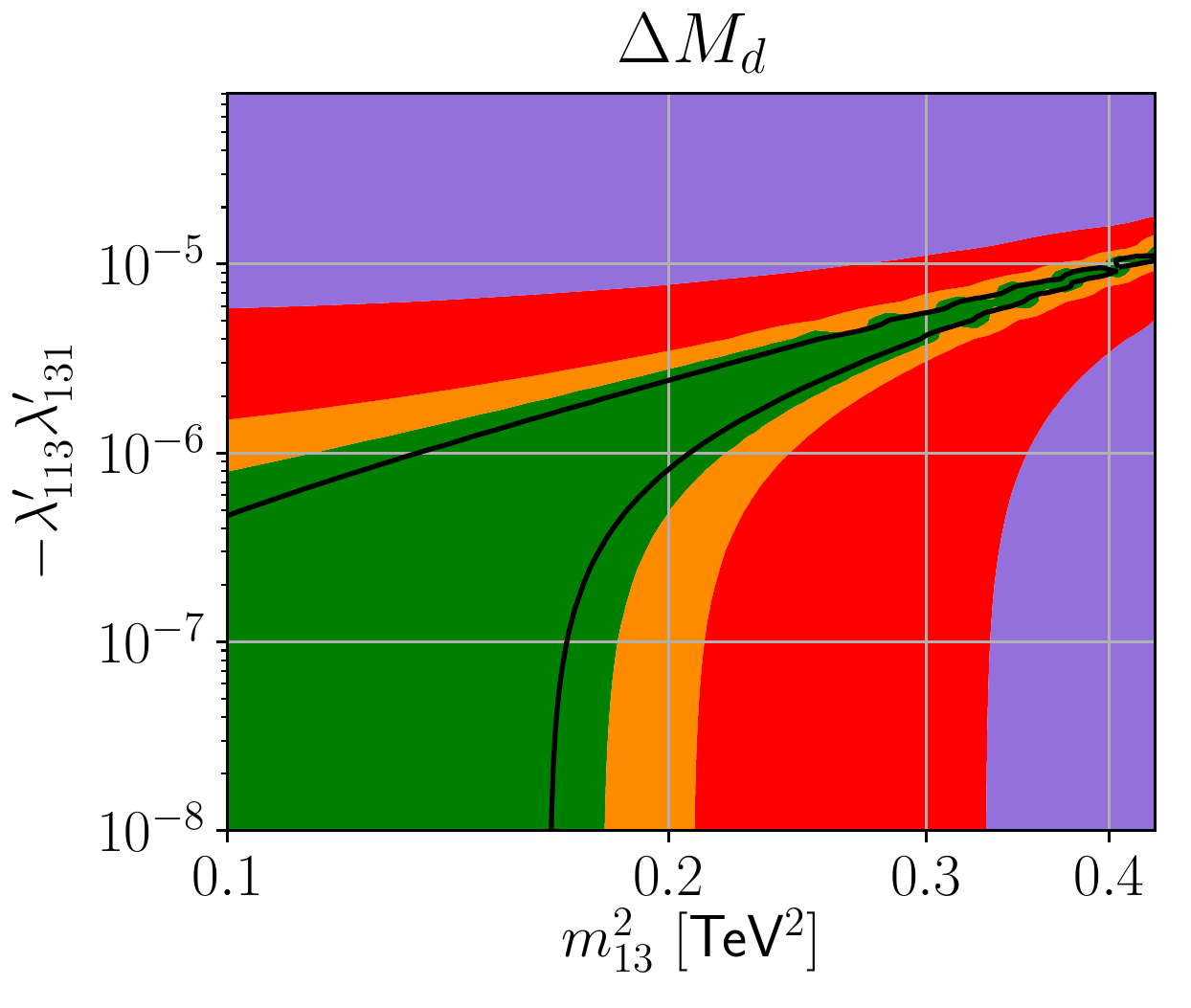}
\includegraphics[width=.45\linewidth]{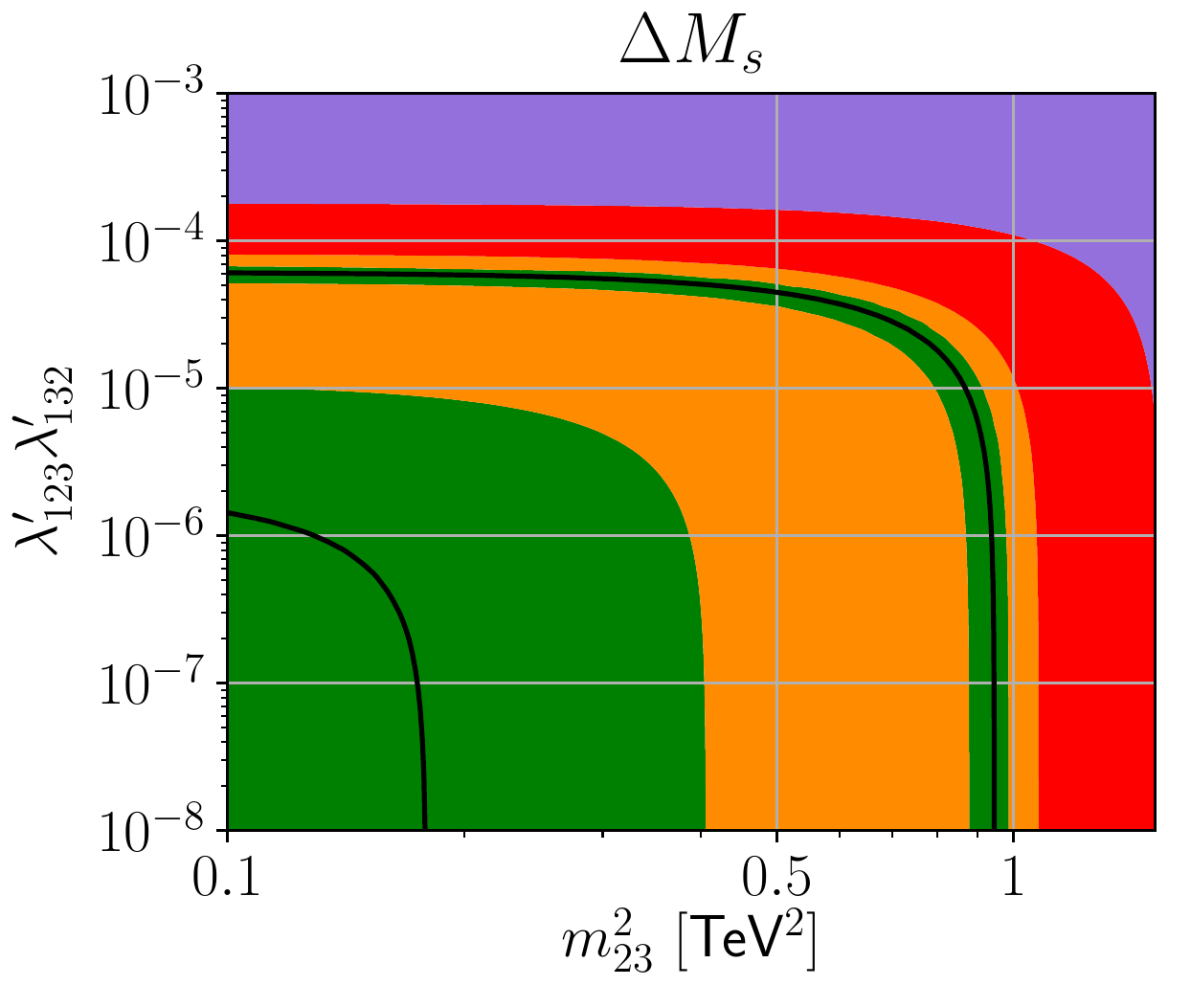}
\includegraphics[width=.45\linewidth]{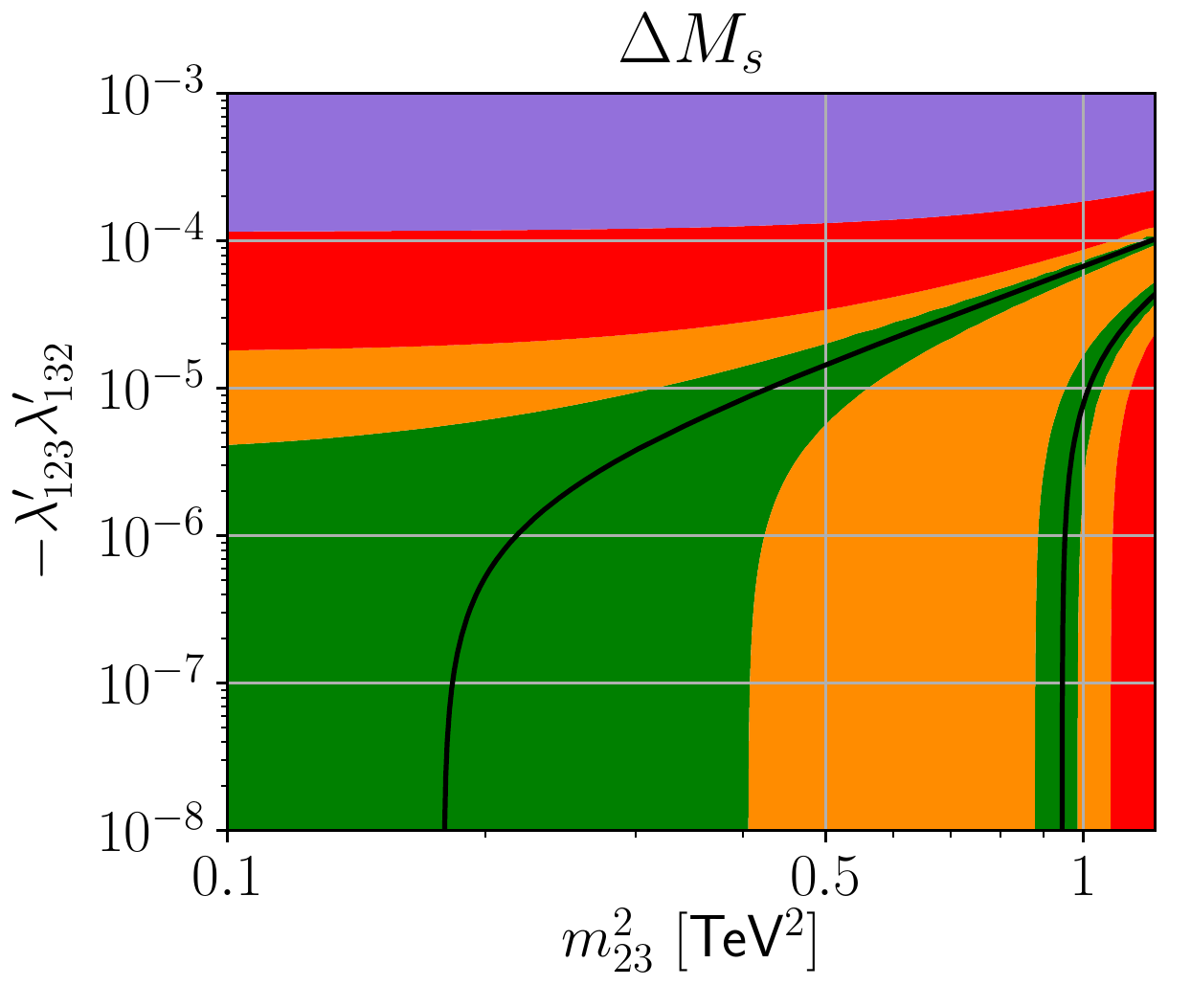}
\includegraphics[width=.45\linewidth]{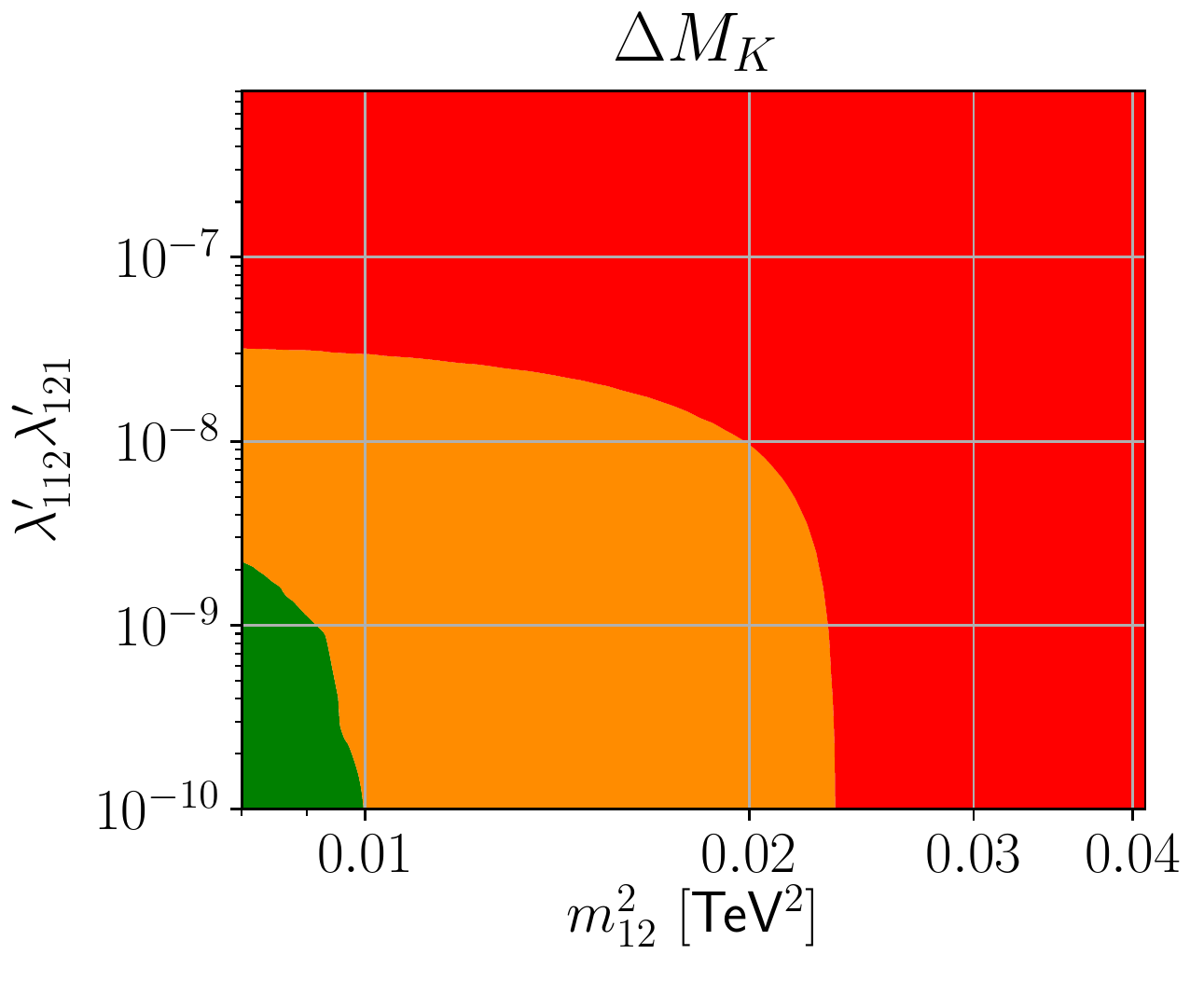}
\includegraphics[width=.45\linewidth]{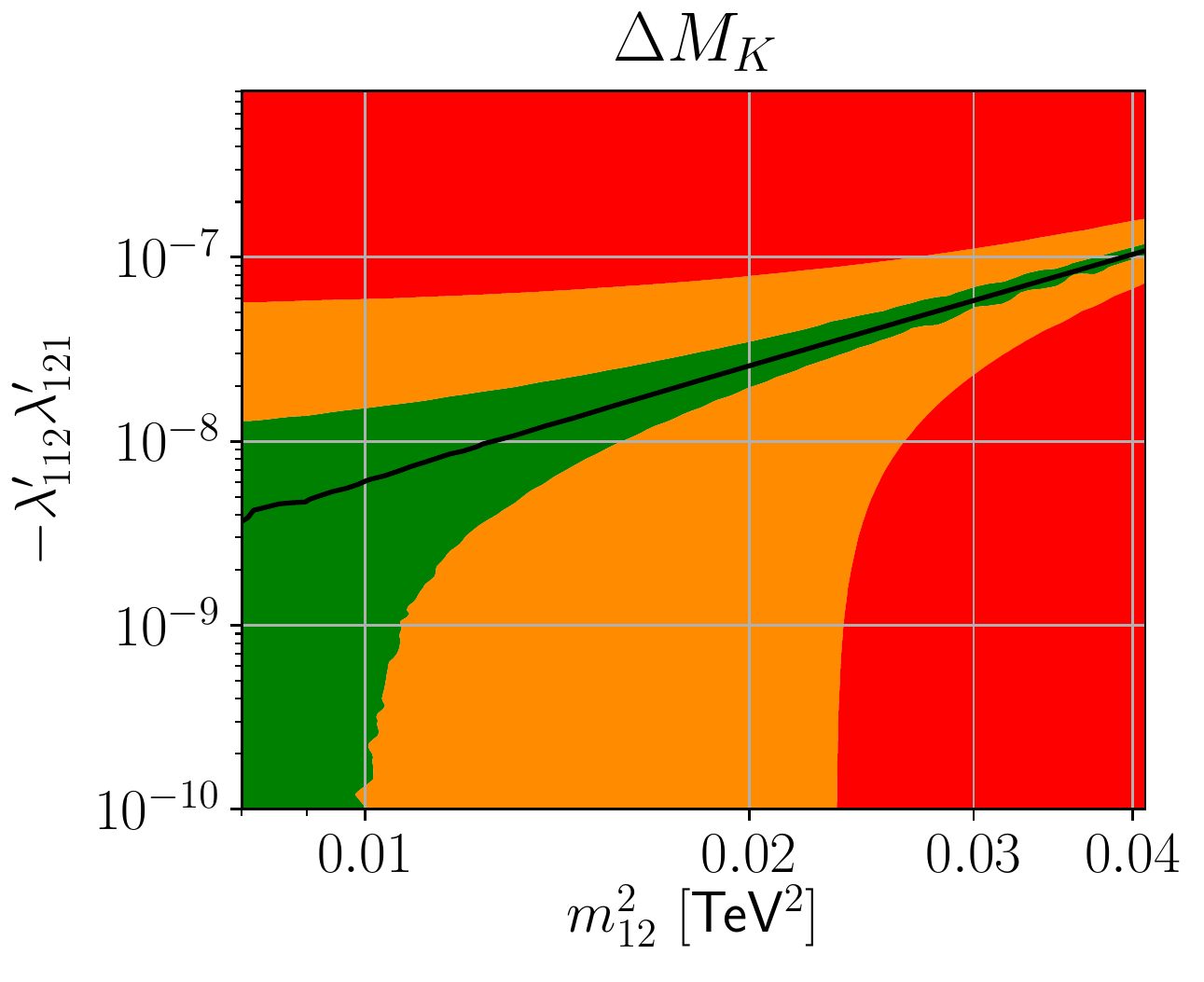}
\caption{Constraints from the meson-oscillation parameters in the presence of both flavor-violating $LQ\bar D$-couplings and ($R_p$-conserving) 
flavor-violating mixing in the squark sector. The spectrum is set to the scenario SUSY-RpV(a) of Table~\ref{Tbl:parameters_scenarios}, with the slepton/sneutrino mass 
at $1.5$\,TeV. The flavor-violating quadratic soft mass parameters in the squark sector, $m^2_{ij}$, are chosen to be degenerate for left-handed and right-handed squarks. 
The color code follows the same conventions as before.
\label{fig:squark_mixing_vs_tree_couplings}}
\end{figure}

 \begin{figure}[htbp]
\centering
\includegraphics[width=.45\linewidth]{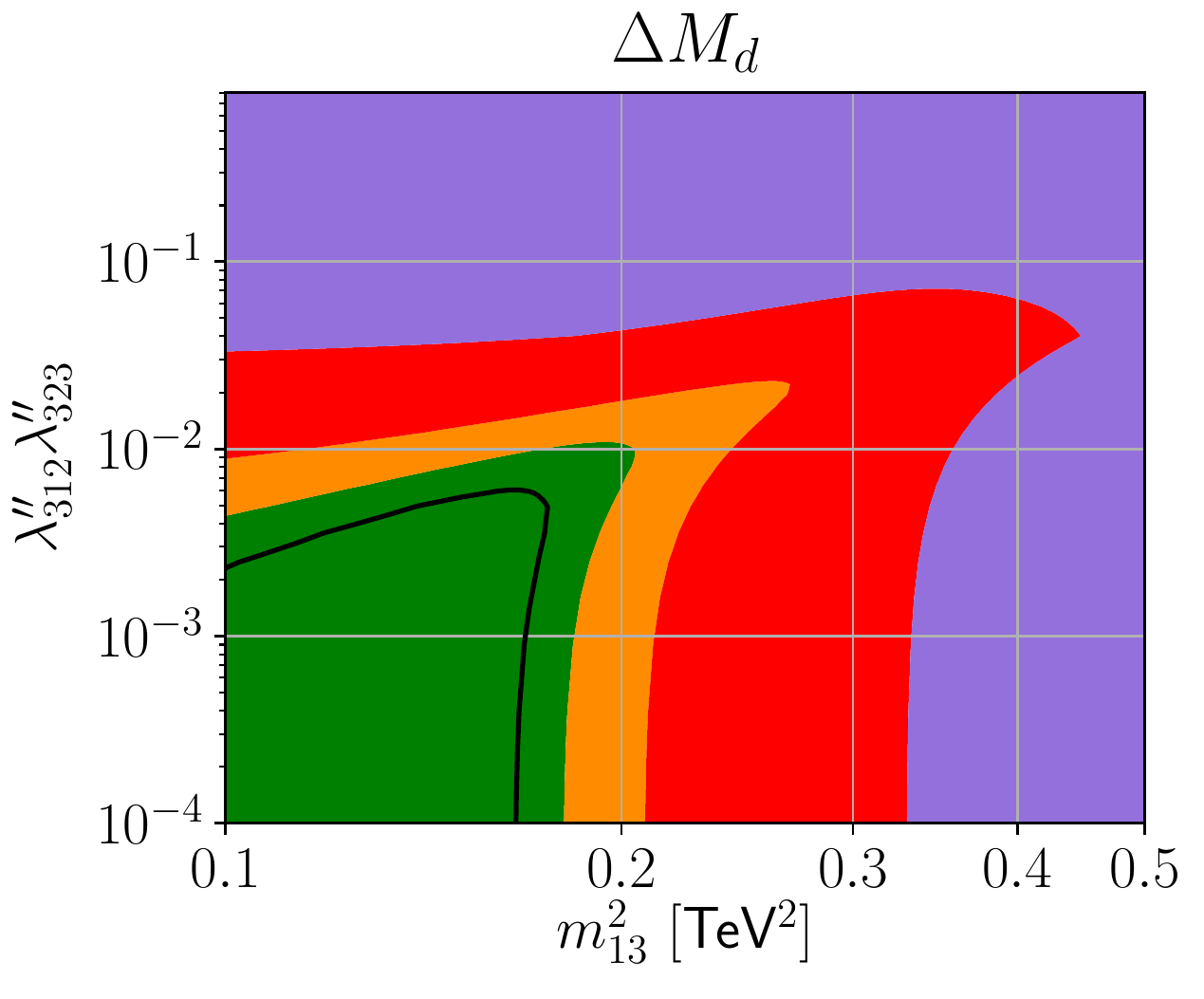}
\includegraphics[width=.45\linewidth]{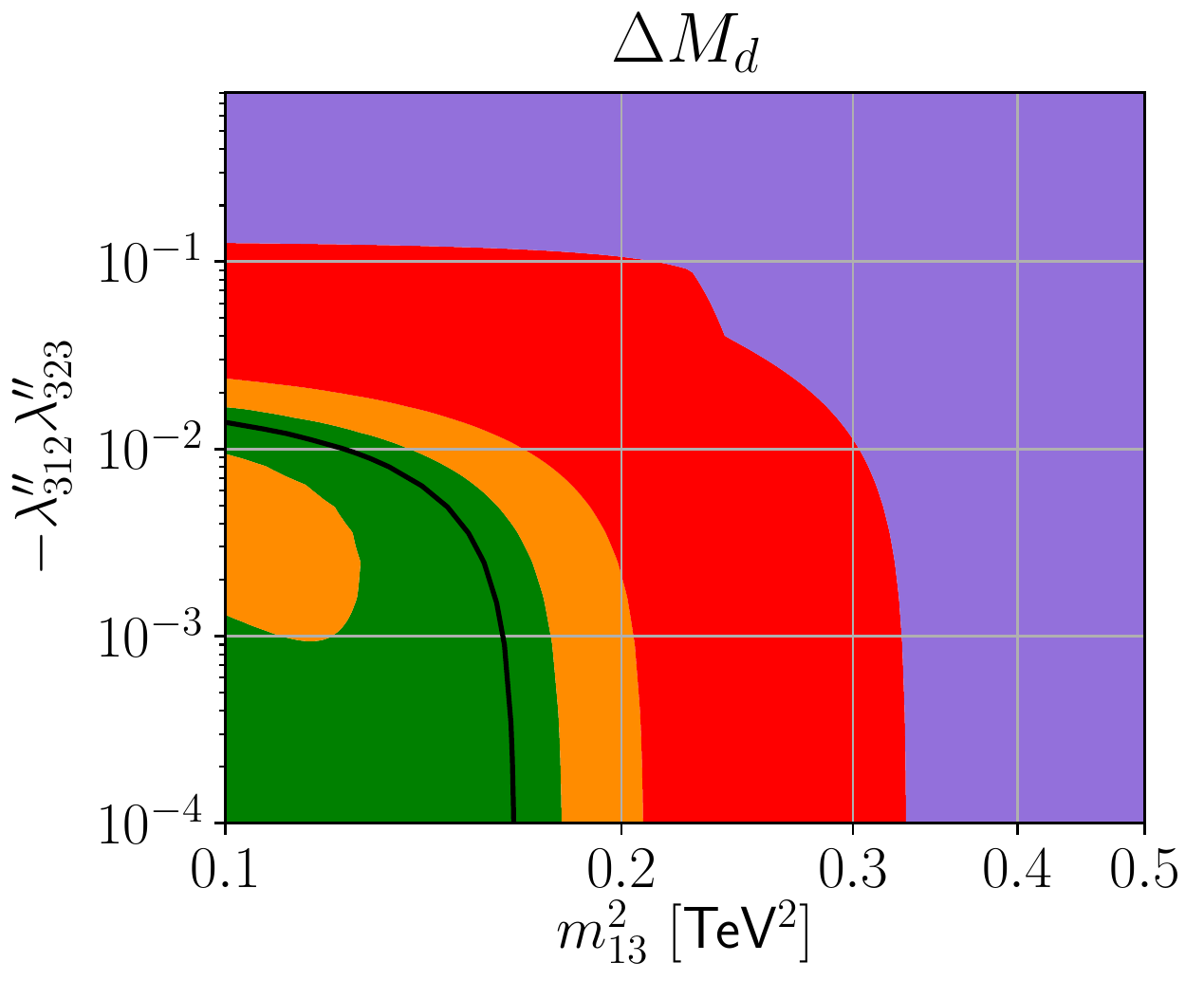}
\includegraphics[width=.45\linewidth]{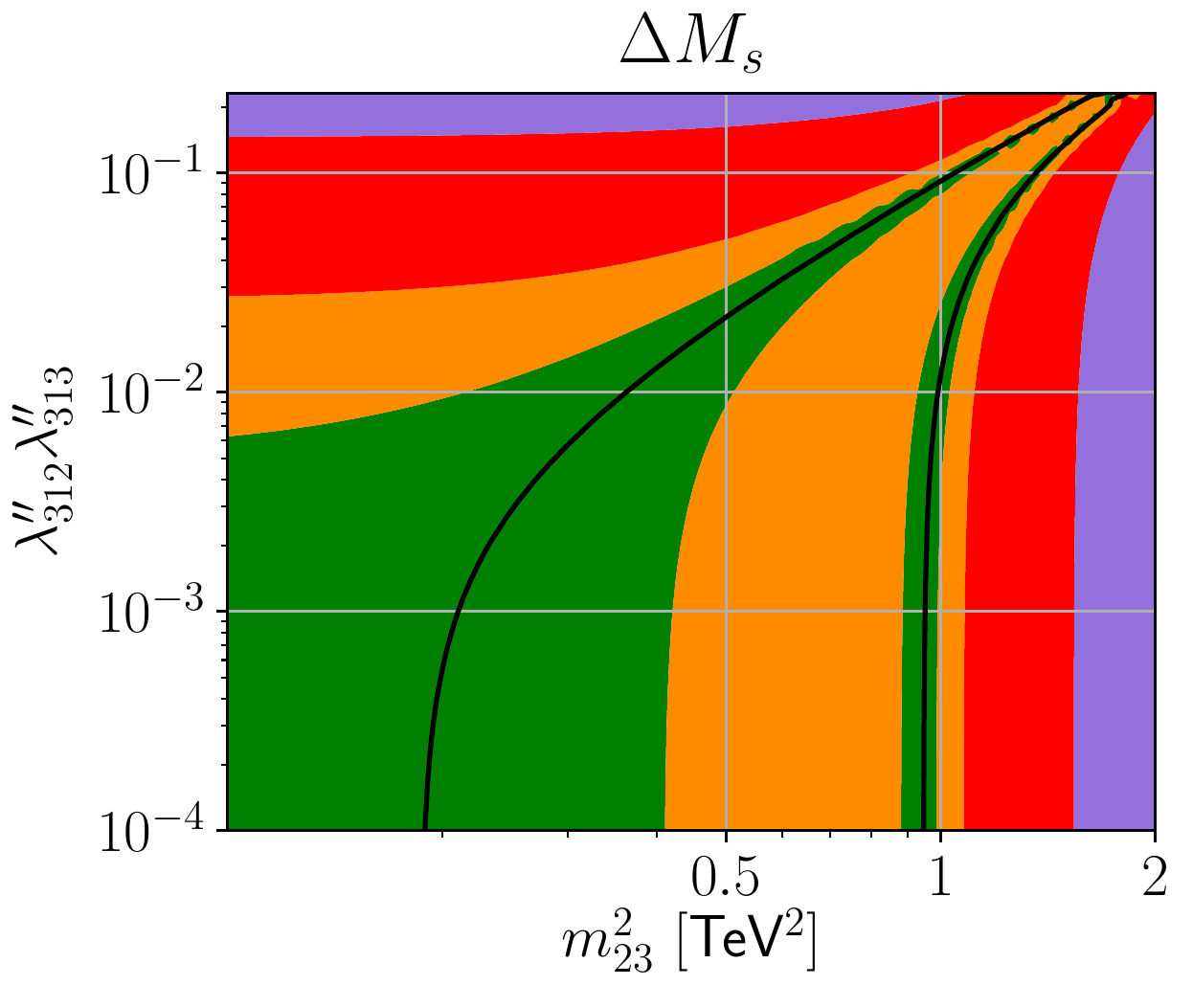}
\includegraphics[width=.45\linewidth]{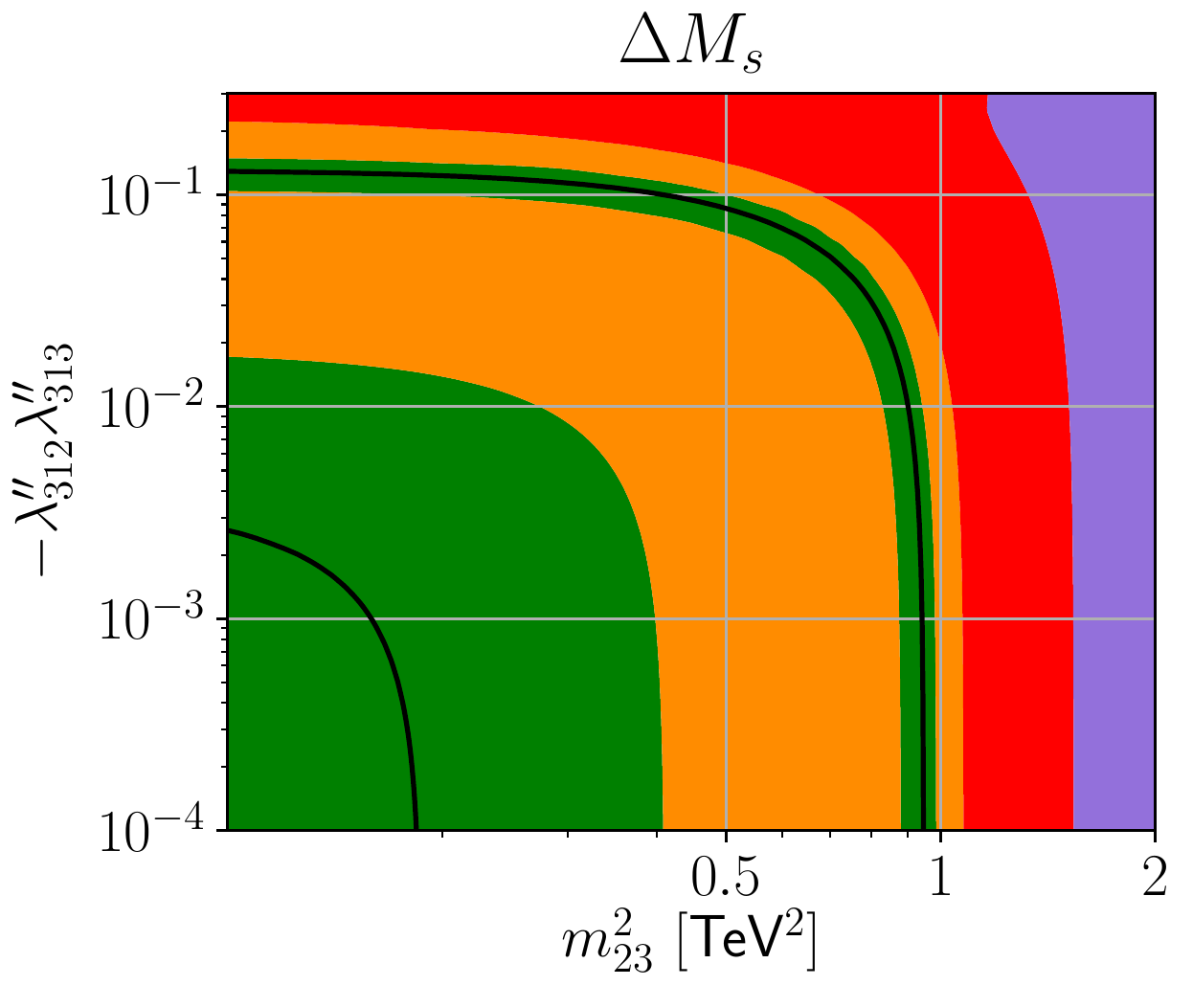}
\includegraphics[width=.45\linewidth]{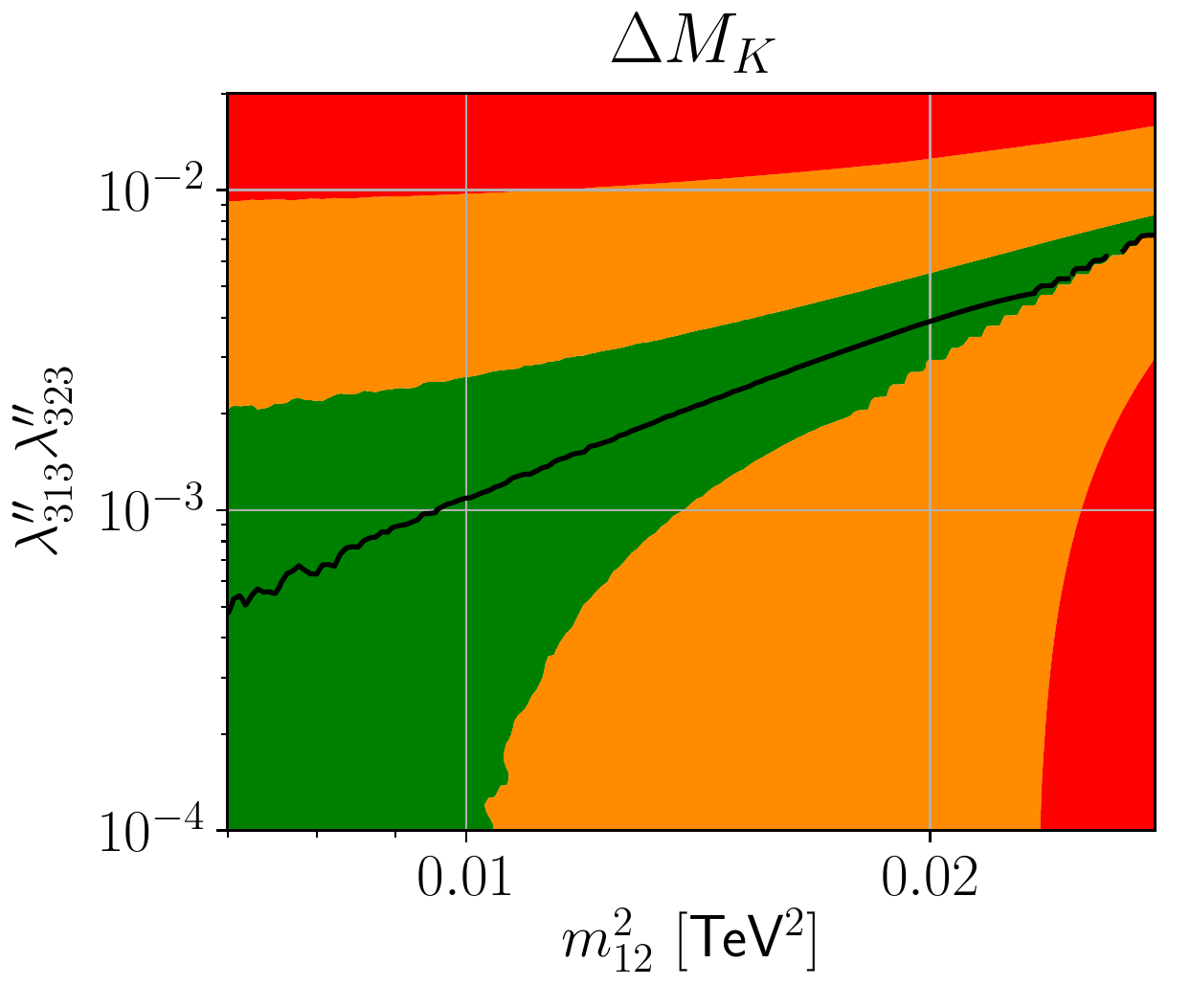}
\includegraphics[width=.45\linewidth]{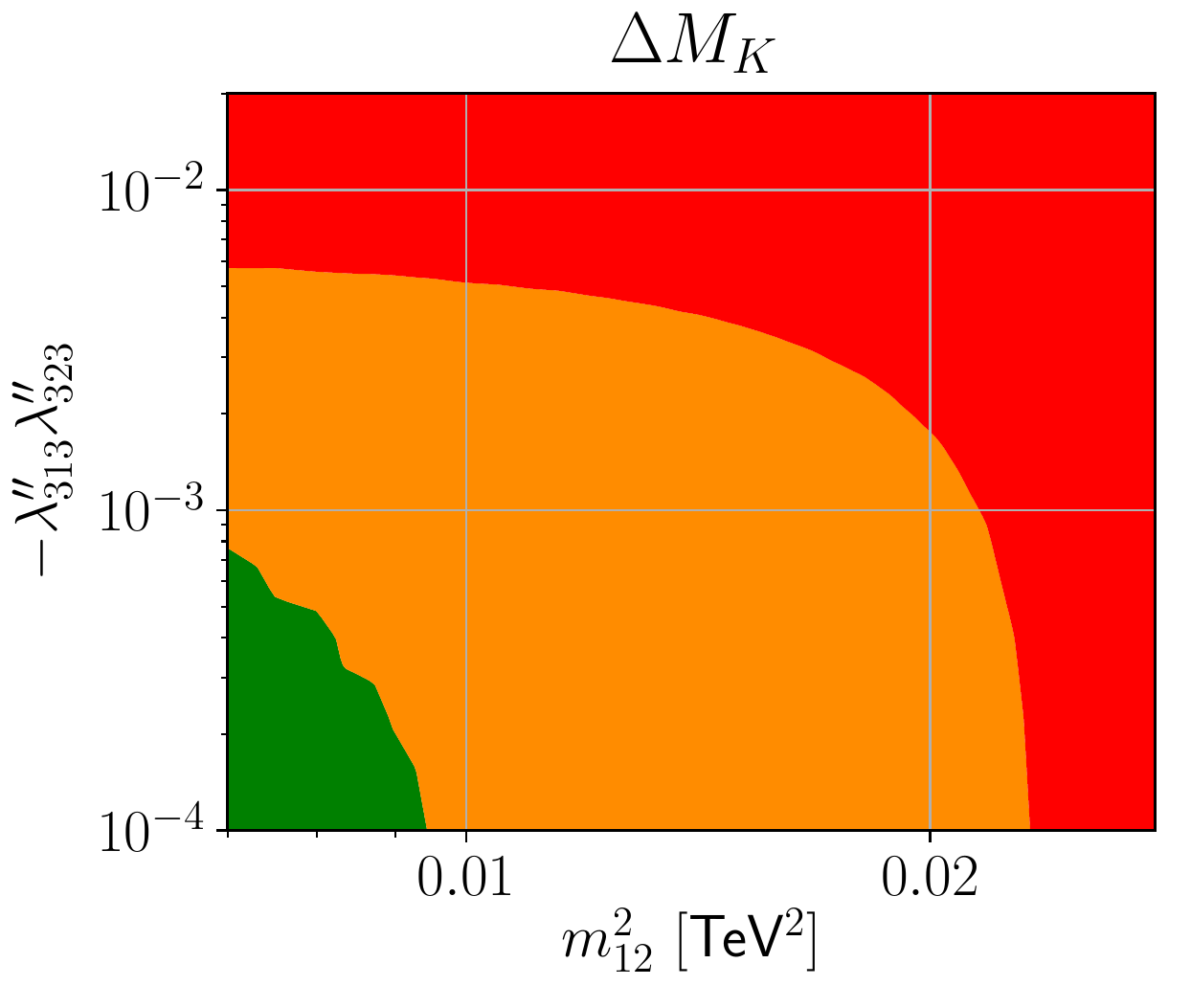}
\caption{Constraints from the meson-oscillation parameters in the presence of both flavor-violating $\bar U\bar D\bar D$-couplings and flavor-violating
squark mixing. The parameters are set to the scenario SUSY-RpV(a) from Table~\ref{Tbl:parameters_scenarios}. The color code is unchanged compared to previous plots.
\label{fig:squark_mixing_vs_tree_couplings_UDD}}
\end{figure}

RpV-couplings are not the only new sources of flavor violation in SUSY-inspired models. In fact, the large number of possible flavor-violating parameters of the 
$R_p$-conserving soft-SUSY-breaking Lagrangian is often perceived as a weakness for this class of model, known as the SUSY Flavor Problem. In particular,
the soft quadratic mass-terms in the squark sector $m^2_{Q,\bar U,\bar D}$ and the trilinear soft terms $A_{U,D}$ are matrices in flavor-space that are not 
necessarily aligned with the flavor-structure of the Yukawa/CKM matrices. In this case, flavor-violation is generated in $L-L$, $R-R$ (for $\tilde m^2$) or $L-R$ (for $A$) squark mixing. Correspondingly, flavor-changing-neutral gluinos or neutralinos, as well as new flavor-changing chargino couplings, 
could contribute to $\Delta M_{K,d,s}$ in \textit{e.g.}\  diagrams of the form of Fig.~\ref{fig:topologies_boxes}, (b--d) --- see \textit{e.g.}\ Ref.~\cite{Saha:2003tq}. 
Here, we wish to illustrate the potential interplay of $R_p$-conserving and RpV flavor violation. In particular, we note that the presence of flavor-violating effects 
in RpV-couplings would likely mediate flavor-violation in the squark sector via the RGE's \cite{Allanach:1999mh}.

We will focus on $R_p$-conserving flavor-violation in the quadratic squark mass parameters $m^2_{ij}$, where we assume the diagonal terms to be degenerate for squarks of left-handed and 
right-handed type (for simplicity): $m^2_{\bar D}= m^2_Q\equiv m^2$. Flavor-violation in the trilinear soft terms would lead to comparable effects at the level of the meson-oscillation
parameters. However, large $A$-terms easily produce new (\textit{e.g.}\ color- and charge-violating) minima in the scalar potential, that lead to instability of the usual vacuum,
with possibly short-time tunnelling. In fact, we find that such stability considerations
typically constrain the $A$-terms much more efficiently than the $\Delta M$'s.

In Fig.~\ref{fig:squark_mixing_vs_tree_couplings}, we allow for non-vanishing $m^2_{13}$, $m^2_{23}$ or $m^2_{12}$, simultaneously with non-zero 
$\lambda'_{113}\lambda'_{131}$, $\lambda'_{123}\lambda'_{132}$ and $\lambda'_{112}\lambda'_{121}$. The former induce contributions to $\Delta M_d$, $\Delta M_s$
and $\Delta M_K$ through $R_p$-conserving squark mixing, while the latter provide RpV tree-level contributions to the same $\Delta M$'s. The parameters 
are set to the scenario SUSY-RpV(a) of Table~\ref{Tbl:parameters_scenarios}, with the slepton/sneutrino mass at $1.5$\,TeV. In analogy with the results of 
section~\ref{subsec:more_than_two_nonzero}, we observe that $R_p$-conserving and RpV contributions may interfere destructively or constructively.
Thus, allowed funnels with comparatively large values of the RpV-couplings open. In particular, we note that a tiny $m^2_{12}$ is sufficient for relaxing 
limits from $\Delta M_K$, while the typical values of $m^2_{13}$ and $m^2_{23}$ affecting $\Delta M_{d}$ and $\Delta M_s$ are significantly larger.

A similar analysis can be performed with RpV of the $\bar U\bar D\bar D$-type. This is shown in Fig.~\ref{fig:squark_mixing_vs_tree_couplings_UDD}.

In this subsection, we have stressed that the limits originating from meson-oscillation parameters are quite sensitive to the possible existence of 
flavor-violating sources beyond that of the RpV-couplings. A full analysis of these effects thus appears necessary when testing a complete model.

\section{Conclusions}\label{sec:conclusion}
In this paper, we have analyzed the meson-mixing parameters $\Delta M_{d,s}$ and $\Delta M_K$ at the full one-loop order in the RpV-MSSM.
In particular, we have completed earlier calculations in the literature, in which only tree-level and box diagrams were usually considered. 
We also 
performed a numerical study based on our results and employing recent experimental and lattice data. The tighter limits that we derive --- as compared to older works --- illustrate the improvement of the precision in experimental measurements, but also the relevance of some of the new contributions that 
we consider. In particular, the interplay of SM-like and $LQ\bar D$-type flavor-violation modifies the scaling of the bounds with the sneutrino/slepton mass for a
whole class of couplings.
Finally, we have emphasized the possibility of interference effects amongst new sources of flavor violation, either exclusively in the RpV-sector
or in association with $R_p$-conserving squark mixing. While the appearance of allowed directions with comparatively large couplings largely intervenes as a fine-tuned 
curiosity in the low-energy perspective of our work, it also stresses the relevance of a detailed analysis of the observables when considering a complete high-energy model,
since accidental relations among parameters could affect the picture of low-energy limits.

\vspace*{1cm}
\section*{Acknowledgements}
H.~K.~D. and Z.S.W.~are supported by the Sino-German DFG grant SFB CRC 110 ``Symmetries and the Emergence of Structure in QCD''. 
 M.E.K.~is supported by the DFG Research Unit 2239 ``New Physics
at the LHC''.
The work of~F.~D. was supported in part by the MEINCOP~(Spain) under contract
\mbox{FPA2016-78022-P}, in part by the ``Spanish Agencia Estatal de
Investigaci\'on''~(AEI) and the EU~``Fondo Europeo de Desarrollo
Regional''~(FEDER) through the project \mbox{FPA2016-78022-P}, and in
part by the~AEI through the grant IFT~Centro de Excelencia Severo
Ochoa \mbox{SEV-2016-0597}. V.M.L. acknowledges support of the BMBF under project 05H18PDCA1.
We thank the authors of {\tt FlavorKit}, and in particular Avelino Vicente, for vivid exchanges and invaluable assistance in cross-checking our results 
with the {\tt FlavorKit} routines. Likewise, we thank David Straub for very helpful correspondence concerning 
{\tt Flavio}. We also thank Martin Hirsch and Toby Opferkuch for useful discussions. Z.S.W. thanks the IFIC for hospitality, and thanks the COST Action CA15108 for the financial support during his research stay at the IFIC. We thank Alexander Lenz for comments concerning the current SM evaluation of $\Delta M_s$. 

\appendix
\section{Notations}\label{Appendix:Notations}
\subsection{Mixing matrices}\label{Appendix:Mixing matrices}
\begin{itemize}
 \item The squark mass matrices mix left- and right-handed components. We define the mass-eigenstates in terms of a unitary rotation of the gauge/flavor-eigenstates:
 \begin{equation}
  \begin{cases}
   U_{\alpha}=X^{U_L}_{\alpha f}\,U_L^f+X^{U_R}_{\alpha f}\,\bar{U}_R^{f\,*}\\
   D_{\alpha}=X^{D_L}_{\alpha f}\,D_L^f+X^{D_R}_{\alpha f}\,\bar{D}_R^{f\,*}
  \end{cases}
 \end{equation}
 Here, $U_{\alpha}$ (resp.\ $D_{\alpha}$) represents the scalar-up (resp.\ sdown) mass state with mass $m_{U_{\alpha}}$ (resp.\ $m_{D_{\alpha}}$). Summation over the 
 generation index $f$ is implicit.
 \item R-parity violation leads to a mixing of charged-Higgs and slepton fields. We define the mass-eigenstates $H_{\alpha}^{\pm}$ with mass $m_{H_{\alpha}}$ as:
 \begin{equation}
  H_{\alpha}^-=X_{\alpha u}^C\,H_u^-+X_{\alpha d}^C\,H_d^-+X^C_{\alpha E_L^f}\,E_L^f+X^C_{\alpha E_R^f}\,\bar{E}_R^{f\,*}\,.
 \end{equation}
 \item Similarly, the neutral Higgs mass-states involve both the doublet-Higgs, $H_u^0=v_u+\frac{h_u^0+\imath\,a_u^0}{\sqrt{2}}$ and $H_d^0=v_d+\frac{h_d^0+\imath\,a_d^0}{\sqrt{2}}$, 
and the sneutrino fields, $N^f_L=\frac{h^0_{N_f}+\imath\,a^0_{N_f}}{\sqrt{2}}$; in the CP-violating case, CP-even and CP-odd components mix as well.
 \begin{equation}
  S_{\alpha}=X_{\alpha u}^R\,h_u^0+X_{\alpha d}^R\,h_d^0+X^R_{\alpha N_f}\,h_{N_f}^0+X_{\alpha u}^I\,a_u^0+X_{\alpha d}^I\,a_d^0+X^I_{\alpha N_f}\,a_{N_f}^0\,.
 \end{equation}
 $S_{\alpha}$ denotes the mass-eigenstate associated with the mass $m_{S_{\alpha}}$.
 \item The charged winos $\tilde{w}^+$, $\tilde{w}^-$, higgsinos $\tilde{h}_u^+$, $\tilde{h}_d^-$ and lepton fields $e_L^f$, $\bar{e}_R^{f}$ define the chargino sector.
 For the mass $m_{\chi^{\pm}_k}$, the associated eignstate is given by:
 \begin{equation}
  \begin{cases}
  \chi_k^+=V_{kw}\,\tilde{w}^++V_{ku}\,\tilde{h}_u^++V_{ke_f}\,\bar{e}_R^{f}\,,\\
  \chi_k^-=U_{kw}\,\tilde{w}^-+U_{kd}\,\tilde{h}_d^-+U_{ke_f}\,e_L^f\,.
  \end{cases}
 \end{equation}
 \item The violation of R-parity also mixes neutrino and neutralino states. The eigenstate with mass $m_{\chi^0_k}$ reads:
  \begin{equation}
  \chi_k^0=N_{kb}\,\tilde{b}^0+N_{kw}\,\tilde{w}^0+N_{ku}\,\tilde{h}_u^0+N_{kd}\,\tilde{h}_d^0+N_{k\nu_f}\,\nu_L^f\,.
 \end{equation}
\end{itemize}

\subsection{Feynman rules}
Here, we list the various couplings that are relevant in our calculation. The combinatorial factors appearing in the lagrangian density in the case of identical coupling particles have been explicitly factored out, e.g.~${\cal L}\ni -\frac{g^{S_{\alpha}ZZ}}{2}S_{\alpha}ZZ$.

\begin{itemize}
 \item Neutral-Higgs-sneutrinos / down quarks:
 \begin{equation}\label{coup:H0d}
  g_L^{S_{\alpha}d_kd_i}=-\frac{1}{\sqrt{2}}\left[Y_d^i\delta_{ki}(X_{\alpha d}^R+\imath X_{\alpha d}^I)+\lambda'_{fik}(X_{\alpha N_L^f}^R+\imath X_{\alpha N_L^f}^I)\right]=(g_R^{S_{\alpha}d_id_k})^*
 \end{equation}
 \item Charged-Higgs-sleptons / quarks:
 \begin{equation}\label{coup:Hcud}
  g_L^{H_{\alpha}u_kd_i}=-Y_u^kV^{CKM}_{ki}X^C_{\alpha u}\, ;\ \ \ g_R^{H_{\alpha}u_kd_i}=-Y_d^iV^{CKM}_{ki}X^C_{\alpha d}-\lambda'^*_{fli}V^{CKM}_{kl}X^C_{\alpha E_L^f}
 \end{equation}
 \item sdowns / neutralino-neutrinos / down quarks:
 \begin{align}\label{coup:Dneud}
  & g_L^{D_{\alpha}\chi_kd_i}=-\frac{1}{\sqrt{2}}\left(\frac{g'}{3}N_{k\tilde{b}}^*-gN_{k\tilde{w}}^*\right)X_{\alpha i}^{D_L}-Y_d^iN_{kd}^*X_{\alpha i}^{D_R}-\lambda'_{fi\beta}N_{k\nu_f}^*X_{\alpha \beta}^{D_R}\nonumber\\
  & g_R^{D_{\alpha}\chi_kd_i}=-\frac{\sqrt{2}}{3}g'N_{k\tilde{b}}X_{\alpha i}^{D_R}-Y_d^iN_{kd}X_{\alpha i}^{D_L}-\lambda'^*_{f\beta i}N_{k\nu_f}X_{\alpha \beta}^{D_L}
 \end{align}
 \item sdowns / gluinos / down quarks ($T^A$ are the colour Gell-Mann matrices):
 \begin{equation}\label{coup:Dglud}
  g_L^{D_{\alpha}^a\tilde{g}^{A}d^b_i}=-\sqrt{2}g_se^{-\imath\phi_{M_3}/2}X^{D_L}_{\alpha i}T^A_{ab}\ \ \ ;\ \ \ g_R^{D_{\alpha}^a\tilde{g}^{A}d^b_i}=\sqrt{2}g_se^{\imath\phi_{M_3}/2}X^{D_R}_{\alpha i}T^A_{ab}
 \end{equation}
 \item scalar-ups / chargino-leptons / down quarks:
 \begin{align}\label{coup:Uchad}
  & g_L^{U_{\alpha}\chi_kd_i}=V^{CKM}_{\beta i}\left[Y_u^{\beta}V_{ku}^*X_{\alpha \beta}^{U_R}-gV_{k\tilde{w}}^*X_{\alpha \beta}^{U_L}\right]\nonumber\\
  & g_R^{U_{\alpha}\chi_kd_i}=V^{CKM}_{\beta f}\left[Y_d^{i}\delta_{if}U_{kd}X_{\alpha \beta}^{U_L}+\lambda'^*_{lfi}U_{k e_l}X_{\alpha\beta}^{U_L}\right]
 \end{align}
 \item scalar-ups / down quarks ($a,b,c$: colour-indices):
 \begin{equation}\label{coup:Udd}
  g_L^{U^a_{\alpha}d^b_kd^c_i}=0\ \ \ ;\ \ \ g_R^{U^a_{\alpha}d^b_kd^c_i}=\varepsilon_{abc}\lambda''^*_{fki}X^{U_R}_{\alpha f}
 \end{equation}
 \item sdowns / up / down quarks ($a,b,c$: colour-indices):
 \begin{equation}\label{coup:Dud}
  g_L^{D^a_{\alpha}u^b_kd^c_i}=0\ \ \ ;\ \ \ g_R^{D^a_{\alpha}u^b_kd^c_i}=\varepsilon_{bac}\lambda''^*_{kfi}X^{D_R}_{\alpha f}
 \end{equation}
 \item W / up / down quarks:
 \begin{equation}\label{coup:Wud}
  g_L^{Wu_kd_i}=\frac{g}{\sqrt{2}}V^{CKM}_{ki}\ \ \ ;\ \ \ g_R^{Wu_kd_i}=0
 \end{equation}
 \item Z / down quarks:
 \begin{equation}\label{coup:Zdd}
  g_L^{Zd_kd_i}=\frac{\sqrt{g'^2+g^2}}{2}\left(-1+\frac{2}{3}s^2_W\right)\delta_{ik}\ \ \ ;\ \ \ g_R^{Zd_kd_i}=\frac{\sqrt{g'^2+g^2}}{3}s^2_W\delta_{ik}
 \end{equation}
 \item Neutral-Higgs-sneutrinos / up quarks:
 \begin{equation}\label{coup:H0u}
  g_L^{S_{\alpha}u_ju_k}=-\frac{Y_u^j}{\sqrt{2}}\delta_{jk}\left(X^R_{\alpha u}+\imath X^I_{\alpha u}\right)=\left(g_R^{S_{\alpha}u_ku_j}\right)^*
 \end{equation}
 \item Neutral-Higgs-sneutrinos / charginos-leptons:
 \begin{multline}\label{coup:H0cha}
  g_L^{S_{\alpha}\chi^+_j\chi^-_k}=-\frac{1}{\sqrt{2}}\left\{Y_e^f\left[\left(X^R_{\alpha d}+\imath X^I_{\alpha d}\right)V^*_{je_f}U^*_{ke_f}-\left(X^R_{\alpha \tilde{N}_f}+\imath X^I_{\alpha \tilde{N}_f}\right)V^*_{je_f}U^*_{kd}\right]\right.\\
  +g\left[\left(X^R_{\alpha u}-\imath X^I_{\alpha u}\right)V^*_{ju}U^*_{kw}+\left(X^R_{\alpha d}-\imath X^I_{\alpha d}\right)V^*_{jw}U^*_{kd}+\left(X^R_{\alpha \tilde{N}_f}-\imath X^I_{\alpha \tilde{N}_f}\right)V^*_{jw}U^*_{ke_f}\right]\\
  \left.+\lambda_{fmn}\left(X^R_{\alpha \tilde{N}_f}+\imath X^I_{\alpha \tilde{N}_f}\right)V^*_{j e_n}U^*_{ke_m}\right\}=\left(g_R^{S_{\alpha}\chi^+_k\chi^-_j}\right)^*
 \end{multline}
 \item Neutral-Higgs-sneutrinos / neutrino-neutralinos:
 \begin{multline}\label{coup:H0neu}
  g_L^{S_{\alpha}\chi^0_j\chi^0_k}=-\frac{g'}{2}\left[\left(X^R_{\alpha u}-\imath X^I_{\alpha u}\right)(N^*_{ju}N^*_{kb}+N^*_{jb}N^*_{ku})-\left(X^R_{\alpha d}-\imath X^I_{\alpha d}\right)(N^*_{jd}N^*_{kb}+N^*_{kd}N^*_{jb})\right.\\
  \left.-\left(X^R_{\alpha \tilde{N}_f}-\imath X^I_{\alpha \tilde{N}_f}\right)(N^*_{j\nu_f}N^*_{kb}+N^*_{k\nu_f}N^*_{jb})\right]\\
  +\frac{g}{2}\left[\left(X^R_{\alpha u}-\imath X^I_{\alpha u}\right)(N^*_{ju}N^*_{kw}+N^*_{jw}N^*_{ku})-\left(X^R_{\alpha d}-\imath X^I_{\alpha d}\right)(N^*_{jd}N^*_{kw}+N^*_{kd}N^*_{jw})\right.\\
  \left.-\left(X^R_{\alpha \tilde{N}_f}-\imath X^I_{\alpha \tilde{N}_f}\right)(N^*_{j\nu_f}N^*_{kw}+N^*_{k\nu_f}N^*_{jw})\right]
  =\left(g_R^{S_{\alpha}\chi^0_k\chi^0_j}\right)^*
 \end{multline}
 \item Neutral-Higgs-sneutrinos / W's:
 \begin{equation}\label{coup:H0W}
  g^{S_{\alpha}WW}=\frac{g^2}{\sqrt{2}}\left(v_uX^R_{\alpha u}+v_dX^R_{\alpha d}\right)
 \end{equation}
 \item Neutral-Higgs-sneutrinos / Z's:
 \begin{equation}\label{coup:H0Z}
  g^{S_{\alpha}ZZ}=\frac{g'^2+g^2}{\sqrt{2}}\left(v_uX^R_{\alpha u}+v_dX^R_{\alpha d}\right)
 \end{equation}
 \item Neutral-Higgs-sneutrinos / W-ghosts $g_W^{\pm}$'s:
 \begin{equation}\label{coup:H0gW}
  g^{S_{\alpha}g_Wg_W}=-\frac{g^2}{2\sqrt{2}}\left[v_u(X^R_{\alpha u}+\imath X^I_{\alpha u})+v_d(X^R_{\alpha d}-\imath X^I_{\alpha d})\right]
 \end{equation}
 \item Neutral-Higgs-sneutrinos / Z-ghosts $g_Z$'s:
 \begin{equation}\label{coup:H0gZ}
  g^{S_{\alpha}g_Zg_Z}=-\frac{g'^2+g^2}{2\sqrt{2}}\left[v_uX^R_{\alpha u}+v_dX^R_{\alpha d}\right]
 \end{equation}
 \item Neutral-Higgs-sneutrinos / $W$ / Charged-Higgs-sleptons:
 \begin{equation}\label{coup:H0WHc}
  g^{S_{\alpha}WH_{k}}=\frac{g}{2}\left[(X^R_{\alpha d}-\imath X^I_{\alpha d})X_{kd}^{C\,*}-(X^R_{\alpha u}+\imath X^I_{\alpha u})X_{ku}^{C\,*}+(X^R_{\alpha \tilde{N}_f}-\imath X^I_{\alpha \tilde{N}_f})X_{k\tilde{E}^f_L}^{C\,*}\right]
 \end{equation}
 \item Neutral-Higgs-sneutrinos / $Z$ / Neutral-Higgs-sneutrinos:
 \begin{equation}\label{coup:H0ZH0}
  g^{S_{\alpha}ZS_{k}}=\imath\frac{\sqrt{g'^2+g^2}}{2}\left[X_{\alpha d}^RX_{kd}^I-X_{\alpha d}^IX_{kd}^R-X_{\alpha u}^RX_{ku}^I+X_{\alpha u}^IX_{ku}^R+X_{\alpha \tilde{N}_{f}}^RX_{k\tilde{N}_{f}}^I-X_{\alpha \tilde{N}_{f}}^IX_{k\tilde{N}_{f}}^R\right]
 \end{equation}
 \item Neutral-Higgs-sneutrinos / scalar-ups:
\begin{multline}\label{coup:H0UU}
 g^{\tilde{U}_k\tilde{U}_lS_{\alpha}}=-\sqrt{2}\left[Y_u^{f\,2}v_uX_{\alpha u}^R+\frac{1}{4}\left(\frac{g'^2}{3}-g^2\right)(v_uX^R_{\alpha u}-v_dX^R_{\alpha d})\right]X^{\tilde{U}_f}_{kL}X^{\tilde{U}_f\,*}_{lL}\\
 -\sqrt{2}\left[Y_u^{f\,2}v_uX_{\alpha u}^R-\frac{g'^2}{3}(v_uX^R_{\alpha u}-v_dX^R_{\alpha d})\right]X^{\tilde{U}_f}_{kR}X^{\tilde{U}_f\,*}_{lR}\\
 -\frac{1}{\sqrt{2}}\left[A_u^{ff'}(X_{\alpha u}^R+\imath X_{\alpha u}^I)-\mu^* Y_u^f\delta_{ff'}(X_{\alpha d}^R-\imath X_{\alpha d}^I)\right]X^{\tilde{U}_{f'}}_{kR}X^{\tilde{U}_f\,*}_{lL}\\
 -\frac{1}{\sqrt{2}}\left[A_u^{ff'\,*}(X_{\alpha u}^R-\imath X_{\alpha u}^I)-\mu Y_u^f\delta_{ff'}(X_{\alpha d}^R+\imath X_{\alpha d}^I)\right]X^{\tilde{U}_f}_{kL}X^{\tilde{U}_{f'}\,*}_{lR}
\end{multline}
 \item Neutral-Higgs-sneutrinos / sdowns:
{\small\begin{multline}\label{coup:H0DD}
 g^{\tilde{D}_k\tilde{D}_lS_{\alpha}}=-\sqrt{2}\left[Y_d^{f\,2}v_dX_{\alpha d}^R+\frac{1}{4}\left(\frac{g'^2}{3}+g^2\right)(v_uX^R_{\alpha u}-v_dX^R_{\alpha d})\right]X^{\tilde{D}_f}_{kL}X^{\tilde{D}_f\,*}_{lL}\\
 -\frac{v_d}{\sqrt{2}}\left[Y_d^f\lambda'^*_{ghf}(X^R_{\alpha \tilde{N}_g}-\imath X^I_{\alpha \tilde{N}_g})+Y_d^h\lambda'_{gfh}(X^R_{\alpha \tilde{N}_g}+\imath X^I_{\alpha \tilde{N}_g})\right]X^{\tilde{D}_h}_{kL}X^{\tilde{D}_f\,*}_{lL}\\
 -\sqrt{2}\left[Y_d^{f\,2}v_dX_{\alpha d}^R+\frac{g'^2}{6}(v_uX^R_{\alpha u}-v_dX^R_{\alpha d})\right]X^{\tilde{D}_f}_{kR}X^{\tilde{D}_f\,*}_{lR}\\
 -\frac{v_d}{\sqrt{2}}\left[Y_d^f\lambda'^*_{gfh}(X^R_{\alpha \tilde{N}_g}-\imath X^I_{\alpha \tilde{N}_g})+Y_d^h\lambda'_{ghf}(X^R_{\alpha \tilde{N}_g}+\imath X^I_{\alpha \tilde{N}_g})\right]X^{\tilde{D}_f}_{kR}X^{\tilde{D}_h\,*}_{lR}\\
 -\frac{1}{\sqrt{2}}\left[A_d^{ff'}(X_{\alpha d}^R+\imath X_{\alpha d}^I)-\mu^* Y_d^f\delta_{ff'}(X_{\alpha u}^R-\imath X_{\alpha u}^I)+A'_{gff'}(X_{\alpha \tilde{N}_g}^R+\imath X_{\alpha \tilde{N}_g}^I)\right]X^{\tilde{D}_{f'}}_{kR}X^{\tilde{D}_f\,*}_{lL}\\
 -\frac{1}{\sqrt{2}}\left[A_d^{ff'\,*}(X_{\alpha d}^R-\imath X_{\alpha d}^I)-\mu Y_d^f\delta_{ff'}(X_{\alpha u}^R+\imath X_{\alpha u}^I)+A'^*_{gff'}(X_{\alpha \tilde{N}_g}^R-\imath X_{\alpha \tilde{N}_g}^I)\right]X^{\tilde{D}_f}_{kL}X^{\tilde{D}_{f'}\,*}_{lR}
\end{multline}}
 \item Neutral-Higgs-sneutrinos / Charged Higgs-sleptons
{\small\begin{multline}\label{coup:H0HcHc}
 g^{H_kH_lS_{\alpha}}=-\sqrt{2}\left\{\left[Y_e^{f\,2}v_dX_{\alpha d}^R+\frac{1}{4}\left(-g'^2+g^2\right)(v_uX^R_{\alpha u}-v_dX^R_{\alpha d})\right]\delta_{ff'}\right.\\
 \left.\null\hspace{3cm}-\frac{v_d}{2}\left[Y_e^{f'}\lambda^*_{fgf'}(X^R_{\alpha \tilde{N}_g}-\imath X^I_{\alpha \tilde{N}_g})+Y_e^f\lambda_{f'gf}(X^R_{\alpha \tilde{N}_g}+\imath X^I_{\alpha \tilde{N}_g})\right]\right\}X^{C}_{k\tilde{E}^f_L}X^{C\,*}_{l\tilde{E}^{f'}_L}\\
 -\sqrt{2}\left\{\left[Y_e^{f\,2}v_dX_{\alpha d}^R+\frac{g'^2}{2}(v_uX^R_{\alpha u}-v_dX^R_{\alpha d})\right]\delta_{ff'}\right.\\
 \left.\null\hspace{3cm}-\frac{v_d}{2}\left[Y_e^{f}\lambda^*_{fgf'}(X^R_{\alpha \tilde{N}_g}-\imath X^I_{\alpha \tilde{N}_g})+Y_e^{f'}\lambda_{f'gf}(X^R_{\alpha \tilde{N}_g}+\imath X^I_{\alpha \tilde{N}_g})\right]\right\}X^{C}_{k\tilde{E}^f_R}X^{C\,*}_{l\tilde{E}^{f'}_R}\\
 -\frac{1}{\sqrt{2}}\left[A_e^{f'f}(X_{\alpha d}^R+\imath X_{\alpha d}^I)-\mu^* Y_e^f\delta_{ff'}(X_{\alpha u}^R-\imath X_{\alpha u}^I)+A_{gf'f}(X_{\alpha \tilde{N}_g}^R+\imath X_{\alpha \tilde{N}_g}^I)\right]X^{C}_{k\tilde{E}^f_R}X^{C\,*}_{l\tilde{E}^{f'}_L}\\
 -\frac{1}{\sqrt{2}}\left[A_e^{ff'\,*}(X_{\alpha d}^R-\imath X_{\alpha d}^I)-\mu Y_e^f\delta_{ff'}(X_{\alpha u}^R+\imath X_{\alpha u}^I)+A^*_{gff'}(X_{\alpha \tilde{N}_g}^R-\imath X_{\alpha \tilde{N}_g}^I)\right]X^{C}_{k\tilde{E}^f_L}X^{C\,*}_{l\tilde{E}^{f'}_R}\\
 -\frac{1}{2\sqrt{2}}\left[g'^2(v_uX_{\alpha u}^R-v_dX_{\alpha d}^R)+g^2(v_uX_{\alpha u}^R+v_dX_{\alpha d}^R)\right]X^{C}_{ku}X^{C\,*}_{lu}\\
 -\frac{1}{2\sqrt{2}}\left[g'^2(v_dX_{\alpha d}^R-v_uX_{\alpha u}^R)+g^2(v_uX_{\alpha u}^R+v_dX_{\alpha d}^R)\right]X^{C}_{kd}X^{C\,*}_{ld}\\
 -\frac{g^2}{2\sqrt{2}}\left[v_u(X_{\alpha d}^R-\imath X_{\alpha d}^I)+v_d(X_{\alpha u}^R-\imath X_{\alpha u}^I)\right]X^{C}_{ku}X^{C\,*}_{ld}\\
 -\frac{g^2}{2\sqrt{2}}\left[v_u(X_{\alpha d}^R+\imath X_{\alpha d}^I)+v_d(X_{\alpha u}^R+\imath X_{\alpha u}^I)\right]X^{C}_{kd}X^{C\,*}_{lu}\\
 +\frac{1}{\sqrt{2}}\left[A_e^{ff'}(X_{\alpha \tilde{N}_f}^R+\imath X_{\alpha \tilde{N}_f}^I)X^{C}_{k\tilde{E}^{f'}_R}X^{C\,*}_{ld}+A_e^{ff'\,*}(X_{\alpha \tilde{N}_f}^R-\imath X_{\alpha \tilde{N}_f}^I)X^{C}_{kd}X^{C\,*}_{l\tilde{E}^{f'}_R}\right]\\
 +\frac{Y_e^{f\,2}v_d}{\sqrt{2}}\left[(X_{\alpha \tilde{N}_f}^R-\imath X_{\alpha \tilde{N}_f}^I)X^{C}_{kd}X^{C\,*}_{l\tilde{E}^f_L}+(X_{\alpha \tilde{N}_f}^R+\imath X_{\alpha \tilde{N}_f}^I)X^{C}_{k\tilde{E}^f_L}X^{C\,*}_{ld}\right]\\
 +\frac{Y_e^f}{\sqrt{2}}\left[\mu^*(X_{\alpha \tilde{N}_f}^R+\imath X_{\alpha \tilde{N}_f}^I)X^{C}_{k\tilde{E}^f_R}X^{C\,*}_{lu}+\mu(X_{\alpha \tilde{N}_f}^R-\imath X_{\alpha \tilde{N}_f}^I)X^{C}_{ku}X^{C\,*}_{l\tilde{E}^f_R}\right]\\
 -\frac{g^2}{2\sqrt{2}}\left[(X_{\alpha \tilde{N}_f}^R+\imath X_{\alpha \tilde{N}_f}^I)X^{C}_{k\tilde{E}^f_L}(v_uX^{C\,*}_{lu}+v_dX^{C\,*}_{ld})+(X_{\alpha \tilde{N}_f}^R-\imath X_{\alpha \tilde{N}_f}^I)(v_uX^{C}_{ku}+v_dX^{C}_{kd})X^{C\,*}_{l\tilde{E}^f_L}\right]
\end{multline}}
 \item Cubic Neutral-Higgs-sneutrinos:
\begin{align}\label{coup:H03}
 g^{S_{\alpha}S_{\beta}S_{\gamma}}= \frac{g'^2+g^2}{4\sqrt{2}}&\left[v_u\left(\Pi^{S\,uuu}_{\alpha\beta\gamma}+\Pi^{A\,uuu}_{\alpha\beta\gamma}-\Pi^{S\,udd}_{\alpha\beta\gamma}-\Pi^{A\,udd}_{\alpha\beta\gamma}-\Pi^{S\,u\tilde{N}_f\tilde{N}_f}_{\alpha\beta\gamma}-\Pi^{A\,u\tilde{N}_f\tilde{N}_f}_{\alpha\beta\gamma}\right)\right.\nonumber\\
 &\left.+v_d\left(\Pi^{S\,ddd}_{\alpha\beta\gamma}+\Pi^{A\,ddd}_{\alpha\beta\gamma}-\Pi^{S\,duu}_{\alpha\beta\gamma}-\Pi^{A\,duu}_{\alpha\beta\gamma}-\Pi^{S\,d\tilde{N}_f\tilde{N}_f}_{\alpha\beta\gamma}-\Pi^{A\,d\tilde{N}_f\tilde{N}_f}_{\alpha\beta\gamma}\right)\right]
\end{align}
where:
\begin{align*}
 \Pi^{S\,abc}_{\alpha\beta\gamma} =&X^{R}_{\alpha a}X^{R}_{\beta b}X^{R}_{\gamma c}+X^{R}_{\alpha b}X^{R}_{\beta c}X^{R}_{\gamma a}+X^{R}_{\alpha c}X^{R}_{\beta a}X^{R}_{\gamma b}+X^{R}_{\alpha a}X^{R}_{\beta c}X^{R}_{\gamma b}+X^{R}_{\alpha c}X^{R}_{\beta b}X^{R}_{\gamma a}+X^{R}_{\alpha b}X^{R}_{\beta a}X^{R}_{\gamma c}\\
 \Pi^{A\,abc}_{\alpha\beta\gamma} =&X^{R}_{\alpha a}\left(X^{I}_{\beta b}X^{I}_{\gamma c}+X^{I}_{\beta c}X^{I}_{\gamma b}\right)+X^{R}_{\beta a}\left(X^{I}_{\alpha b}X^{I}_{\gamma c}+X^{I}_{\alpha c}X^{I}_{\gamma b}\right)+X^{R}_{\gamma a}\left(X^{I}_{\alpha b}X^{I}_{\beta c}+X^{I}_{\alpha c}X^{I}_{\beta b}\right)
\end{align*}
 \item Neutral-Higgs-sneutrinos / W quartic:
 \begin{equation}\label{coup:H02WW}
g^{WWS_{\alpha}S_{\beta}}=\frac{g^2}{2}\left[X^R_{\alpha u}X^R_{\beta u}+X^I_{\alpha u}X^I_{\beta u}+X^R_{\alpha d}X^R_{\beta d}
+X^I_{\alpha d}X^I_{\beta d}+X^R_{\alpha \tilde{N}_f}X^R_{\beta \tilde{N}_f}+X^I_{\alpha \tilde{N}_f}X^I_{\beta \tilde{N}_f}\right]
 \end{equation}
 \item Neutral-Higgs-sneutrinos / Z quartic:
 \begin{equation}\label{coup:H02ZZ}
g^{ZZS_{\alpha}S_{\beta}}=\frac{g'^2+g^2}{2}\left[X^R_{\alpha u}X^R_{\beta u}+X^I_{\alpha u}X^I_{\beta u}+X^R_{\alpha d}X^R_{\beta d}
+X^I_{\alpha d}X^I_{\beta d}+X^R_{\alpha \tilde{N}_f}X^R_{\beta \tilde{N}_f}+X^I_{\alpha \tilde{N}_f}X^I_{\beta \tilde{N}_f}\right]
 \end{equation}
 \item Neutral-Higgs-sneutrinos / scalar-ups quartic:
\begin{multline}\label{coup:H02UU}
 g^{\tilde{U}_k\tilde{U}_lS_{\alpha}S_{\beta}}=-Y_u^{f\,2}\left(X_{\alpha u}^RX_{\beta u}^R+X_{\alpha u}^IX_{\beta u}^I\right)\left(X^{\tilde{U}_f}_{kL}X^{\tilde{U}_f\,*}_{lL}+X^{\tilde{U}_f}_{kR}X^{\tilde{U}_f\,*}_{lR}\right)\\
 -\left[\frac{1}{4}\left(\frac{g'^2}{3}-g^2\right)X^{\tilde{U}_f}_{kL}X^{\tilde{U}_f\,*}_{lL}-\frac{g'^2}{3}X^{\tilde{U}_f}_{kR}X^{\tilde{U}_f\,*}_{lR}\right]\\
\times\left(X_{\alpha u}^RX_{\beta u}^R+X_{\alpha u}^IX_{\beta u}^I-X_{\alpha d}^RX_{\beta d}^R-X_{\alpha d}^IX_{\beta d}^I-X_{\alpha \tilde{N}_{f'}}^RX_{\beta \tilde{N}_{f'}}^R-X_{\alpha \tilde{N}_{f'}}^IX_{\beta \tilde{N}_{f'}}^I\right)
\end{multline}
 \item Neutral-Higgs-sneutrinos / sdowns quartic:
\begin{multline}\label{coup:H02DD}
 g^{\tilde{D}_k\tilde{D}_lS_{\alpha}S_{\beta}}=-Y_d^{f\,2}\left(X_{\alpha d}^RX_{\beta d}^R+X_{\alpha d}^IX_{\beta d}^I\right)\left(X^{\tilde{D}_f}_{kL}X^{\tilde{D}_f\,*}_{lL}+X^{\tilde{D}_f}_{kR}X^{\tilde{D}_f\,*}_{lR}\right)\\
 -\left[\frac{1}{4}\left(\frac{g'^2}{3}+g^2\right)X^{\tilde{D}_f}_{kL}X^{\tilde{D}_f\,*}_{lL}+\frac{g'^2}{6}X^{\tilde{D}_f}_{kR}X^{\tilde{D}_f\,*}_{lR}\right]\\
\times\left(X_{\alpha u}^RX_{\beta u}^R+X_{\alpha u}^IX_{\beta u}^I-X_{\alpha d}^RX_{\beta d}^R-X_{\alpha d}^IX_{\beta d}^I-X_{\alpha \tilde{N}_{f'}}^RX_{\beta \tilde{N}_{f'}}^R-X_{\alpha \tilde{N}_{f'}}^IX_{\beta \tilde{N}_{f'}}^I\right)\\
-\frac{Y_d^f}{2}\left(\lambda'^*_{ghf}X^{\tilde{D}_h}_{kL}X^{\tilde{D}_f\,*}_{lL}+\lambda'^*_{gfh}X^{\tilde{D}_f}_{kR}X^{\tilde{D}_h\,*}_{lR}\right)\left[(X^R_{\alpha d}+\imath X^I_{\alpha d})(X^R_{\beta \tilde{N}_g}-\imath X^I_{\beta \tilde{N}_g})+(\alpha\leftrightarrow\beta)\right]\\
-\frac{Y_d^f}{2}\left(\lambda'_{ghf}X^{\tilde{D}_f}_{kL}X^{\tilde{D}_h\,*}_{lL}+\lambda'_{gfh}X^{\tilde{D}_h}_{kR}X^{\tilde{D}_f\,*}_{lR}\right)\left[(X^R_{\alpha d}-\imath X^I_{\alpha d})(X^R_{\beta \tilde{N}_g}+\imath X^I_{\beta \tilde{N}_g})+(\alpha\leftrightarrow\beta)\right]\\
-\frac{1}{2}\left(\lambda'_{ghf}\lambda'^*_{mnf}X^{\tilde{D}_n}_{kL}X^{\tilde{D}_h\,*}_{lL}+\lambda'_{gfh}\lambda'^*_{mfn}X^{\tilde{D}_h}_{kR}X^{\tilde{D}_n\,*}_{lR}\right)\left[(X^R_{\alpha \tilde{N}_g}+\imath X^I_{\alpha \tilde{N}_g})(X^R_{\beta \tilde{N}_m}-\imath X^I_{\beta \tilde{N}_m})+(\alpha\leftrightarrow\beta)\right]
\end{multline}
 \item Neutral-Higgs-sneutrinos / Charged Higgs-sleptons quartic:
\begin{multline}\label{coup:H02HcHc}
 {\cal L}\ni-Y_e^{f\,2}\left[|H_d^0|^2\left(|E_L^f|^2+|E_R^{c\,f}|^2\right)+|N_L^f|^2H_d^+H_d^--H_d^0N_L^{f\,*}H_d^+E_L^f-H_d^{0*}N_L^fE_L^{f\,*}H_d^-\right]\\
 -\lambda_{jki}\lambda^*_{mni}N_L^jN_L^{m\,*}E_L^{n\,*}E_L^k-\lambda_{ijk}\lambda^*_{imn}N_L^jN_L^{m\,*}E_R^{c\,k}E_R^{c\,n\,*}-Y_e^fY_e^{f'}N_L^fN_L^{f'\,*}E_R^{c\,f}E_R^{c\,f'\,*}\\
 +Y_e^f\left[\lambda^*_{fij}H_d^0N_L^{i\,*}E_R^{c\,f}E_R^{c\,j\,*}+\lambda^*_{ijf}H_d^0N_L^{j\,*}E_L^{i\,*}E_L^f+\lambda^*_{ijf}N_L^fN_L^{i\,*}E_L^{j\,*}H_d^-+cc\right]\\
 -\frac{g'^2}{4}\left[|H_u^0|^2-|H_d^0|^2-|N_L^f|^2\right]\left[H_u^+H_u^--H_d^+H_d^--|E_L^{f'}|^2+2|E_R^{c\,f'}|^2\right]\\
 -\frac{g^2}{4}\left[\left(|H_u^0|^2+|H_d^0|^2+|N_L^f|^2\right)H_u^+H_u^-+\left(|H_d^0|^2+|H_u^0|^2-|N_L^f|^2\right)H_d^+H_d^-\right.\\
 +2N_L^fN_L^{f'\,*}E_L^{f\,*}E_L^{f'}+\left(|H_u^0|^2-|H_d^0|^2-|N_L^f|^2\right)|E_L^f|^2+2H_u^{0\,*}H_d^{0\,*}H_u^+H_d^-+2H_u^0H_d^0H_d^+H_u^-\\
 \left.+2N_L^{f\,*}H_u^{0\,*}H_u^+E_L^f+2N_L^fH_u^{0}E_L^{f\,*}H_u^-+2N_L^{f\,*}H_d^{0}H_d^+E_L^f+2N_L^fH_d^{0\,*}E_L^{f\,*}H_d^-\right]
\end{multline}
The coupling $g^{H_kH_lS_{\alpha}S_{\beta}}$ is obtained through the replacements $H_u^+\to X^C_{ku}$, $H_d^+\to X^C_{kd}$, $E_L^{f\,*}\to X^C_{k\tilde{E}_L^f}$, $E_R^{c\,f}\to X^C_{k\tilde{E}_R^f}$,
 $H_u^-\to X^{C\,*}_{lu}$, $H_d^-\to X^{C\,*}_{ld}$, $E_L^{f}\to X^{C\,*}_{l\tilde{E}_L^f}$, $E_R^{c\,f\,*}\to X^{C\,*}_{l\tilde{E}_R^f}$, $H_u^0\to X^R_{.u}+\imath X^I_{.u}$, $H_d^0\to X^R_{.d}+\imath X^I_{.d}$,
 and $N_L^f\to X^R_{.\tilde{N}_f}+\imath X^I_{.\tilde{N}_f}$ ($.=\alpha,\ \beta$ indifferently, such that the coupling is symmetric over the exchange $\alpha\leftrightarrow\beta$ in the end).
 \item Neutral-Higgs-sneutrinos quartic:
\begin{multline}\label{coup:H04}
 g^{S_{\alpha}S_{\beta}S_{\gamma}S_{\delta}}=\frac{g'^2+g^2}{32}\left[\Pi_{\alpha\beta\gamma\delta}^{S\,uuuu}+\Pi_{\alpha\beta\gamma\delta}^{S\,dddd}-2\Pi_{\alpha\beta\gamma\delta}^{S\,uudd}
 -2\Pi_{\alpha\beta\gamma\delta}^{S\,uu\tilde{N}_f\tilde{N_f}}+2\Pi_{\alpha\beta\gamma\delta}^{S\,dd\tilde{N}_f\tilde{N}_f}\right.\\
\null\hspace{2cm} +\Pi_{\alpha\beta\gamma\delta}^{S\,\tilde{N}_f\tilde{N_f}\tilde{N}_{f'}\tilde{N}_{f'}} +\Pi_{\alpha\beta\gamma\delta}^{P\,uuuu}+\Pi_{\alpha\beta\gamma\delta}^{P\,dddd}-2\Pi_{\alpha\beta\gamma\delta}^{P\,uudd}
 -2\Pi_{\alpha\beta\gamma\delta}^{P\,uu\tilde{N}_f\tilde{N_f}}+2\Pi_{\alpha\beta\gamma\delta}^{P\,dd\tilde{N}_f\tilde{N}_f}\\
  \null\hspace{1cm}+\Pi_{\alpha\beta\gamma\delta}^{P\,\tilde{N}_f\tilde{N_f}\tilde{N}_{f'}\tilde{N}_{f'}}+2\Pi_{\alpha\beta\gamma\delta}^{S\,uu\,P\,uu}+2\Pi_{\alpha\beta\gamma\delta}^{S\,dd\,P\,dd}-2\Pi_{\alpha\beta\gamma\delta}^{S\,uu\,P\,dd}-2\Pi_{\alpha\beta\gamma\delta}^{S\,dd\,P\,uu}\\
 \null\hspace{1.9cm} \left.-2\Pi_{\alpha\beta\gamma\delta}^{S\,uu\,P\,\tilde{N}_f\tilde{N_f}}-2\Pi_{\alpha\beta\gamma\delta}^{S\,\tilde{N}_f\tilde{N_f}\,P\,uu}
+2\Pi_{\alpha\beta\gamma\delta}^{S\,dd\,P\,\tilde{N}_f\tilde{N}_f}+2\Pi_{\alpha\beta\gamma\delta}^{S\,\tilde{N}_f\tilde{N}_f\,P\,dd}+2\Pi_{\alpha\beta\gamma\delta}^{S\,\tilde{N}_f\tilde{N_f}\,P\,\tilde{N}_{f'}\tilde{N}_{f'}}\right]
\end{multline}
where:
\begin{align*}
 &\Pi^{S\,abcd}_{ijkl}=\sum_{\sigma\in S_4}X^{R}_{\sigma(i)a}X^{R}_{\sigma(j)b}X^{R}_{\sigma(k)c}X^{R}_{\sigma(l)d}\ \ \ \ \ ;\ \ \ \Pi^{P\,abcd}_{ijkl}=\sum_{\sigma\in S_4}X^{I}_{\sigma(i)a}X^{I}_{\sigma(j)b}X^{I}_{\sigma(k)c}X^{I}_{\sigma(l)d}\\
 &\Pi^{S\,ab\,P\,cd}_{ijkl}=\sum_{\sigma\in S_4}X^{R}_{\sigma(i)a}X^{R}_{\sigma(j)b}X^{I}_{\sigma(k)c}X^{I}_{\sigma(l)d}
\end{align*}
\end{itemize}

\subsection{Loop-functions}\label{subsec:LoopFunctions}
The loop functions relevant for our computations are
\begin{itemize}
 \item $A_0(m)=-16\pi^2\imath\int{\frac{d^Dk}{(2\pi)^D}\frac{1}{k^2-m^2}}\,$.
 \item $B_0(p,m_1,m_2)=-16\pi^2\imath\int{\frac{d^Dk}{(2\pi)^D}\frac{1}{[k^2-m_1^2][(k+p)^2-m_2^2]}}\,$.
 \item $p^{\mu}B_1(p,m_1,m_2)=-16\pi^2\imath\int{\frac{d^Dk}{(2\pi)^D}\frac{k^{\mu}}{[k^2-m_1^2][(k+p)^2-m_2^2]}}\,$.
 \item $\left[g^{\mu\nu}B_{22}+p^{\mu}p^{\nu}B_{21}\right](p,m_1,m_2)=-16\pi^2\imath\int{\frac{d^Dk}{(2\pi)^D}\frac{k^{\mu}k^{\nu}}{[k^2-m_1^2][(k+p)^2-m_2^2]}}\,$.
 \item $C_0(p_1,p_2,m_1,m_2,m_3)=-16\pi^2\imath\int{\frac{d^Dk}{(2\pi)^D}\frac{1}{[k^2-m_1^2][(k+p_1)^2-m_2^2][(k+p_1+p_2)^2-m_3^2]}}\,$.
 \item $\left[p_1^{\mu}C_{11}+p_2^{\mu}C_{12}\right](p_1,p_2,m_1,m_2,m_3)=-16\pi^2\imath\int{\frac{d^Dk}{(2\pi)^D}\frac{k^{\mu}}{[k^2-m_1^2][(k+p_1)^2-m_2^2][(k+p_1+p_2)^2-m_3^2]}}\,$.
 \item $\left[g^{\mu\nu}C_{24}+p_1^{\mu}p_1^{\nu}C_{21}+p_2^{\mu}p_2^{\nu}C_{22}+(p_1^{\mu}p_2^{\nu}+p_2^{\mu}p_1^{\nu})C_{23}\right](p_1,p_2,m_1,m_2,m_3)=\\
 -16\pi^2\imath\int{\frac{d^Dk}{(2\pi)^D}\frac{k^{\mu}k^{\nu}}{[k^2-m_1^2][(k+p_1)^2-m_2^2][(k+p_1+p_2)^2-m_3^2]}}\,$.
 \item $D_0(m_1,m_2,m_3,m_4)=-16\pi^2\imath\int{\frac{d^Dk}{(2\pi)^D}\frac{1}{[k^2-m_1^2][k^2-m_2^2][k^2-m_3^2][k^2-m_4^2]}}\,$.
 \item $D_2(m_1,m_2,m_3,m_4)=-16\pi^2\imath\int{\frac{d^Dk}{(2\pi)^D}\frac{k^2}{[k^2-m_1^2][k^2-m_2^2][k^2-m_3^2][k^2-m_4^2]}}\,$.
\end{itemize}
Explicit expressions for these functions in the limit of vanishing external momenta can \textit{e.g.}\ be found in Ref.~\cite{Abada:2014kba}.

\section{Tree level contributions}\label{appendix:tree_level}
The tree-level contribution to the $d_i\bar{d}_j\to d_j\bar{d}_i$ amplitudes corresponds to the topology of Fig.\ref{dia:Tree} and is mediated by a sneutrino 
internal line. It generates the following terms in the EFT:
\begin{equation}
 {\cal L}_{\mbox{\tiny EFT}}\ni\frac{1}{2m^2_{S_{\alpha}}}\left[\left(g_L^{S_{\alpha}d_jd_i}\right)^2O_2+\left(g_R^{S_{\alpha}d_jd_i}\right)^2\tilde{O}_2+2g_L^{S_{\alpha}d_jd_i}g_R^{S_{\alpha}d_jd_i}O_4\right]
\end{equation}
where the couplings $g_{L,R}^{S_{\alpha}d_jd_i}$ are defined in Eq.(\ref{coup:H0d}). The sum over sneutrino/neutral-Higgs mixed states $S_{\alpha}$ with mass
$m_{S_{\alpha}}$ is implicit. The operators $O_2$, $\tilde{O}_2$, etc, are defined in Eq.(\ref{eqn:EffOps}).

\section{\boldmath $d_i-d_j$ self-energy contributions}\label{appendix:quarkSE}

Loop corrections on the external $d$-fermion legs are determined by the LSZ reduction. Defining the matrix of renormalized $d_i-d_j$ self energies as:
$\hat{\Sigma}^{ij}(p\!\!\!/)=\hat{\Sigma}^{ij}_L(p\!\!\!/)P_L+\hat{\Sigma}^{ij}_R(p\!\!\!/)P_R=P_L\tilde{\Sigma}^{ij}_L(p\!\!\!/)+P_R\tilde{\Sigma}^{ij}_R(p\!\!\!/)$, 
we derive the contribution to the EFT:
{\scriptsize\begin{multline}
  {\cal L}_{\mbox{\tiny EFT}}\ni\frac{1}{2m^2_{S_{\alpha}}}\left\{g_L^{S_{\alpha}d_jd_i}\left[\frac{1}{2}g_L^{S_{\alpha}d_jd_i}\left(\left.\frac{d\hat{\Sigma}_L^{jj}}{dp\!\!\!/}\right|_{p\!\!\!/_{d_j}}+\left.\frac{d\hat{\Sigma}_L^{jj}}{dp\!\!\!/}\right|_{p\!\!\!/'_{d_j}}+\left.\frac{d\tilde{\Sigma}_{L}^{ii}}{dp\!\!\!/}\right|_{p\!\!\!/_{d_i}}+\left.\frac{d\tilde{\Sigma}_{L}^{ii}}{dp\!\!\!/}\right|_{p\!\!\!/'_{d_i}}\right)\right.\right.\\
 +\sum_{k\neq j}g_L^{S_{\alpha}d_kd_i}\left(\left.\frac{m_{d_k}\hat{\Sigma}_L^{jk}+p\!\!\!/_{d_j}\hat{\Sigma}_R^{jk}}{m^2_{d_j}-m^2_{d_k}}\right|_{p\!\!\!/_{d_j}}+\left.\frac{m_{d_k}\hat{\Sigma}_L^{jk}+p\!\!\!/'_{d_j}\hat{\Sigma}_R^{jk}}{m^2_{d_j}-m^2_{d_k}}\right|_{p\!\!\!/'_{d_j}}\right)\\
 \left.+\sum_{k\neq i}g_L^{S_{\alpha}d_jd_k}\left(\left.\frac{m_{d_k}\tilde{\Sigma}_L^{ki}+p\!\!\!/_{d_i}\tilde{\Sigma}_R^{ki}}{m_{d_i}^2-m_{d_k}^2}\right|_{p\!\!\!/_{d_i}}+\left.\frac{m_{d_k}\tilde{\Sigma}_L^{ki}+p\!\!\!/'_{d_i}\tilde{\Sigma}_R^{ki}}{m_{d_i}^2-m_{d_k}^2}\right|_{p\!\!\!/'_{d_i}}\right)\right]O_2\\
 \null\hspace{0cm}+g_R^{S_{\alpha}d_jd_i}\left[\frac{1}{2}g_R^{S_{\alpha}d_jd_i}\left(\left.\frac{d\hat{\Sigma}_R^{jj}}{dp\!\!\!/}\right|_{p\!\!\!/_{d_j}}+\left.\frac{d\hat{\Sigma}_R^{jj}}{dp\!\!\!/}\right|_{p\!\!\!/'_{d_j}}+\left.\frac{d\tilde{\Sigma}_{R}^{ii}}{dp\!\!\!/}\right|_{p\!\!\!/_{d_i}}+\left.\frac{d\tilde{\Sigma}_{R}^{ii}}{dp\!\!\!/}\right|_{p\!\!\!/'_{d_i}}\right)\right.\\
 +\sum_{k\neq j}g_R^{S_{\alpha}d_kd_i}\left(\left.\frac{m_{d_k}\hat{\Sigma}_R^{jk}+p\!\!\!/_{d_j}\hat{\Sigma}_L^{jk}}{m^2_{d_j}-m^2_{d_k}}\right|_{p\!\!\!/_{d_j}}+\left.\frac{m_{d_k}\hat{\Sigma}_R^{jk}+p\!\!\!/'_{d_j}\hat{\Sigma}_L^{jk}}{m^2_{d_j}-m^2_{d_k}}\right|_{p\!\!\!/'_{d_j}}\right)\\
 \left.+\sum_{k\neq i}g_R^{S_{\alpha}d_jd_k}\left(\left.\frac{m_{d_k}\tilde{\Sigma}_R^{ki}+p\!\!\!/_{d_i}\tilde{\Sigma}_L^{ki}}{m_{d_i}^2-m_{d_k}^2}\right|_{p\!\!\!/_{d_i}}+\left.\frac{m_{d_k}\tilde{\Sigma}_R^{ki}+p\!\!\!/'_{d_i}\tilde{\Sigma}_L^{ki}}{m_{d_i}^2-m_{d_k}^2}\right|_{p\!\!\!/'_{d_i}}\right)\right]\tilde{O}_2\\
 +\left(g_L^{S_{\alpha}d_jd_i}\left[\frac{1}{2}g_R^{S_{\alpha}d_jd_i}\left(\left.\frac{d\hat{\Sigma}_R^{jj}}{dp\!\!\!/}\right|_{p\!\!\!/_{d_j}}+\left.\frac{d\hat{\Sigma}_R^{jj}}{dp\!\!\!/}\right|_{p\!\!\!/'_{d_j}}+\left.\frac{d\tilde{\Sigma}_{R}^{ii}}{dp\!\!\!/}\right|_{p\!\!\!/_{d_i}}+\left.\frac{d\tilde{\Sigma}_{R}^{ii}}{dp\!\!\!/}\right|_{p\!\!\!/'_{d_i}}\right)\right.\right.\\
 +\sum_{k\neq j}g_R^{S_{\alpha}d_kd_i}\left(\left.\frac{m_{d_k}\hat{\Sigma}_R^{jk}+p\!\!\!/_{d_j}\hat{\Sigma}_L^{jk}}{m^2_{d_j}-m^2_{d_k}}\right|_{p\!\!\!/_{d_j}}+\left.\frac{m_{d_k}\hat{\Sigma}_R^{jk}+p\!\!\!/'_{d_j}\hat{\Sigma}_L^{jk}}{m^2_{d_j}-m^2_{d_k}}\right|_{p\!\!\!/'_{d_j}}\right)\\
 \left.+\sum_{k\neq i}g_R^{S_{\alpha}d_jd_k}\left(\left.\frac{m_{d_k}\tilde{\Sigma}_R^{ki}+p\!\!\!/_{d_i}\tilde{\Sigma}_L^{ki}}{m_{d_i}^2-m_{d_k}^2}\right|_{p\!\!\!/_{d_i}}+\left.\frac{m_{d_k}\tilde{\Sigma}_R^{ki}+p\!\!\!/'_{d_i}\tilde{\Sigma}_L^{ki}}{m_{d_i}^2-m_{d_k}^2}\right|_{p\!\!\!/'_{d_i}}\right)\right]\\
 +g_R^{S_{\alpha}d_jd_i}\left[\frac{1}{2}g_L^{S_{\alpha}d_jd_i}\left(\left.\frac{d\hat{\Sigma}_L^{jj}}{dp\!\!\!/}\right|_{p\!\!\!/_{d_j}}+\left.\frac{d\hat{\Sigma}_L^{jj}}{dp\!\!\!/}\right|_{p\!\!\!/'_{d_j}}+\left.\frac{d\tilde{\Sigma}_{L}^{ii}}{dp\!\!\!/}\right|_{p\!\!\!/_{d_i}}+\left.\frac{d\tilde{\Sigma}_{L}^{ii}}{dp\!\!\!/}\right|_{p\!\!\!/'_{d_i}}\right)\right.\\
 +\sum_{k\neq j}g_L^{S_{\alpha}d_kd_i}\left(\left.\frac{m_{d_k}\hat{\Sigma}_L^{jk}+p\!\!\!/_{d_j}\hat{\Sigma}_R^{jk}}{m^2_{d_j}-m^2_{d_k}}\right|_{p\!\!\!/_{d_j}}+\left.\frac{m_{d_k}\hat{\Sigma}_L^{jk}+p\!\!\!/'_{d_j}\hat{\Sigma}_R^{jk}}{m^2_{d_j}-m^2_{d_k}}\right|_{p\!\!\!/'_{d_j}}\right)\\
 \left.\left.\left.+\sum_{k\neq i}g_L^{S_{\alpha}d_jd_k}\left(\left.\frac{m_{d_k}\tilde{\Sigma}_L^{ki}+p\!\!\!/_{d_i}\tilde{\Sigma}_R^{ki}}{m_{d_i}^2-m_{d_k}^2}\right|_{p\!\!\!/_{d_i}}+\left.\frac{m_{d_k}\tilde{\Sigma}_L^{ki}+p\!\!\!/'_{d_i}\tilde{\Sigma}_R^{ki}}{m_{d_i}^2-m_{d_k}^2}\right|_{p\!\!\!/'_{d_i}}\right)\right]\right)O_4\right\}\,,
\end{multline}}
where the momenta $p\!\!\!/_{d_j}$, $p\!\!\!/'_{d_j}$, $p\!\!\!/_{d_i}$ and $p\!\!\!/'_{d_i}$ are evaluated at the values $m_{d_j}$, $-m_{d_j}$, $m_{d_i}$ and $-m_{d_i}$.
We list below the contributions to the self-energies.

\subsection{Scalar/fermion loop}
{\scriptsize\begin{multline}
-\imath\Sigma^{S/f}_{d_jd_i}(p\!\!\!/)=
\frac{\imath}{16\pi^2}\left\{-p\!\!\!/\left[g_L^{Sfd_j\,*}g_L^{Sfd_i}P_L+g_R^{Sfd_j\,*}g_R^{Sfd_i}P_R\right]B_1+m_f\left[g_R^{Sfd_j\,*}g_L^{Sfd_i}P_L+g_L^{Sfd_j\,*}g_R^{Sfd_i}P_R\right]B_0\right\}\\(-p,m_f,m_S)
\end{multline}}
The scalar/fermion pair $(S/f)$ is summed over the following list of particles:
\begin{itemize}
 \item Higgs-sneutrino/down: couplings from Eq.(\ref{coup:H0d}).
 \item Charged Higgs-slepton/up: couplings from Eq.(\ref{coup:Hcud}).
 \item sdown/neutralino-neutrino: couplings from Eq.(\ref{coup:Dneud}).
 \item sdown/gluino: couplings from Eq.(\ref{coup:Dglud}); color-factor $C_2(3)=4/3$.
 \item sup/chargino-lepton: couplings from  Eq.(\ref{coup:Uchad}).
 \item sup/down: couplings from Eq.(\ref{coup:Udd});
  color factor: $\varepsilon_{abc}\varepsilon_{abd}=2\delta_{cd}$.
 \item sdown/up: couplings from Eq.(\ref{coup:Dud});
  color factor: $\varepsilon_{abc}\varepsilon_{abd}=2\delta_{cd}$.
\end{itemize}

\subsection{Vector/fermion loop}
{\scriptsize\begin{multline}
-\imath\Sigma^{V/f}_{d_jd_i}(p)=
-\frac{\imath}{16\pi^2}\left\{(D-2)p\!\!\!/\left[g_L^{Vfd_j\,*}g_L^{Vfd_i}P_L+g_R^{Vfd_j\,*}g_R^{Vfd_i}P_R\right]B_1\right.\\\left.+Dm_f\left[g_R^{Vfd_j\,*}g_L^{Vfd_i}P_L+g_L^{Vfd_j\,*}g_R^{Vfd_i}P_R\right]B_0\right\}(-p,m_f,m_V)
\end{multline}}
The vector/fermion pair $(S/f)$ is summed over the following list of particles:
\begin{itemize}
 \item $W$/up: Eq.(\ref{coup:Wud}).
 \item $Z$/down: Eq.(\ref{coup:Zdd}).
\end{itemize}

\subsection{Counterterm}
Defining the generic $d$-mass counterterm $\delta m_{d\,ji}=\delta m^L_{d\,ji}P_L+\delta m^R_{d\,ji}P_R$ as well as the $d$-wave-function counterterm 
$\delta Z_{d\,ji}=\delta Z^L_{d\,ji}P_L+\delta Z^R_{d\,ji}P_R$, we arrive at the following contribution:
{\scriptsize\begin{multline}
-\imath\Sigma^{CT}_{d_jd_i}(p)=\imath\frac{p\!\!\!/}{2}\left[\left(\delta Z^L_{d\,ji}+\delta Z^{L\,*}_{d\,ij}\right)P_L+\left(\delta Z^R_{d\,ji}+\delta Z^{R\,*}_{d\,ij}\right)P_R\right]\\
\null\hspace{2cm}-\imath\left[\left(\delta m^L_{d\,ji}+\frac{1}{2}\left(m_{d_i}\delta Z^{R\,*}_{d\,ij}+m_{d_j}\delta Z^{L}_{d\,ji}\right)\right)P_L+\left(\delta m^R_{d\,ji}+\frac{1}{2}\left(m_{d_i}\delta Z^{L\,*}_{d\,ij}+m_{d_j}\delta Z^{R}_{d\,ji}\right)\right)P_R\right]
\end{multline}}
In principle, $\delta m^L_{d\,ji}=\left(\delta m^R_{d\,ij}\right)^*=\delta Y^L_{d\,ji}v_d+Y_d^i\delta_{ij}\delta v_d$.

\section{Sneutrino-Higgs self-energies}\label{appendix:scalarSE}
We assume that the tadpoles (Higgs, gauge bosons) vanish, which supposes certain relations at the loop-level between vevs and tree-level parameters. Then, defining 
the renormalized neutral-scalar self-energy matrix $\hat{\Sigma}^S_{\alpha\beta}$, we derive the following contribution to the EFT:
\begin{equation}
 {\cal L}_{\mbox{\tiny EFT}}\ni\frac{-1}{2m^2_{S_{\alpha}}m^2_{S_{\beta}}}\left[g_L^{S_{\alpha}d_jd_i}\hat{\Sigma}^S_{\alpha\beta}g_L^{S_{\beta}d_jd_i}\,O_2+g_R^{S_{\alpha}d_jd_i}\hat{\Sigma}^S_{\alpha\beta}g_R^{S_{\beta}d_jd_i}\,\tilde{O}_2+2g_L^{S_{\alpha}d_jd_i}\hat{\Sigma}^S_{\alpha\beta}g_R^{S_{\beta} d_jd_i}O_4\right]\,.
\end{equation}
The various contributions to the neutral-scalar self-energies are listed below.

\subsection{Scalar $A_0$-loop}
\begin{equation}
 -\imath\Sigma_{\alpha\beta}^{S\,A_S}=
 -\frac{\imath}{16\pi^2}g^{\tilde{S}\tilde{S}S_{\alpha}S_{\beta}}A_0(m_{\tilde{S}}) \,.
\end{equation}
This contribution is summed over the scalar $\tilde{S}$, taking value in the following list of particles:
\begin{itemize}
 \item scalar-ups: couplings from Eq.(\ref{coup:H02UU}). 3 colors contributing.
 \item sdowns: couplings from Eq.(\ref{coup:H02DD}). 3 colors contributing.
 \item Charged Higgs-sleptons: couplings from Eq.(\ref{coup:H02HcHc}).
 \item Higgs-sneutrinos: couplings from Eq.(\ref{coup:H04}); symmetry-factor $1/2$.
\end{itemize}

\subsection{Vector $A_0$-loop}
\begin{equation}
 -\imath\Sigma_{\alpha\beta}^{S\,A_V}=
 \frac{\imath}{16\pi^2}g^{VVS_{\alpha}S_{\beta}}D\,A_0(m_{V})
\end{equation}
The vector $V$ belongs to the following list of particles:
\begin{itemize}
 \item W's: couplings from Eq.(\ref{coup:H02WW}).
 \item Z's: couplings from Eq.(\ref{coup:H02ZZ}); symmetry-factor $1/2$.
\end{itemize}

\subsection{Scalar $B$-loop}
\begin{equation}
 -\imath\Sigma_{\alpha\beta}^{S\,B_S}=
 =\frac{\imath}{16\pi^2}g^{S_{\delta}S_{\gamma}S_{\alpha}}g^{S_{\gamma}S_{\delta}S_{\beta}}B_0(m_{S_{\gamma}},m_{S_{\delta}})
\end{equation}
The scalar pair $(S_{\gamma},S_{\delta})$ is summed over the particles:
\begin{itemize}
 \item scalar-ups: couplings from Eq.(\ref{coup:H0UU}). 3 colors contributing.
 \item sdowns: couplings from Eq.(\ref{coup:H0DD}). 3 colors contributing.
 \item Charged Higgs-sleptons: couplings from Eq.(\ref{coup:H0HcHc}).
 \item Higgs-sneutrinos: couplings from Eq.(\ref{coup:H03}).
\end{itemize}

\subsection{Fermion $B$-loop}
\begin{multline}
 -\imath\Sigma_{\alpha\beta}^{S\,B_f}
 =\frac{-2\imath}{16\pi^2}\left\{\left[g_L^{S_{\alpha}\tilde{f}f}g_L^{S_{\beta}\tilde{f}f\,*}+g_R^{S_{\alpha}\tilde{f}f}g_R^{S_{\beta}\tilde{f}f\,*}\right]DB_{22}\right.\\
\null\left.+\left[g_L^{S_{\alpha}\tilde{f}f}g_R^{S_{\beta}\tilde{f}f\,*}+g_R^{S_{\alpha}\tilde{f}f}g_L^{S_{\beta}\tilde{f}f\,*}\right]m_fm_{\tilde{f}}B_0\right\}(m_f,m_{\tilde{f}})\hspace{1.8cm}\null
\end{multline}
List of particles for the fermion pair $(f,\tilde{f})$:
\begin{itemize}
 \item ups: couplings of Eq.(\ref{coup:H0u}). 3 colors contributing.
 \item downs: couplings of Eq.(\ref{coup:H0d}). 3 colors contributing.
 \item charginos-leptons: couplings of Eq.(\ref{coup:H0cha}).
 \item neutrino-neutralinos: couplings of Eq.(\ref{coup:H0neu}); symmetry-factor $1/2$.
\end{itemize}

\subsection{Vector $B$-loop}
\begin{equation}
 -\imath\Sigma_{\alpha\beta}^{S\,B_V}
 =\frac{\imath}{16\pi^2}g^{S_{\alpha}VV}g^{S_{\beta}VV}DB_0(m_{V},m_{V})
\end{equation}
The vector $V$ is summed over:
\begin{itemize}
 \item W's: couplings of Eq.(\ref{coup:H0W}).
 \item Z's: couplings of Eq.(\ref{coup:H0Z}); symmetry-factor $1/2$
\end{itemize}

\subsection{Ghost $B$-loop}
\begin{equation}
 -\imath\Sigma_{\alpha\beta}^{S\,B_g} 
 =-\frac{\imath}{16\pi^2}g^{S_{\alpha}gg}g^{S_{\beta}gg}B_0(m_g,m_g)
\end{equation}
The contribution is summed over the ghost fields $g$:
\begin{itemize}
 \item $g_W$'s: couplings of Eq.(\ref{coup:H0gW}).
 \item $g_Z$: couplings of Eq.(\ref{coup:H0gZ}).
\end{itemize}

\subsection{Scalar/vector $B$-loop}
\begin{equation}
 -\imath\Sigma_{\alpha\beta}^{S\,B_{SV}} 
 =\frac{\imath}{16\pi^2}g^{S_{\alpha}VS\,*}g^{S_{\beta}VS}DB_{22}(m_V,m_S)
\end{equation}
List of particles for the scalar/vector pair $(S/V)$:
\begin{itemize}
 \item Charged Higgs-slepton / $W$: couplings of Eq.(\ref{coup:H0WHc}).
 \item Higgs - sneutrino / $Z$: couplings of Eq.(\ref{coup:H0ZH0}).
\end{itemize}

\subsection{Counterterms}
Defining the neutral scalar mass and wave-function counterterms $\delta m^2_{\alpha\beta}$ and $\delta Z^{S}_{\alpha\beta}$:
\begin{equation}
 -\imath\Sigma_{\alpha\beta}^{S\,CT}=-\imath\left[\delta m^2_{\alpha\beta}+\frac{1}{2}\delta Z^{S}_{\alpha\beta}\left(m^2_{S_{\alpha}}+m^2_{S_{\beta}}\right)\right]
\end{equation}

\section{Vertex corrections}\label{appendix:VertexCorrection}
The vertex corrections to the EFT are obtained as:
\begin{equation}
 {\cal L}_{\mbox{\tiny EFT}}\ni\frac{1}{2m^2_{S_{\alpha}}}\left[g_L^{S_{\alpha}d_jd_i}\hat{V}_L^{S_{\alpha}d_jd_i}\,O_2+g_R^{S_{\alpha}d_jd_i}\hat{V}_R^{S_{\alpha}d_jd_i}\,\tilde{O}_2+\left(g_R^{S_{\alpha}d_jd_i}\hat{V}_L^{S_{\alpha}d_jd_i}+g_L^{S_{\alpha}d_jd_i}\hat{V}_R^{S_{\alpha}d_jd_i}\right)O_4\right]
\end{equation}
where the $\bar{d}_jd_i$-neutral-Higgs renormalized vertex function $\hat{V}^{S_{\alpha}d_jd_i}=\hat{V}_L^{S_{\alpha}d_jd_i}P_L+\hat{V}_L^{S_{\alpha}d_jd_i}P_R$ receives the contributions listed below.

\subsection{Scalar/fermion loop with cubic scalar coupling}
{\scriptsize\begin{multline}
-\imath\hat{V}^{S_{\alpha}d_jd_i}[Sff,S^3] 
=-\frac{\imath}{16\pi^2}g^{S_{\alpha}S_{k}S_{l}}\left[g_R^{S_{l}fd_j\,*}g_L^{S_{k}fd_i}P_L+g_L^{S_{l}fd_j\,*}g_R^{S_{k}fd_i}P_R\right]m_f C_0(m_f,m_{S_{k}},m_{S_{l}})
\end{multline}}
List of particles for the scalar/fermion triplet $(S_k,S_l/f)$:
\begin{itemize}
 \item Higgs-sneutrino/down: couplings from Eqs.(\ref{coup:H0d}),(\ref{coup:H03}).
 \item Charged Higgs-slepton/up: couplings from Eqs.(\ref{coup:Hcud}),(\ref{coup:H0HcHc}).
 \item sdown/neutralino-neutrino: couplings from Eqs.(\ref{coup:Dneud}),(\ref{coup:H0DD}).
 \item sdown/gluino: couplings from Eqs.(\ref{coup:Dglud}),(\ref{coup:H0DD}); color-factor $C_2(3)=4/3$.
 \item sup/chargino-lepton: couplings from Eqs.(\ref{coup:Uchad}),(\ref{coup:H0UU}).
 \item sup/down: couplings from Eqs.(\ref{coup:Udd}),(\ref{coup:H0UU}).
 \item sdown/up: couplings from Eqs.(\ref{coup:Dud}),(\ref{coup:H0DD}).
\end{itemize}

\subsection{Scalar/fermion loop without cubic scalar coupling}
{\scriptsize\begin{multline}
-\imath\hat{V}^{S_{\alpha}d_jd_i}[Sff]=-\frac{\imath}{16\pi^2}\left\{\left[g_R^{Sf_{l}d_j\,*}g_R^{S_{\alpha}f_{l}f_{k}}g_L^{Sf_{k}d_i}P_L+g_L^{Sf_{l}d_j\,*}g_L^{S_{\alpha}f_{l}f_{k}}g_R^{Sf_{k}d_i}P_R\right]D\,C_{24}\right.\\
+\left.\left[g_R^{Sf_{l}d_j\,*}g_L^{S_{\alpha}f_{l}f_{k}}g_L^{Sf_{k}d_i}P_L+g_L^{Sf_{l}d_j\,*}g_R^{S_{\alpha}f_{l}f_{k}}g_R^{Sf_{k}d_i}P_R\right]m_{f_{k}}m_{f_{l}} C_0\right\}(m_S,m_{f_{k}},m_{f_{l}})
\end{multline}}
List of particles for the scalar/fermion triplet $(S/f_k,f_l)$:
\begin{itemize}
 \item Higgs-sneutrino/down: couplings from Eq.(\ref{coup:H0d}).
 \item Charged Higgs-slepton/up: couplings from Eqs.(\ref{coup:Hcud}),(\ref{coup:H0u}).
 \item sdown/neutralino-neutrino: couplings from Eqs.(\ref{coup:Dneud}),(\ref{coup:H0neu}).
 \item sup/chargino-lepton: couplings from Eqs.(\ref{coup:Uchad}),(\ref{coup:H0cha}).
 \item sup/down: couplings from Eqs.(\ref{coup:Udd}),(\ref{coup:H0d}).
 \item sdown/up: couplings from Eqs.(\ref{coup:Dud}),(\ref{coup:H0u}).
\end{itemize}

\subsection{Vector/fermion loop with scalar-vector coupling}
{\scriptsize\begin{multline}
-\imath\hat{V}^{S_{\alpha}d_jd_i}[SVV,Vff]=-\frac{\imath}{16\pi^2}g^{S_{\alpha}V_{k}V_{l}}\left[g_R^{V_{l}fd_j\,*}g_L^{V_{k}fd_i}P_L+g_L^{V_{l}fd_j\,*}g_R^{V_{k}fd_i}P_R\right]Dm_f\,C_0(m_f,m_{V_{k}},m_{V_{l}})
\end{multline}}
The vector/fermion triplet $(V_k,V_l/f)$ takes the following values:
\begin{itemize}
 \item $W$/up: couplings from Eqs.(\ref{coup:Wud}),(\ref{coup:H0W}).
 \item $Z$/down: couplings from Eqs.(\ref{coup:Zdd}),(\ref{coup:H0Z}).
\end{itemize}

\subsection{Vector/fermion loop with scalar-fermion coupling}
{\scriptsize\begin{multline}
-\imath\hat{V}^{S_{\alpha}d_jd_i}[SVV,Sff]=\frac{\imath}{16\pi^2}\left\{\left[g_R^{Vf_{l}d_j\,*}g_L^{S_{\alpha}f_{l}f_{k}}g_L^{Vf_{k}d_i}P_L+g_L^{Vf_{l}d_j\,*}g_R^{S_{\alpha}f_{l}f_{k}}g_R^{Vf_{k}d_i}P_R\right]D^2\,C_{24}\right.\\
+\left.\left[g_R^{Vf_{l}d_j\,*}g_R^{S_{\alpha}f_{l}f_{k}}g_L^{Vf_{k}d_i}P_L+g_L^{Vf_{l}d_j\,*}g_L^{S_{\alpha}f_{l}f_{k}}g_R^{Vf_{k}d_i}P_R\right]Dm_{f_{k}}m_{f_{l}} C_0\right\}(m_V,m_{f_{k}},m_{f_{l}})
\end{multline}}
The vector/fermion triplet $(V/f_k,f_l)$ takes the following values:
\begin{itemize}
 \item $W$/up: couplings from Eqs.(\ref{coup:Wud}),(\ref{coup:H0u}).
 \item $Z$/down: couplings from Eqs.(\ref{coup:Zdd}),(\ref{coup:H0d}).
\end{itemize}

\subsection{Vector/Scalar/fermion loops}
{\scriptsize\begin{multline}
-\imath\hat{V}^{S_{\alpha}d_jd_i}[VSf]=-\frac{\imath}{16\pi^2}\left\{g^{VSS_{\alpha}}\left[g_R^{Sfd_j\,*}g_L^{Vfd_i}P_L+g_L^{Sfd_j\,*}g_R^{Vfd_i}P_R\right]+g^{SVS_{\alpha}}\left[g_R^{Vfd_j\,*}g_L^{Sfd_i}P_L+g_L^{Vfd_j\,*}g_R^{Sfd_i}P_R\right]\right\}\\ \times D\,C_{24}(m_f,m_S,m_V)
\end{multline}}
List of particles for the scalar/vector/fermion triplet $(S/V/f)$:
\begin{itemize}
 \item charged-Higgs-slepton/$W$/up: couplings from Eqs.(\ref{coup:Wud}),(\ref{coup:Hcud}),(\ref{coup:H0WHc}).
 \item neutral-Higgs-sneutrino/$Z$/down: couplings from Eqs.(\ref{coup:Zdd}),(\ref{coup:H0d}),(\ref{coup:H0ZH0}).
\end{itemize}

\subsection{Counterterms}
The counterterm contribution $-\imath\hat{V}^{S_{\alpha}d_jd_i}[CT]$ reads:
{\scriptsize\begin{multline}
\imath\left\{-\frac{1}{\sqrt{2}}\left[\delta Y_{d\,ji}^L(X^R_{k d}+\imath X^I_{k d})+\delta\lambda'^L_{fij}(X^R_{k \tilde{N}_f}+\imath X^I_{k \tilde{N}_f})\right]+\frac{1}{2}\left[\delta Z^{R\,*}_{d\, jl}g_L^{S_{\alpha}d_ld_i}+\delta Z^L_{d\, il}g_L^{S_{\alpha}d_jd_l}+\delta Z^S_{k\alpha}g_L^{S_{\alpha}d_jd_i}\right]\right\}P_L     \\
+\imath\left\{-\frac{1}{\sqrt{2}}\left[\delta Y_{d\,ji}^R(X^R_{k d}-\imath X^I_{k d})+\delta\lambda'^R_{fji}(X^R_{k \tilde{N}_f}-\imath X^I_{k \tilde{N}_f})\right]+\frac{1}{2}\left[\delta Z^{L\,*}_{d\, jl}g_R^{S_{\alpha}d_ld_i}+\delta Z^R_{d\, il}g_R^{S_{\alpha}d_jd_l}+\delta Z^S_{k\alpha}g_R^{S_{\alpha}d_jd_i}\right]\right\}P_R
\end{multline}}
where $\delta Y_{d\,ji}^R=\left(\delta Y_{d\,ij}^L\right)^*$ is the counterterm to the Yukawa coupling and $\delta\lambda'^R_{fji}=\left(\delta\lambda'^L_{fji}\right)^*$
is the counterterm to the $\lambda'$ coupling.

\section{Box diagrams}\label{Appendix:Box_diagrams}
Here, we collect the box-diagram contributions to the $d_i\bar{d}_j\to d_j\bar{d}_i$ amplitude. The results are listed according to the topologies of Fig.\ref{fig:topologies_boxes}.
\subsection{Vector/fermion/vector/fermion ``straight'' box}\label{subsec:vfvf-stright}

\paragraph{Case $V_{\alpha,\beta}$ colour-singlets}
{\scriptsize\begin{multline}
{\cal L}_{\mbox{\tiny EFT}}\ni\frac{1}{32\pi^2}\left\{g_L^{V_{\alpha}f_kd_j\,*}g_L^{V_{\beta}f_kd_i}g_L^{V_{\beta}f_ld_j\,*}g_L^{V_{\alpha}f_ld_i}\,D_2\,O_1+g_R^{V_{\alpha}f_kd_j\,*}g_R^{V_{\beta}f_kd_i}g_R^{V_{\beta}f_ld_j\,*}g_R^{V_{\alpha}f_ld_i}\,D_2\,\tilde{O}_1\right.\\
+16g_R^{V_{\alpha}f_kd_j\,*}g_L^{V_{\beta}f_kd_i}g_R^{V_{\beta}f_ld_j\,*}g_L^{V_{\alpha}f_ld_i}m_{f_k}m_{f_l}D_0\,O_2+16g_L^{V_{\alpha}f_kd_j\,*}g_R^{V_{\beta}f_kd_i}g_L^{V_{\beta}f_ld_j\,*}g_R^{V_{\alpha}f_ld_i}m_{f_k}m_{f_l}D_0\,\tilde{O}_2\\
+16\left[g_R^{V_{\alpha}f_kd_j\,*}g_L^{V_{\beta}f_kd_i}g_L^{V_{\beta}f_ld_j\,*}g_R^{V_{\alpha}f_ld_i}+g_L^{V_{\alpha}f_kd_j\,*}g_R^{V_{\beta}f_kd_i}g_R^{V_{\beta}f_ld_j\,*}g_L^{V_{\alpha}f_ld_i}\right]m_{f_k}m_{f_l}D_0\,O_4\\
\left.-2\left[g_L^{V_{\alpha}f_kd_j\,*}g_L^{V_{\beta}f_kd_i}g_R^{V_{\beta}f_ld_j\,*}g_R^{V_{\alpha}f_ld_i}+g_R^{V_{\alpha}f_kd_j\,*}g_R^{V_{\beta}f_kd_i}g_L^{V_{\beta}f_ld_j\,*}g_L^{V_{\alpha}f_ld_i}\right]\,D_2\,O_5\right\}(m_{S_{\alpha}},m_{f_k},m_{S_{\beta}},m_{f_l})
\end{multline}}
List of particles:
\begin{itemize}
 \item W / up: couplings from Eq.(\ref{coup:Wud}).
\end{itemize}

\subsection{Scalar/fermion/scalar/fermion ``straight'' box}\label{subsec:sfsf-stright}

\paragraph{Case 1: $S_{\alpha,\beta}$ colour-singlets}
{\scriptsize\begin{multline}
{\cal L}_{\mbox{\tiny EFT}}\ni\frac{1}{32\pi^2}\left\{g_L^{S_{\alpha}f_kd_j\,*}g_L^{S_{\beta}f_kd_i}g_L^{S_{\beta}f_ld_j\,*}g_L^{S_{\alpha}f_ld_i}\frac{D_2}{4}\,O_1+g_R^{S_{\alpha}f_kd_j\,*}g_R^{S_{\beta}f_kd_i}g_R^{S_{\beta}f_ld_j\,*}g_R^{S_{\alpha}f_ld_i}\frac{D_2}{4}\,\tilde{O}_1\right.\\
+g_R^{S_{\alpha}f_kd_j\,*}g_L^{S_{\beta}f_kd_i}g_R^{S_{\beta}f_ld_j\,*}g_L^{S_{\alpha}f_ld_i}m_{f_k}m_{f_l}D_0\,O_2+g_L^{S_{\alpha}f_kd_j\,*}g_R^{S_{\beta}f_kd_i}g_L^{S_{\beta}f_ld_j\,*}g_R^{S_{\alpha}f_ld_i}m_{f_k}m_{f_l}D_0\,\tilde{O}_2\\
+\left[g_R^{S_{\alpha}f_kd_j\,*}g_L^{S_{\beta}f_kd_i}g_L^{S_{\beta}f_ld_j\,*}g_R^{S_{\alpha}f_ld_i}+g_L^{S_{\alpha}f_kd_j\,*}g_R^{S_{\beta}f_kd_i}g_R^{S_{\beta}f_ld_j\,*}g_L^{S_{\alpha}f_ld_i}\right]m_{f_k}m_{f_l}D_0\,O_4\\
\left.-\left[g_L^{S_{\alpha}f_kd_j\,*}g_L^{S_{\beta}f_kd_i}g_R^{S_{\beta}f_ld_j\,*}g_R^{S_{\alpha}f_ld_i}+g_R^{S_{\alpha}f_kd_j\,*}g_R^{S_{\beta}f_kd_i}g_L^{S_{\beta}f_ld_j\,*}g_L^{S_{\alpha}f_ld_i}\right]\frac{D_2}{2}\,O_5\right\}(m_{S_{\alpha}},m_{f_k},m_{S_{\beta}},m_{f_l})
\end{multline}}
List of particles:
\begin{itemize}
 \item Higgs-sneutrino / down: couplings from Eq.(\ref{coup:H0d}).
 \item Charged Higgs-slepton / up: couplings from Eq.(\ref{coup:Hcud}).
\end{itemize}

\paragraph{Case 2: $f_{k,l}$ colour-singlets}
{\scriptsize\begin{multline}
{\cal L}_{\mbox{\tiny EFT}}\ni\frac{1}{32\pi^2}\left\{g_L^{S_{\alpha}f_kd_j\,*}g_L^{S_{\beta}f_kd_i}g_L^{S_{\beta}f_ld_j\,*}g_L^{S_{\alpha}f_ld_i}\frac{D_2}{4}\,O_1+g_R^{S_{\alpha}f_kd_j\,*}g_R^{S_{\beta}f_kd_i}g_R^{S_{\beta}f_ld_j\,*}g_R^{S_{\alpha}f_ld_i}\frac{D_2}{4}\,\tilde{O}_1\right.\\
+g_R^{S_{\alpha}f_kd_j\,*}g_L^{S_{\beta}f_kd_i}g_R^{S_{\beta}f_ld_j\,*}g_L^{S_{\alpha}f_ld_i}m_{f_k}m_{f_l}D_0\,O_3+g_L^{S_{\alpha}f_kd_j\,*}g_R^{S_{\beta}f_kd_i}g_L^{S_{\beta}f_ld_j\,*}g_R^{S_{\alpha}f_ld_i}m_{f_k}m_{f_l}D_0\,\tilde{O}_3\\
-\left[g_L^{S_{\alpha}f_kd_j\,*}g_L^{S_{\beta}f_kd_i}g_R^{S_{\beta}f_ld_j\,*}g_R^{S_{\alpha}f_ld_i}+g_R^{S_{\alpha}f_kd_j\,*}g_R^{S_{\beta}f_kd_i}g_L^{S_{\beta}f_ld_j\,*}g_L^{S_{\alpha}f_ld_i}\right]\frac{D_2}{2}\,O_4\\
\left.+\left[g_R^{S_{\alpha}f_kd_j\,*}g_L^{S_{\beta}f_kd_i}g_L^{S_{\beta}f_ld_j\,*}g_R^{S_{\alpha}f_ld_i}+g_L^{S_{\alpha}f_kd_j\,*}g_R^{S_{\beta}f_kd_i}g_R^{S_{\beta}f_ld_j\,*}g_L^{S_{\alpha}f_ld_i}\right]m_{f_k}m_{f_l}D_0\,O_5\right\}(m_{S_{\alpha}},m_{f_k},m_{S_{\beta}},m_{f_l})
\end{multline}}
List of particles:
\begin{itemize}
 \item sdown / neutrino-neutralino: couplings from Eq.(\ref{coup:Dneud}).
 \item sup / chargino-lepton: couplings from Eq.(\ref{coup:Uchad}).
\end{itemize}

\paragraph{Case 3: all fields colour-triplets}
{\scriptsize\begin{multline}
{\cal L}_{\mbox{\tiny EFT}}\ni\frac{1}{32\pi^2}\left\{g_L^{S_{\alpha}f_kd_j\,*}g_L^{S_{\beta}f_kd_i}g_L^{S_{\beta}f_ld_j\,*}g_L^{S_{\alpha}f_ld_i}\frac{D_2}{2}\,O_1+g_R^{S_{\alpha}f_kd_j\,*}g_R^{S_{\beta}f_kd_i}g_R^{S_{\beta}f_ld_j\,*}g_R^{S_{\alpha}f_ld_i}\frac{D_2}{2}\,\tilde{O}_1\right.\\
+g_R^{S_{\alpha}f_kd_j\,*}g_L^{S_{\beta}f_kd_i}g_R^{S_{\beta}f_ld_j\,*}g_L^{S_{\alpha}f_ld_i}m_{f_k}m_{f_l}D_0(O_2+O_3)+g_L^{S_{\alpha}f_kd_j\,*}g_R^{S_{\beta}f_kd_i}g_L^{S_{\beta}f_ld_j\,*}g_R^{S_{\alpha}f_ld_i}m_{f_k}m_{f_l}D_0(\tilde{O}_2+\tilde{O}_3)\\
+(O_4+O_5)\left(\left[g_R^{S_{\alpha}f_kd_j\,*}g_L^{S_{\beta}f_kd_i}g_L^{S_{\beta}f_ld_j\,*}g_R^{S_{\alpha}f_ld_i}+g_L^{S_{\alpha}f_kd_j\,*}g_R^{S_{\beta}f_kd_i}g_R^{S_{\beta}f_ld_j\,*}g_L^{S_{\alpha}f_ld_i}\right]m_{f_k}m_{f_l}D_0\right.\\
\left.\left.-\left[g_L^{S_{\alpha}f_kd_j\,*}g_L^{S_{\beta}f_kd_i}g_R^{S_{\beta}f_ld_j\,*}g_R^{S_{\alpha}f_ld_i}+g_R^{S_{\alpha}f_kd_j\,*}g_R^{S_{\beta}f_kd_i}g_L^{S_{\beta}f_ld_j\,*}g_L^{S_{\alpha}f_ld_i}\right]\frac{D_2}{2}\right)\right\}(m_{S_{\alpha}},m_{f_k},m_{S_{\beta}},m_{f_l})
\end{multline}}
List of particles:
\begin{itemize}
 \item sdown / up: couplings from Eq.(\ref{coup:Dud}).
 \item sup / down: couplings from Eq.(\ref{coup:Udd}).
\end{itemize}

\paragraph{Case 4: $f_{k,l}$ colour-octets}
{\scriptsize\begin{multline}
{\cal L}_{\mbox{\tiny EFT}}\ni\frac{1}{32\pi^2}\left\{\frac{11}{18}g_L^{S_{\alpha}f_kd_j\,*}g_L^{S_{\beta}f_kd_i}g_L^{S_{\beta}f_ld_j\,*}g_L^{S_{\alpha}f_ld_i}\frac{D_2}{4}\,O_1+\frac{11}{18}g_R^{S_{\alpha}f_kd_j\,*}g_R^{S_{\beta}f_kd_i}g_R^{S_{\beta}f_ld_j\,*}g_R^{S_{\alpha}f_ld_i}\frac{D_2}{4}\,\tilde{O}_1\right.\\
+g_R^{S_{\alpha}f_kd_j\,*}g_L^{S_{\beta}f_kd_i}g_R^{S_{\beta}f_ld_j\,*}g_L^{S_{\alpha}f_ld_i}m_{f_k}m_{f_l}D_0\left(\frac{7}{12}O_2+\frac{1}{36}O_3\right)+g_L^{S_{\alpha}f_kd_j\,*}g_R^{S_{\beta}f_kd_i}g_L^{S_{\beta}f_ld_j\,*}g_R^{S_{\alpha}f_ld_i}m_{f_k}m_{f_l}D_0\left(\frac{7}{12}\tilde{O}_2+\frac{1}{36}\tilde{O}_3\right)\\
+\left[g_R^{S_{\alpha}f_kd_j\,*}g_L^{S_{\beta}f_kd_i}g_L^{S_{\beta}f_ld_j\,*}g_R^{S_{\alpha}f_ld_i}+g_L^{S_{\alpha}f_kd_j\,*}g_R^{S_{\beta}f_kd_i}g_R^{S_{\beta}f_ld_j\,*}g_L^{S_{\alpha}f_ld_i}\right]m_{f_k}m_{f_l}D_0\left(\frac{7}{12}O_4+\frac{1}{36}O_5\right)\\
\left.-\left[g_L^{S_{\alpha}f_kd_j\,*}g_L^{S_{\beta}f_kd_i}g_R^{S_{\beta}f_ld_j\,*}g_R^{S_{\alpha}f_ld_i}+g_R^{S_{\alpha}f_kd_j\,*}g_R^{S_{\beta}f_kd_i}g_L^{S_{\beta}f_ld_j\,*}g_L^{S_{\alpha}f_ld_i}\right]\frac{D_2}{4}\left(\frac{1}{18}O_4+\frac{7}{6}O_5\right)\right\}(m_{S_{\alpha}},m_{f_k},m_{S_{\beta}},m_{f_l})
\end{multline}}
List of particles:
\begin{itemize}
 \item sdown / gluino: couplings from Eq.(\ref{coup:Dglud}) (stripped from Gell-Mann matrix element).
\end{itemize}

\paragraph{Case 5: $f_{k,l}$ colour-octet+singlet}
{\scriptsize\begin{multline}
{\cal L}_{\mbox{\tiny EFT}}\ni\frac{1}{32\pi^2}\left\{\frac{1}{3}g_L^{S_{\alpha}f_kd_j\,*}g_L^{S_{\beta}f_kd_i}g_L^{S_{\beta}f_ld_j\,*}g_L^{S_{\alpha}f_ld_i}\frac{D_2}{4}\,O_1+\frac{1}{3}g_R^{S_{\alpha}f_kd_j\,*}g_R^{S_{\beta}f_kd_i}g_R^{S_{\beta}f_ld_j\,*}g_R^{S_{\alpha}f_ld_i}\frac{D_2}{4}\,\tilde{O}_1\right.\\
+g_R^{S_{\alpha}f_kd_j\,*}g_L^{S_{\beta}f_kd_i}g_R^{S_{\beta}f_ld_j\,*}g_L^{S_{\alpha}f_ld_i}m_{f_k}m_{f_l}D_0\frac{1}{2}\left(O_2-\frac{1}{3}O_3\right)+g_L^{S_{\alpha}f_kd_j\,*}g_R^{S_{\beta}f_kd_i}g_L^{S_{\beta}f_ld_j\,*}g_R^{S_{\alpha}f_ld_i}m_{f_k}m_{f_l}D_0\frac{1}{2}\left(\tilde{O}_2-\frac{1}{3}\tilde{O}_3\right)\\
+\left[g_R^{S_{\alpha}f_kd_j\,*}g_L^{S_{\beta}f_kd_i}g_L^{S_{\beta}f_ld_j\,*}g_R^{S_{\alpha}f_ld_i}+g_L^{S_{\alpha}f_kd_j\,*}g_R^{S_{\beta}f_kd_i}g_R^{S_{\beta}f_ld_j\,*}g_L^{S_{\alpha}f_ld_i}\right]m_{f_k}m_{f_l}D_0\left(O_4-\frac{1}{3}O_5\right)\\
\left.+\left[g_L^{S_{\alpha}f_kd_j\,*}g_L^{S_{\beta}f_kd_i}g_R^{S_{\beta}f_ld_j\,*}g_R^{S_{\alpha}f_ld_i}+g_R^{S_{\alpha}f_kd_j\,*}g_R^{S_{\beta}f_kd_i}g_L^{S_{\beta}f_ld_j\,*}g_L^{S_{\alpha}f_ld_i}\right]\frac{D_2}{4}\left(\frac{1}{3}O_4-O_5\right)\right\}(m_{S_{\alpha}},m_{f_k},m_{S_{\beta}},m_{f_l})
\end{multline}}
List of particles:
\begin{itemize}
 \item sdown / gluino / sdown / neutralino-neutrino: couplings from Eqs.(\ref{coup:Dneud}),(\ref{coup:Dglud}) (stripped from Gell-Mann matrix element); $\times 2$ ($\pi$-rotated diagram).
\end{itemize}

\subsection{Scalar/fermion/scalar/fermion ``scalar-cross'' box}\label{subsec:sfsf-scross}

\paragraph{Case 1: $S_{\alpha,\beta}$ colour-singlets}
{\scriptsize\begin{multline}
{\cal L}_{\mbox{\tiny EFT}}\ni\frac{1}{32\pi^2}\left\{-g_L^{S_{\alpha}f_kd_j\,*}g_L^{S_{\beta}f_kd_i}g_L^{S_{\alpha}f_ld_j\,*}g_L^{S_{\beta}f_ld_i}\frac{D_2}{4}\,O_1-g_R^{S_{\alpha}f_kd_j\,*}g_R^{S_{\beta}f_kd_i}g_R^{S_{\alpha}f_ld_j\,*}g_R^{S_{\beta}f_ld_i}\frac{D_2}{4}\,\tilde{O}_1\right.\\
+g_R^{S_{\alpha}f_kd_j\,*}g_L^{S_{\beta}f_kd_i}g_R^{S_{\alpha}f_ld_j\,*}g_L^{S_{\beta}f_ld_i}m_{f_k}m_{f_l}D_0\,O_2+g_L^{S_{\alpha}f_kd_j\,*}g_R^{S_{\beta}f_kd_i}g_L^{S_{\alpha}f_ld_j\,*}g_R^{S_{\beta}f_ld_i}m_{f_k}m_{f_l}D_0\,\tilde{O}_2\\
+\left[g_R^{S_{\alpha}f_kd_j\,*}g_L^{S_{\beta}f_kd_i}g_L^{S_{\alpha}f_ld_j\,*}g_R^{S_{\beta}f_ld_i}+g_L^{S_{\alpha}f_kd_j\,*}g_R^{S_{\beta}f_kd_i}g_R^{S_{\alpha}f_ld_j\,*}g_L^{S_{\beta}f_ld_i}\right]m_{f_k}m_{f_l}D_0\,O_4\\
\left.+\left[g_L^{S_{\alpha}f_kd_j\,*}g_L^{S_{\beta}f_kd_i}g_R^{S_{\alpha}f_ld_j\,*}g_R^{S_{\beta}f_ld_i}+g_R^{S_{\alpha}f_kd_j\,*}g_R^{S_{\beta}f_kd_i}g_L^{S_{\alpha}f_ld_j\,*}g_L^{S_{\beta}f_ld_i}\right]\frac{D_2}{2}\,O_5\right\}(m_{S_{\alpha}},m_{f_k},m_{S_{\beta}},m_{f_l})
\end{multline}}
List of particles:
\begin{itemize}
 \item Higgs-sneutrino / down: couplings from Eq.(\ref{coup:H0d}).
\end{itemize}

\paragraph{Case 2: $f_k$ colour-singlet}
{\scriptsize\begin{multline}
{\cal L}_{\mbox{\tiny EFT}}\ni\frac{1}{32\pi^2}\left\{g_R^{S_{\alpha}f_kd_j\,*}g_L^{S_{\beta}f_kd_i}g_R^{S_{\alpha}f_ld_j\,*}g_L^{S_{\beta}f_ld_i}m_{f_k}m_{f_l}D_0(O_2-O_3)+g_L^{S_{\alpha}f_kd_j\,*}g_R^{S_{\beta}f_kd_i}g_L^{S_{\alpha}f_ld_j\,*}g_R^{S_{\beta}f_ld_i}m_{f_k}m_{f_l}D_0(\tilde{O}_2-\tilde{O}_3)\right.\\
+(O_4-O_5)\left(\left[g_R^{S_{\alpha}f_kd_j\,*}g_L^{S_{\beta}f_kd_i}g_L^{S_{\alpha}f_ld_j\,*}g_R^{S_{\beta}f_ld_i}+g_L^{S_{\alpha}f_kd_j\,*}g_R^{S_{\beta}f_kd_i}g_R^{S_{\alpha}f_ld_j\,*}g_L^{S_{\beta}f_ld_i}\right]m_{f_k}m_{f_l}D_0\right.\\
\left.\left.-\left[g_L^{S_{\alpha}f_kd_j\,*}g_L^{S_{\beta}f_kd_i}g_R^{S_{\alpha}f_ld_j\,*}g_R^{S_{\beta}f_ld_i}+g_R^{S_{\alpha}f_kd_j\,*}g_R^{S_{\beta}f_kd_i}g_L^{S_{\alpha}f_ld_j\,*}g_L^{S_{\beta}f_ld_i}\right]\frac{D_2}{2}\right)\right\}(m_{S_{\alpha}},m_{f_k},m_{S_{\beta}},m_{f_l})
\end{multline}}
List of particles:
\begin{itemize}
 \item sup / chargino-lepton / sup / down: couplings from Eqs.(\ref{coup:Uchad}),(\ref{coup:Udd}).
 \item sdown / neutralino-neutrino / sdown / up: couplings from Eqs.(\ref{coup:Dneud}),(\ref{coup:Dud}).
\end{itemize}

\paragraph{Case 3: $f_k$ colour-triplet}
{\scriptsize\begin{multline}
{\cal L}_{\mbox{\tiny EFT}}\ni\frac{1}{32\pi^2}\left\{-g_L^{S_{\alpha}f_kd_j\,*}g_L^{S_{\beta}f_kd_i}g_L^{S_{\alpha}f_ld_j\,*}g_L^{S_{\beta}f_ld_i}\frac{D_2}{4}\,O_1-g_R^{S_{\alpha}f_kd_j\,*}g_R^{S_{\beta}f_kd_i}g_R^{S_{\alpha}f_ld_j\,*}g_R^{S_{\beta}f_ld_i}\frac{D_2}{4}\,\tilde{O}_1\right.\\
+g_R^{S_{\alpha}f_kd_j\,*}g_L^{S_{\beta}f_kd_i}g_R^{S_{\alpha}f_ld_j\,*}g_L^{S_{\beta}f_ld_i}m_{f_k}m_{f_l}D_0\frac{1}{6}(5O_2+O_3)+g_L^{S_{\alpha}f_kd_j\,*}g_R^{S_{\beta}f_kd_i}g_L^{S_{\alpha}f_ld_j\,*}g_R^{S_{\beta}f_ld_i}m_{f_k}m_{f_l}D_0\frac{1}{6}(5\tilde{O}_2+\tilde{O}_3)\\
+\left[g_R^{S_{\alpha}f_kd_j\,*}g_L^{S_{\beta}f_kd_i}g_L^{S_{\alpha}f_ld_j\,*}g_R^{S_{\beta}f_ld_i}+g_L^{S_{\alpha}f_kd_j\,*}g_R^{S_{\beta}f_kd_i}g_R^{S_{\alpha}f_ld_j\,*}g_L^{S_{\beta}f_ld_i}\right]m_{f_k}m_{f_l}D_0\frac{1}{6}(5O_4+O_5)\\
\left.+\left[g_L^{S_{\alpha}f_kd_j\,*}g_L^{S_{\beta}f_kd_i}g_R^{S_{\alpha}f_ld_j\,*}g_R^{S_{\beta}f_ld_i}+g_R^{S_{\alpha}f_kd_j\,*}g_R^{S_{\beta}f_kd_i}g_L^{S_{\alpha}f_ld_j\,*}g_L^{S_{\beta}f_ld_i}\right]\frac{D_2}{4}\frac{1}{3}(O_4+5O_5)\right\}(m_{S_{\alpha}},m_{f_k},m_{S_{\beta}},m_{f_l})
\end{multline}}
List of particles:
\begin{itemize}
 \item sdown / gluino / sdown / up: couplings from Eqs.(\ref{coup:Dud}),(\ref{coup:Dglud}) (stripped from Gell-Mann matrix element); $\times2$ ($\pi$-rotated diagram).
\end{itemize}

\subsection{Scalar/fermion/scalar/fermion ``fermion-cross'' box}\label{subsec:sfsf-fcross}

\paragraph{Case 1: $f_k$ colour-singlet}

{\scriptsize\begin{multline}
{\cal L}_{\mbox{\tiny EFT}}\ni\frac{1}{32\pi^2}\left\{g_L^{S_{\alpha}f_kd_j\,*}g_L^{S_{\beta}f_kd_j\,*}g_L^{S_{\alpha}f_ld_i}g_L^{S_{\beta}f_ld_i}\frac{m_{f_k}m_{f_l}}{2}D_0\,O_1+g_R^{S_{\alpha}f_kd_j\,*}g_R^{S_{\beta}f_kd_j\,*}g_R^{S_{\alpha}f_ld_i}g_R^{S_{\beta}f_ld_i}\frac{m_{f_k}m_{f_l}}{2}D_0\,\tilde{O}_1\right.\\
-g_R^{S_{\alpha}f_kd_j\,*}g_R^{S_{\beta}f_kd_j\,*}g_L^{S_{\alpha}f_ld_i}g_L^{S_{\beta}f_ld_i}m_{f_k}m_{f_l}D_0(O_2+O_3)-g_L^{S_{\alpha}f_kd_j\,*}g_L^{S_{\beta}f_kd_j\,*}g_R^{S_{\alpha}f_ld_i}g_R^{S_{\beta}f_ld_i}m_{f_k}m_{f_l}D_0(\tilde{O}_2+\tilde{O}_3)\\
-\left[g_L^{S_{\alpha}f_kd_j\,*}g_R^{S_{\beta}f_kd_j\,*}g_R^{S_{\alpha}f_ld_i}g_L^{S_{\beta}f_ld_i}+g_R^{S_{\alpha}f_kd_j\,*}g_L^{S_{\beta}f_kd_j\,*}g_L^{S_{\alpha}f_ld_i}g_R^{S_{\beta}f_ld_i}\right]\frac{D_2}{2}\,O_4\\
\left.+\left[g_R^{S_{\alpha}f_kd_j\,*}g_L^{S_{\beta}f_kd_j\,*}g_R^{S_{\alpha}f_ld_i}g_L^{S_{\beta}f_ld_i}+g_L^{S_{\alpha}f_kd_j\,*}g_R^{S_{\beta}f_kd_j\,*}g_L^{S_{\alpha}f_ld_i}g_R^{S_{\beta}f_ld_i}\right]\frac{D_2}{2}\,O_5\right\}(m_{S_{\alpha}},m_{f_k},m_{S_{\beta}},m_{f_l})
\end{multline}}
List of particles:
\begin{itemize}
 \item sdown / neutrino-neutralino: couplings from Eq.(\ref{coup:Dneud}).
\end{itemize}

\paragraph{Case 2: $S_{\alpha}$ colour-singlet}

{\scriptsize\begin{multline}
{\cal L}_{\mbox{\tiny EFT}}\ni\frac{1}{32\pi^2}\left\{
-g_R^{S_{\alpha}f_kd_j\,*}g_R^{S_{\beta}f_kd_j\,*}g_L^{S_{\alpha}f_ld_i}g_L^{S_{\beta}f_ld_i}m_{f_k}m_{f_l}D_0\,O_3-g_L^{S_{\alpha}f_kd_j\,*}g_L^{S_{\beta}f_kd_j\,*}g_R^{S_{\alpha}f_ld_i}g_R^{S_{\beta}f_ld_i}m_{f_k}m_{f_l}D_0\,\tilde{O}_3\right.\\
-(O_4-O_5)\left(\left[g_R^{S_{\alpha}f_kd_j\,*}g_L^{S_{\beta}f_kd_j\,*}g_R^{S_{\alpha}f_ld_i}g_L^{S_{\beta}f_ld_i}+g_L^{S_{\alpha}f_kd_j\,*}g_R^{S_{\beta}f_kd_j\,*}g_L^{S_{\alpha}f_ld_i}g_R^{S_{\beta}f_ld_i}\right]\right.\\
\left.\left.+\left[g_L^{S_{\alpha}f_kd_j\,*}g_R^{S_{\beta}f_kd_j\,*}g_R^{S_{\alpha}f_ld_i}g_L^{S_{\beta}f_ld_i}+g_R^{S_{\alpha}f_kd_j\,*}g_L^{S_{\beta}f_kd_j\,*}g_L^{S_{\alpha}f_ld_i}g_R^{S_{\beta}f_ld_i}\right]\right)\frac{D_2}{2}\right\}(m_{S_{\alpha}},m_{f_k},m_{S_{\beta}},m_{f_l})
\end{multline}}
List of particles:
\begin{itemize}
 \item Charged Higgs-slepton / up / sdown / up: couplings from Eqs.(\ref{coup:Hcud}),(\ref{coup:Dud}).
 \item Higgs-sneutrino / down / sup / down: couplings from Eqs.(\ref{coup:H0d}),(\ref{coup:Udd}).
\end{itemize}

\paragraph{Case 3: $f_{k,l}$ colour-octets}

{\scriptsize\begin{multline}
{\cal L}_{\mbox{\tiny EFT}}\ni\frac{1}{32\pi^2}\left\{\frac{1}{18}g_L^{S_{\alpha}f_kd_j\,*}g_L^{S_{\beta}f_kd_j\,*}g_L^{S_{\alpha}f_ld_i}g_L^{S_{\beta}f_ld_i}m_{f_k}m_{f_l}D_0\,O_1+\frac{1}{18}g_R^{S_{\alpha}f_kd_j\,*}g_R^{S_{\beta}f_kd_j\,*}g_R^{S_{\alpha}f_ld_i}g_R^{S_{\beta}f_ld_i}m_{f_k}m_{f_l}D_0\,\tilde{O}_1\right.\\
-\frac{1}{9}g_R^{S_{\alpha}f_kd_j\,*}g_R^{S_{\beta}f_kd_j\,*}g_L^{S_{\alpha}f_ld_i}g_L^{S_{\beta}f_ld_i}m_{f_k}m_{f_l}D_0(O_2+O_3)-\frac{1}{9}g_L^{S_{\alpha}f_kd_j\,*}g_L^{S_{\beta}f_kd_j\,*}g_R^{S_{\alpha}f_ld_i}g_R^{S_{\beta}f_ld_i}m_{f_k}m_{f_l}D_0(\tilde{O}_2+\tilde{O}_3)\\
-\frac{1}{9}\left[g_R^{S_{\alpha}f_kd_j\,*}g_L^{S_{\beta}f_kd_j\,*}g_R^{S_{\alpha}f_ld_i}g_L^{S_{\beta}f_ld_i}+g_L^{S_{\alpha}f_kd_j\,*}g_R^{S_{\beta}f_kd_j\,*}g_L^{S_{\alpha}f_ld_i}g_R^{S_{\beta}f_ld_i}\right]\frac{D_2}{4}(5O_4-3O_5)\\
\left.-\frac{1}{9}\left[g_L^{S_{\alpha}f_kd_j\,*}g_R^{S_{\beta}f_kd_j\,*}g_R^{S_{\alpha}f_ld_i}g_L^{S_{\beta}f_ld_i}+g_R^{S_{\alpha}f_kd_j\,*}g_L^{S_{\beta}f_kd_j\,*}g_L^{S_{\alpha}f_ld_i}g_R^{S_{\beta}f_ld_i}\right]\frac{D_2}{4}(3O_4-5O_5)\right\}(m_{S_{\alpha}},m_{f_k},m_{S_{\beta}},m_{f_l})
\end{multline}}
List of particles:
\begin{itemize}
 \item sdown / gluinos: couplings from Eq.(\ref{coup:Dglud}) (stripped from Gell-Mann matrix element).
\end{itemize}

\paragraph{Case 4: $f_{k,l}$ colour-octet+singlet}

{\scriptsize\begin{multline}
{\cal L}_{\mbox{\tiny EFT}}\ni\frac{1}{32\pi^2}\left\{\frac{1}{6}g_L^{S_{\alpha}f_kd_j\,*}g_L^{S_{\beta}f_kd_j\,*}g_L^{S_{\alpha}f_ld_i}g_L^{S_{\beta}f_ld_i}m_{f_k}m_{f_l}D_0\,O_1+\frac{1}{6}g_R^{S_{\alpha}f_kd_j\,*}g_R^{S_{\beta}f_kd_j\,*}g_R^{S_{\alpha}f_ld_i}g_R^{S_{\beta}f_ld_i}m_{f_k}m_{f_l}D_0\,\tilde{O}_1\right.\\
-\frac{1}{3}g_R^{S_{\alpha}f_kd_j\,*}g_R^{S_{\beta}f_kd_j\,*}g_L^{S_{\alpha}f_ld_i}g_L^{S_{\beta}f_ld_i}m_{f_k}m_{f_l}D_0(O_2+O_3)-\frac{1}{3}g_L^{S_{\alpha}f_kd_j\,*}g_L^{S_{\beta}f_kd_j\,*}g_R^{S_{\alpha}f_ld_i}g_R^{S_{\beta}f_ld_i}m_{f_k}m_{f_l}D_0(\tilde{O}_2+\tilde{O}_3)\\
+\frac{1}{3}\left[g_R^{S_{\alpha}f_kd_j\,*}g_L^{S_{\beta}f_kd_j\,*}g_R^{S_{\alpha}f_ld_i}g_L^{S_{\beta}f_ld_i}+g_L^{S_{\alpha}f_kd_j\,*}g_R^{S_{\beta}f_kd_j\,*}g_L^{S_{\alpha}f_ld_i}g_R^{S_{\beta}f_ld_i}\right]\frac{D_2}{4}(O_4-3O_5)\\
\left.+\frac{1}{3}\left[g_L^{S_{\alpha}f_kd_j\,*}g_R^{S_{\beta}f_kd_j\,*}g_R^{S_{\alpha}f_ld_i}g_L^{S_{\beta}f_ld_i}+g_R^{S_{\alpha}f_kd_j\,*}g_L^{S_{\beta}f_kd_j\,*}g_L^{S_{\alpha}f_ld_i}g_R^{S_{\beta}f_ld_i}\right]\frac{D_2}{4}(3O_4-O_5)\right\}(m_{S_{\alpha}},m_{f_k},m_{S_{\beta}},m_{f_l})
\end{multline}}
List of particles:
\begin{itemize}
 \item sdown / gluino / sdown / neutralino-neutrino: couplings from Eqs.(\ref{coup:Dneud}),(\ref{coup:Dglud}) (stripped from Gell-Mann matrix element); + diagram with $\chi^0\leftrightarrow\tilde{g}$.
\end{itemize}

\subsection{Vector/fermion/scalar/fermion ``straight'' box}\label{subsec:vfsf-stright}

\paragraph{Case $S$ colour-singlet}
{\scriptsize\begin{multline}
{\cal L}_{\mbox{\tiny EFT}}\ni\frac{1}{32\pi^2}\left\{
-g_L^{Vf_kd_j\,*}g_L^{Sf_kd_i}g_L^{Sf_ld_j\,*}g_L^{Vf_ld_i}m_{f_k}m_{f_l}D_0\,O_1-g_R^{Vf_kd_j\,*}g_R^{Sf_kd_i}g_R^{Sf_ld_j\,*}g_R^{Vf_ld_i}m_{f_k}m_{f_l}D_0\,\tilde{O}_1\right.\\
-2g_R^{Vf_kd_j\,*}g_L^{Sf_kd_i}g_R^{Sf_ld_j\,*}g_L^{Vf_ld_i}\,D_2(O_2+O_3)-2g_L^{Vf_kd_j\,*}g_R^{Sf_kd_i}g_L^{Sf_ld_j\,*}g_R^{Vf_ld_i}\,D_2(\tilde{O}_2+\tilde{O}_3)\\
-\left[g_L^{Vf_kd_j\,*}g_R^{Sf_kd_i}g_R^{Sf_ld_j\,*}g_L^{Vf_ld_i}+g_R^{Vf_kd_j\,*}g_L^{Sf_kd_i}g_L^{Sf_ld_j\,*}g_R^{Vf_ld_i}\right]D_2O_4\\
\left.+2\left[g_L^{Vf_kd_j\,*}g_L^{Sf_kd_i}g_R^{Sf_ld_j\,*}g_R^{Vf_ld_i}+g_R^{Vf_kd_j\,*}g_R^{Sf_kd_i}g_L^{Sf_ld_j\,*}g_L^{Vf_ld_i}\right]m_{f_k}m_{f_l}D_0\,O_5\right\}(m_{V},m_{f_k},m_{S},m_{f_l})
\end{multline}}
List of particles:
\begin{itemize}
 \item Z / down / sneutrino-neutral Higgs /down: couplings from Eqs.(\ref{coup:H0d}),(\ref{coup:Zdd}); $\times2$ ($\pi$-rotated diagram).
 \item W / up / charged Higgs-slepton / up: couplings from Eqs.(\ref{coup:Hcud}),(\ref{coup:Wud}); $\times2$ ($\pi$-rotated diagram).
\end{itemize}

\subsection{Vector/fermion/scalar/fermion ``cross'' boxes}\label{subsec:vfsf-cross}

\paragraph{Case $S$ colour-singlet}
{\scriptsize\begin{multline}
{\cal L}_{\mbox{\tiny EFT}}\ni\frac{1}{32\pi^2}\left\{
-\left(g_L^{Vf_kd_j\,*}g_L^{Sf_kd_i}g_L^{Vf_ld_j\,*}g_L^{Sf_ld_i}m_{f_k}m_{f_l}+g_L^{Sf_kd_j\,*}g_L^{Vf_kd_i}g_L^{Sf_ld_j\,*}g_L^{Vf_ld_i}\right)m_{f_k}m_{f_l}D_0\,O_1\right.\\
-\left(g_R^{Vf_kd_j\,*}g_R^{Sf_kd_i}g_R^{Vf_ld_j\,*}g_R^{Sf_ld_i}+g_R^{Sf_kd_j\,*}g_R^{Vf_kd_i}g_R^{Sf_ld_j\,*}g_R^{Vf_ld_i}\right)m_{f_k}m_{f_l}D_0\,\tilde{O}_1\\
-2\left(g_R^{Vf_kd_j\,*}g_L^{Sf_kd_i}g_R^{Vf_ld_j\,*}g_L^{Sf_ld_i}+g_R^{Sf_kd_j\,*}g_L^{Vf_kd_i}g_R^{Sf_ld_j\,*}g_L^{Vf_ld_i}\right)D_2\,O_3\\
-2\left(g_L^{Vf_kd_j\,*}g_R^{Sf_kd_i}g_L^{Vf_ld_j\,*}g_R^{Sf_ld_i}+g_L^{Sf_kd_j\,*}g_R^{Vf_kd_i}g_L^{Sf_ld_j\,*}g_R^{Vf_ld_i}\right)D_2\,\tilde{O}_3\\
+\left[g_L^{Vf_kd_j\,*}g_R^{Sf_kd_i}g_R^{Vf_ld_j\,*}g_L^{Sf_ld_i}+g_R^{Vf_kd_j\,*}g_L^{Sf_kd_i}g_L^{Vf_ld_j\,*}g_R^{Sf_ld_i}\right.\\
\left.\hspace{1cm}+g_L^{Sf_kd_j\,*}g_R^{Vf_kd_i}g_R^{Sf_ld_j\,*}g_L^{Vf_ld_i}+g_R^{Sf_kd_j\,*}g_L^{Vf_kd_i}g_L^{Sf_ld_j\,*}g_R^{Vf_ld_i}\right]D_2\,O_4\\
+2\left[g_L^{Vf_kd_j\,*}g_L^{Sf_kd_i}g_R^{Vf_ld_j\,*}g_R^{Sf_ld_i}+g_R^{Vf_kd_j\,*}g_R^{Sf_kd_i}g_L^{Vf_ld_j\,*}g_L^{Sf_ld_i}\right.\\
\left.\left.\hspace{1cm}+g_L^{Sf_kd_j\,*}g_L^{Vf_kd_i}g_R^{Sf_ld_j\,*}g_R^{Vf_ld_i}+g_R^{Sf_kd_j\,*}g_R^{Vf_kd_i}g_L^{Sf_ld_j\,*}g_L^{Vf_ld_i}\right]m_{f_k}m_{f_l}D_0\,O_5\right\}(m_{V},m_{f_k},m_{S},m_{f_l})
\end{multline}}
List of particles:
\begin{itemize}
 \item Z / down / sneutrino-neutral Higgs / down: couplings from Eqs.(\ref{coup:H0d}),(\ref{coup:Zdd}).
\end{itemize}

\subsection{Vector/fermion/scalar/fermion ``fermion-cross'' box}\label{subsec:vfsf-fcross}

\paragraph{Case $S$ colour-triplet}

{\scriptsize\begin{multline}
{\cal L}_{\mbox{\tiny EFT}}\ni\frac{1}{32\pi^2}\left\{
g_R^{Vf_kd_j\,*}g_R^{Sf_kd_j\,*}g_L^{Sf_ld_i}g_L^{Vf_ld_i}\frac{D_2}{4}(O_2-O_3)-g_L^{Vf_kd_j\,*}g_L^{Sf_kd_j\,*}g_R^{Sf_ld_i}g_R^{Vf_ld_i}\frac{D_2}{4}(\tilde{O}_2-\tilde{O}_3)\right.\\
\left.+2\left(g_L^{Vf_kd_j\,*}g_R^{Sf_kd_j\,*}+g_R^{Vf_kd_j\,*}g_L^{Sf_kd_j\,*}\right)\left(g_R^{Sf_ld_i}g_L^{Vf_ld_i}+g_L^{Sf_ld_i}g_R^{Vf_ld_i}\right)m_{f_k}m_{f_l}D_0(O_4-O_5)\right\}(m_{V},m_{f_k},m_{S},m_{f_l})
\end{multline}}
List of particles:
\begin{itemize}
 \item W / up / sdown / up: couplings from Eqs.(\ref{coup:Wud}),(\ref{coup:Dud}); $\times2$ ($\pi$-rotated diagram).
 \item Z / down / sup / down: couplings from Eqs.(\ref{coup:Zdd}),(\ref{coup:Udd}); vanishes from antisymmetry of $\lambda''$; $\times2$ ($\pi$-rotated diagram).
\end{itemize}



\providecommand{\href}[2]{#2}\begingroup\raggedright\endgroup
\end{document}